\documentclass[5p,aas_macros]{elsarticle}

\usepackage{ifthen}

\newboolean{useminted}
\setboolean{useminted}{false}

\usepackage{array}
\usepackage{lineno}
\usepackage{xcolor}
\usepackage{lineno}
\usepackage[colorlinks=true,urlcolor=black,linkcolor=blue,citecolor=blue]{hyperref}
\ifthenelse{\boolean{useminted}}{\usepackage{minted}}{}
\usepackage{longtable,booktabs,tabu}
\usepackage{amsmath}
\usepackage{multirow}
\usepackage{pbox}
\usepackage{amssymb}
\usepackage{layouts}
\usepackage{subfig}
\usepackage{microtype}
\usepackage{listings}
\usepackage{xspace}
\usepackage[capitalise]{cleveref}
\usepackage[normalem]{ulem}

\usepackage{etoolbox}

\creflabelformat{equation}{#2\textup{#1}#3}

\makeatletter
\patchcmd{\NAT@citex}
  {\@citea\NAT@hyper@{\NAT@nmfmt{\NAT@nm}\NAT@date}}
  {\@citea\NAT@nmfmt{\NAT@nm}\NAT@hyper@{\NAT@date}}
  {}
  {}
\patchcmd{\NAT@citex}
  {\@citea\NAT@hyper@{%
     \NAT@nmfmt{\NAT@nm}%
     \hyper@natlinkbreak{\NAT@aysep\NAT@spacechar}{\@citeb\@extra@b@citeb}%
     \NAT@date}}
  {\@citea\NAT@nmfmt{\NAT@nm}%
   \NAT@aysep\NAT@spacechar%
   \NAT@hyper@{\NAT@date}}
  {}
  {}
\patchcmd{\NAT@citex}
  {\@citea\NAT@hyper@{%
     \NAT@nmfmt{\NAT@nm}%
     \hyper@natlinkbreak{\NAT@spacechar\NAT@@open\if*#1*\else#1\NAT@spacechar\fi}%
       {\@citeb\@extra@b@citeb}%
     \NAT@date}}
  {\@citea\NAT@nmfmt{\NAT@nm}%
   \NAT@spacechar\NAT@@open\if*#1*\else#1\NAT@spacechar\fi%
   \NAT@hyper@{\NAT@date}}
  {}
  {}
\makeatother

\newcommand{\Nc}{\langle N_{\rm c} \rangle}
\newcommand{\Ns}{\langle N_{\rm s} \rangle}
\newcommand{\Nt}{\langle N_{\rm t} \rangle}
\newcommand{\halomod}{\textsc{halomod}\xspace}
\newcommand{\python}{\textsc{python}\xspace}
\newcommand{\thm}{\textsc{TheHaloMod}\xspace}
\newcommand{\hmf}{\textsc{hmf}\xspace}
\newcommand{\framework}{\texttt{Framework}\xspace}
\newcommand{\component}{\texttt{Component}\xspace}
\newcommand{\parameter}{\texttt{parameter}\xspace}
\newcommand{\cached}{\texttt{cached\_quantity}\xspace}

\newcommand{\red}[1]{\textcolor{red}{{#1}}}

\ifthenelse{\boolean{useminted}}{%
	\usemintedstyle{friendly}
	
	\setminted[python]{
	    frame=leftline,
	    framesep=2mm,
	    baselinestretch=1.2,
	    fontsize=\small,
	}
}{}

\definecolor{codegreen}{rgb}{0,0.6,0}
\definecolor{codegray}{rgb}{0.5,0.5,0.5}
\definecolor{codepurple}{rgb}{0.58,0,0.82}
\definecolor{backcolour}{rgb}{0.95,0.95,0.92}

\lstdefinestyle{mystyle}{
	commentstyle=\color{codegreen},
	keywordstyle=\color{magenta},
	numberstyle=\tiny\color{codegray},
	stringstyle=\color{codepurple},
	basicstyle=\ttfamily\small,
	breakatwhitespace=false,         
	breaklines=true,                 
	captionpos=b,                    
	keepspaces=true,                 
	numbersep=5pt,     
	showspaces=false,                
	showstringspaces=false,
	showtabs=false,                  
	tabsize=2
}
\journal{Astronomy and Computing}

\biboptions{authoryear}

\begin{document}

\begin{frontmatter}

\title{\textsc{TheHaloMod}: An online calculator for the halo model}

\author{Steven~G.~Murray$^a$, Benedikt Diemer$^b$, Zhaoting Chen$^c$, Anton Glenn Neuhold Jr.$^d$, Alexander Schnapp$^d$, Tia Peruzzi$^d$, Daniel Blevins$^d$, Trent Engelman$^d$}
\address{$^a$ School of Earth and Space Exploration, Arizona State University, Tempe, AZ, USA \\
$^b$ Department of Astronomy, University of Maryland, College Park, MD 20742, USA\\
$^c$ Jodrell Bank Centre for Astrophysics, School of Physics and Astronomy, The University of Manchester, Manchester M13 9PL, UK \\
$^d$ ASU School of Computing, Informatics, and Decision Systems Engineering, Arizona State University, Tempe, AZ, USA \\
}

\begin{abstract}
The halo model is a successful framework for describing the distribution of matter in the Universe -- from weak lensing observables to galaxy n-point correlation functions. We review the basic formulation of the halo model and several of its components in the context of galaxy 2-point statistics, developing a coherent framework for its application.

We use this framework to motivate the presentation of a new \python tool for simple and efficient calculation of halo model quantities, and their extension to galaxy statistics via a \textit{halo occupation distribution}, called \halomod. This tool is efficient, simple to use, comprehensive and importantly provides a great deal of flexibility in terms of custom extensions. 
This \python tool is complemented by a new web-application at \url{https://thehalomod.app} that supports the generation of many halo model quantities directly from the browser -- useful for educators, students, theorists and observers.

\end{abstract}


\end{frontmatter}

\section{Introduction}


The halo model \citep{Neyman1953,Scherrer1991,Peacock2000,Seljak2000,Ma2000,Cooray2002} is an enormously successful analytical description of the large-scale distribution of matter in our Universe.
It describes the statistics of the dark matter density field well into the nonlinear regime, beyond the reach of perturbation theory. 
It does so by combining linear theory predictions with empirical properties of dark matter halos, via the assumption that the sum total of dark matter resides in these clumps, and that a handful of simple functions based on the mass of these halos -- such as their radial density profile and clustering bias -- can universally describe them.

In combination with a \textit{halo occupation distribution} (HOD) model \citep{Kauffmann1997,Scoccimarro2001,Berlind2003,Zheng2005}, the predictions of the halo model can be extended to galaxy populations, and therefore used to model clustering in large galaxy surveys. 
One of the key advantages of halo model and HOD formalism is that they can predict any clustering statistic on any scale 
\citep{Zehavi2011}, from real-space or projected 2-point correlation functions (2PCFs), to galaxy-galaxy lensing, to higher-order correlations.  

In practice, the HOD formalism has been widely used in the interpretation of galaxy populations in the past decade. Most of these studies have focused on determining the parameters of the HOD (i.e. the galaxy-halo connection) from the 2PCF of galaxies \citep[eg.][]{Moustakas2002,Bullock2002,Zheng2004,Zehavi2005,Blake2008,Zehavi2011,Beutler2013,Skibba2015,Rodriguez-Torres2015,Nicola2020,Zhou2021, Ishikawa2021} using the analytical framework that we present in this paper. However, other observables such as galaxy-galaxy lensing \citep{Mandelbaum2006,Cacciato2012,Clampitt2017,Dvornik2018}, the 2PCF of radio galaxies \citep{Wake2008a,Kim2011,Nusser2015}, galaxy-quasar cross-correlations \citep{Shen2013}, near-UV cross-correlations \citep{Krause2012}, Fast Radio Bursts \citep[FRBs;][]{Rafiei-Ravandi2020} and \textsc{Hi} intensity map cross-correlations \citep{Padmanabhan2016b,Wolz2019,Chen2021} and auto-correlations \citep{Schneider2020,Umeh2021} have also received application through the same framework. Furthermore, several studies have successfully combined observables (either in cross-correlation or via joint likelihoods) to break degeneracies \citep[eg.][]{Leauthaud2011,Leauthaud2012,More2013,Schaan2021}. 
Finally, the combination of the galaxy 2PCF with other statistics, eg. weak-lensing \citep{More2015,Miyatake2020}, cluster counts and auto-correlations \citep{To2021}, or both \citep{To2020}, stellar mass functions \citep{Coupon2015}
 and group mass-to-number ratios \citep{Reddick2014} has resulted in the ability to simultaneously constrain cosmological parameters along with those of the HOD.

Clearly, the halo model (hereafter HM) framework, complemented by an HOD or comparable mechanism, can be of wide utility in the interpretation of large surveys.
Increasingly, observational studies are employing a non-analytic incarnation of the HM, in which simulated halos are ``painted'' with a particular galaxy sample, given detailed semi-analytic models of the galaxy-halo connection \citep[eg.][]{Carretero2015}.
In particular, the \textsc{halotools}\footnote{\url{https://halotools.readthedocs.io}} library has become a popular implementation of this approach.
Nevertheless, a purely analytic construction of the HM is still of great importance and utility: it is our best fundamental model of the non-linear scales of the matter distribution.

The HM is a complex framework as it synthesises many related sub-components (eg. halo profiles, mass functions, bias models, spatial filters, halo exclusion models, concen\-tration--mass relations) to produce spatial statistics. 
These sub-components can often be modelled independently via simulations, and new more accurate models are being produced regularly by the community. 
This highlights the need for an implementation of the HM framework that has the flexibility and modularity to enable easy switching between models for the various sub-components, and rapid development of new models to incorporate into the framework.


There are a few publicly available implementations of the analytic HM, many of which were used in the testing of our code (e.g., \textsc{chomp}\footnote{\url{https://github.com/morriscb/chomp}  -- discontinued.}, \textsc{HMcode}\footnote{\url{https://github.com/alexander-mead/HMcode/}}, \textsc{CCL}\footnote{\url{https://github.com/LSSTDESC/CCL}} and \textsc{AUM}\footnote{\url{https://github.com/surhudm/aum}}).
In addition, methods of computing HM quantities based on emulation from simulations are becoming available (eg. \textsc{DarkQuest}\footnote{\url{https://github.com/DarkQuestCosmology/dark_emulator_public}} and \textsc{CosmicEmu}\footnote{\url{https://github.com/lanl/CosmicEmu}} -- though the latter only supports matter correlations).
However, it is our experience from that a remarkable number of 
practitioners develop their own tools --- whether based heavily on existing (public or private) code or from the ground up. 
Additionally, existing codes -- where still in active development -- do not tend to offer a great deal of flexibility in the composition and extension of 
the various components and models involved in the halo model, rather focusing on a limited set of well-honed models and tuned for maximum performance.



This paper presents a robust new implementation of the analytic HM, called \halomod\footnote{\url{https://github.com/halomod/halomod}},  that aims to fill this important gap, and be as generally useful as possible by adhering to the following principles:
\begin{itemize}
    \item \textbf{Intuitive.}  The API is well-specified and intuitive for the user, and exhaustively documented. We illustrate the simplicity of the usage of \halomod\ in \S\ref{sec:halomod:overview:usage} and note that full online documentation is available at \url{https://halomod.readthedocs.io}, including API specifications and examples/tutorials. In addition, installing \halomod\ is as simple as running \verb|pip install halomod|.
    
    \item \textbf{Simple.} Though many aspects of the calculations are unavoidably non-trivial, a simple layout of the code within a highly structured framework is important. We lay out \halomod's simple code framework in \S\ref{sec:halomod:overview:concept}. This promotes future development, and usage by a broad cross-section of researchers. 
    
    \item \textbf{Efficient. } Though not as immediately important as flexibility, it is important that the code be efficient. This includes both algorithmic and numerical efficiency, but also efficiency of the writing of user-side code.
    We outline our strategies for efficiency in \S\ref{sec:halomod:overview:efficiency}.
    
    \item \textbf{Flexible/Extendible. } The HM is a rapidly evolving framework, with individual components const\-antly improving, and the framework itself being extended. Building a static implementation is therefore non-conducive  to the development of the field. Components need to be as plug-and-play as possible, with new models easily created and inserted on the fly. Our implementation of 
    such a plug-and-play system is outlined in \S\ref{sec:halomod:overview:flexibility}.
    
    \item \textbf{Comprehensive. } \halomod\ acts as an archive for all the modelling that has been done by the community. It collates and compiles the various models and extensions in a cohesive way so that new models can be quickly compared, and insights gained. Our efforts towards this in \halomod\ are evidenced by the numerous tables of models throughout \S\ref{sec:halomod:components}.
    
    \item \textbf{Open. } \halomod\ is open-source, not simply in the sense that it is publicly available. It is developed with many open-source best-practices, such as continuous integration, high test coverage, automated code linting/formatting, formal software versioning, modern version control practices and online documentation. 
\end{itemize}

Both philosophically and technically, \halomod\ inherits from the \hmf\ halo mass function package\footnote{\url{https://github.com/halomod/hmf}} which was first presented in \cite{Murray2013a}.
Many developments have occurred in \hmf\ since its first publication, and the technical framework of \halomod\ presented in this paper is essentially inherited from the updates in \hmf. 
Thus, this paper can also secondarily be considered an update of \hmf.

Our vision is for \halomod\  to be useful as 
(i) a baseline standard for user-specific private codes, 
(ii) a simple interface for those not actively researching in the field, but who may wish to calculate clustering statistics for their data, 
(iii) a tool for fast exploratory analysis, and comparison between models and 
(iv) a stable framework for more rapid development of theoretical extensions to the HM, and modelling of its various components.


In addition to \halomod\ and \hmf\ package, we also present a new web-application, \thm\footnote{\url{ https://thehalomod.app}}, which is able to generate full halo model quantities (eg. 2-point correlation functions and galaxy power spectra) without ever having to install the \python package.
It is a successor to the popular \textsc{HMFcalc} \citep{Murray2013a} web-application, and includes the full range of functionality of \textsc{HMFcalc}.
The presentation of \thm\ completes our vision for \halomod\ with (v) a tool for educators to easily and interactively present cosmologically relevant quantities graphically.

\begin{figure*}
    \centering
    \includegraphics[width=\linewidth,clip=true,trim=0 2cm 0 1cm]{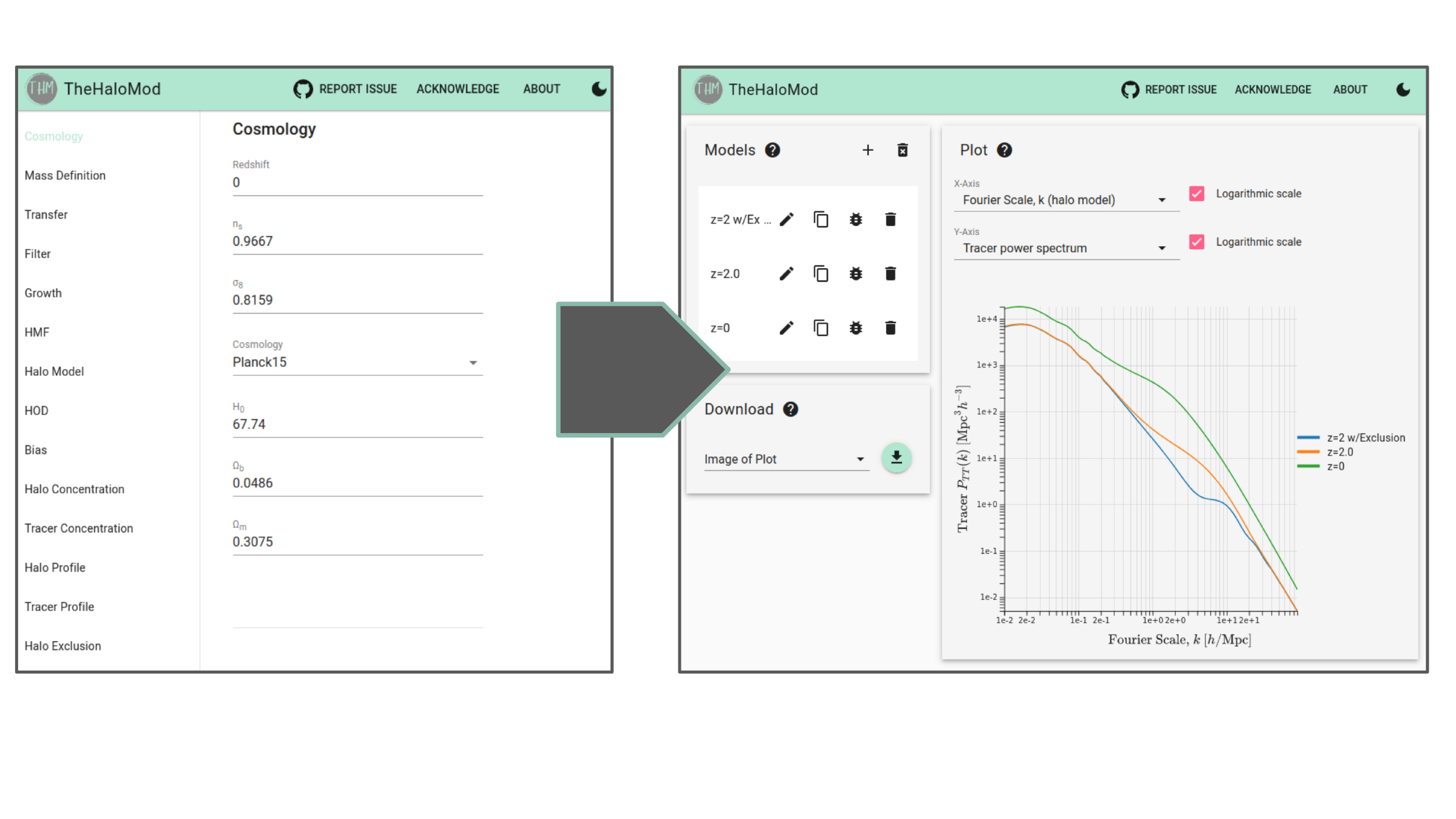} 
    \caption{\textsc{TheHaloMod} in action. The main page (right panel) shows the currently computed models, with a selection of quantities to plot, and a number of actions available for each model (edit/clone/report bug/delete). The plot itself is interactive. The input form (left) provides a wide range of options, including all of the components presented throughout this paper. Each component has its own section. The interface features responsive design and a dark mode.} \label{fig:webpage-image}
\end{figure*}

This paper is structured as follows: \S\ref{sec:DMHaloModel} details the theory of the HM particularly in the context of dark matter 2-point statistics, collating the various components involved in a manner consistent with our implementation. Following this, we describe how to extend the halo model to tracer populations in \S\ref{sec:TracerHaloModel}. Then, we describe our code and its usage in  \S\ref{sec:halomod}, and in \S\ref{sec:applications} we present an illustrative example. In \S\ref{sec:future} we define a prospectus for the future, before summarising and concluding in \S\ref{sec:summary}.

Note that the code to produce all figures of different component models in this paper, as well as the example application, are available publicly as examples in \halomod's documentation\footnote{At \url{https://halomod.readthedocs.io/en/latest/examples/component-showcase.html} and \url{https://halomod.readthedocs.io/en/latest/examples/fitting.html} respectively.}. This paper refers to \textsc{hmf} version 3.4.1 and \textsc{halomod} version 2.1.0.

\section{The Dark Matter Halo Model}
\label{sec:DMHaloModel}
The broad assumption underpinning the HM is that in hierarchical structure formation scenarios, all mass is expected to be bound to halos at some scale\footnote{This assumption is clearly an approximation, since even cold dark matter (CDM) has a free-streaming scale in the early Universe below which we expect non-virialized mass \citep{Frenk2012,Schneider2015}}. If this is the case, then the entire nonlinear density field may be reconstructed by summing contributions from the individual halos. If, in addition, we may describe the average radial density profile of halos as spherically-symmetric, with a shape that depends solely on the mass of the halo itself, we can write:
\begin{equation}
    \label{eq:hm_sum}
    \rho(\mathbf{x}) = \sum \rho_{h,i}(|\mathbf{x} - \mathbf{x_i}|,m_i)
\end{equation}
where $\rho_{h}(r|m)$ is the density of a halo with mass $m$ at radius $r$, and $x_i$ are the coordinates of the halo centres. 

The application of the HM rests in converting the prem\-ise of \cref{eq:hm_sum} into a semi-analytic integration. This requires knowledge of four key components:
\begin{enumerate}
    \item The moments of the density profiles of halos, $\langle \rho(r|m) \rangle$, $\langle \rho^2_{h, i}(r|m) \rangle$, etc.
    \item The abundance of halos of a given mass (termed the halo mass function, or HMF), $n(m)$
    \item The expected overdensity of halos of mass $m$, $\Delta_{\rm h}(\mathbf{x},m)$ in a region with a given overdensity of matter $\delta(\mathbf{x})$, called the \textit{halo bias}.
    \item A model for the spatial distribution of matter, $\delta(\mathbf{x})$.
\end{enumerate}
Furthermore, detailed calculations require the following extra components:
\begin{enumerate}
    \setcounter{enumi}{4}
    \item Additional parameters that describe the halo profile, most
    commonly parameterized as a concentration-mass relation, $c(m)$, linking the shape of a halo profile to its mass.
    \item A model for ``halo exclusion", which accounts for double-counting intra-halo correlations.
    \item To extend the calculations to galaxies, a distribution function for the occupation of halos by galaxies, as a function of halo mass, $N(m)$ (the HOD; cf. \S\ref{sec:TracerHaloModel}). 
\end{enumerate}

Although the halo model can be used to describe the density field at any level of the $n$-point hierarchy, its most common application to date has been to compute 2-point statistics. Thus, in this outline, we shall focus on the framework as it pertains to the 2PCF.
In particular, the 2PCF only requires the first two moments of the density profile, and we will furthermore assume that the second moment is just the square of the average density profile (i.e. we only require the first moment, cf. \citealt{Giocoli2010}).
Our aim in this section is to introduce the theory in a manner conducive to our implementation, so we shall cover each of the six components in turn following the presentation of the core framework.

Throughout the section, we discuss isotropic 2-point statistics, which measure the over-density of pairs of points at a certain scalar separation. This can be formulated either in Euclidean space, as the correlation function $\xi(r)$, or in Fourier-space, as the power spectrum $P(k)$. 
The correlation function is defined as the excess probability of locating a particle at separation $\mathbf{r}$ from a particle at position $\mathbf{x}$ in a given spatial distribution, and is written (for a homogeneous universe)
\begin{equation}
    \xi(\mathbf{r}) = \frac{1}{V}\int_V {\rm d}^3\mathbf{x}\ \delta(\mathbf{x}) \delta(\mathbf{x} - \mathbf{r}),
\end{equation}
where $\delta(\mathbf{x})$ is the overdensity
\begin{equation}
    \delta(\mathbf{x}) = \frac{\rho(\mathbf{x}) - \bar{\rho}}{\bar{\rho}}.
\end{equation}
The power spectrum is merely the Fourier transform of the correlation function, and the standard Fourier convention in cosmology renders it
\begin{equation}
    \xi(\mathbf{r}) = \int \frac{{\rm d}^3 \mathbf{k}}{(2\pi)^3} P(\mathbf{k}) e^{i \mathbf{k}\cdot \mathbf{r}}.
\end{equation}
Under the assumption of cosmic isotropy, the correlation function and power spectrum depend only on the amplitude of $\mathbf{r}$ and $\mathbf{k}$, respectively. Transformation from an (isotropic) power spectrum $P(k)$ to an (isotropic) correlation function $\xi(r)$ can thus be performed using an order-$1/2$ \textit{Hankel} transform, which is the 3D Fourier transform of a spherically symmetric distribution:
\begin{equation}
\label{eq:hankel}
\xi(r) = \frac{1}{2\pi^2}\int_0^\infty P(k)k^2 j_0(kr){\rm d}k,
\end{equation}
where $j_0$ is the zeroth-order \textit{spherical} Bessel function,
\begin{equation}
    j_0(x) = \frac{\sin x}{x}.
\end{equation}
We note that the halo model is \textit{not limited} to such 2-point statistics, but our implementation currently is.

\subsection{Clustering Framework}
\label{sec:haloclustering}
It is convenient to formulate the framework of clustering in two regimes \citep{Seljak2000}, intra- and inter-halo pairs, called the 1-halo and 2-halo terms respectively. These approximately correspond to small- and large-scale structure (i.e. the contribution of each in the opposite regime is negligible), where `small' is sub-megaparsec, and `large' is $\gtrsim5h^{-1}$Mpc. 

Throughout the following, the superscripts $1h$ and $2h$ will be used to denote this segregation. Furthermore, a subscript of $DM$ will denote a dark matter statistic, while $T$ will denote statistics of observable tracers (such as optical or HI-selected galaxies). 

The total power spectrum and 2-point correlation function can be written simply as
\begin{equation}
    P(k) = P^{\rm 1h}(k) + P^{\rm 2h}(k),
\end{equation}
and
\begin{equation}
    \xi(r) = [1 + \xi^{\rm 1h}(r)] + \xi^{\rm 2h}(r).
\end{equation}
We will first describe the calculation of the 1-halo and 2-halo terms, each of which combines multiple physical components, which we will proceed to describe in 
\S\ref{sec:powerspec}-\ref{sec:theory:exclusion}.

\subsubsection{1-halo term}
The pair-counts, proportional to $1+\xi$, within a halo of mass $m$, are given by the self-convolution of the halo profile. Thus the total pair counts, normalised by a random field (specified by the constant $\bar{\rho}^2$) is given by
\begin{equation}
\label{eq:dmcorr1}
1 + \xi_{\rm DM}^{\rm 1h}(r) = \frac{1}{\bar{\rho}^2}\int n(m) \lambda(r|m) {\rm d}m,
\end{equation}
where $\lambda$ is the self-convolution of the halo profile (i.e. the 3D integral of the profile multiplied by itself after a shift of length $r$, which has units of $M^2/V$, cf. \S\ref{sec:profilestheory}), and $n(m)$ is the halo mass function (HMF).

It is common to express this in Fourier-space, since the self-convolution becomes a simple multiplication,
\begin{equation}
    \label{eq:dmpower1}
    P_{\rm DM}^{\rm 1h}(k) = \int n(m) \left(\frac{m}{\bar{\rho}}\right)^2 |u(k|m)|^2 {\rm d}m,
\end{equation} 
where $u(k|m)$ is the mass-normalised Fourier transform of the halo profile (which is unitless and approaches unity as $k\rightarrow0$). This form has the distinct advantage that higher-order correlations merely involve an increase in the exponent of $u$, rather than an intractable multi-dimensional convolution. Conversely, for numerical purposes, if an analytic formulation of the self-convolution exists and the intended quantity is the correlation function, then it is more efficient to use \cref{eq:dmcorr1} directly. 

Note that the behaviour as $k\rightarrow0$ is to tend to a constant:
\begin{equation}
    P_{\rm DM}^{\rm 1h}(k\rightarrow0) = \frac{1}{\bar{\rho}^2} \int n(m) m^2\ {\rm d}m.
\end{equation}
This ``shot-noise-like'' term \citep{Cooray2002,Ginzburg2017} arises due to exclusion features of the halo model, and is spurious. 
At the largest scales, it comes to dominate the 2-halo term.
While true shot-noise is inversely proportional to the \textit{number density} of the field (cf. \S\ref{sec:theory-gal:framework}), in this case we are considering the dark matter field to be essentially continuous (i.e. the mass of each DM `particle' approaches zero, and its number density approaches infinity), and it therefore should have no shot-noise. 
We thus offer an option \texttt{force\_1halo\_turnover} in \textsc{halomod} that forcibly removes this large-scale structure from the 1-halo term.

\subsubsection{2-halo term}
\label{sec:theory:2halo}
The two-halo term is most elegantly expressed in Fourier-space, 
\begin{align}
    P_{\rm DM}^{\rm 2h}(k,r) =  \int \int & {\rm d}m_1 {\rm d}m_2\  I_{\rm DM}(k,m_1) I_{\rm DM}(k,m_2) \nonumber \\ 
    &\times P_{\rm hh}(k,r|m_1,m_2),
\end{align}
where
\begin{equation}
    I_{\rm DM}(k,m) = n(m) u(k|m) \frac{m}{\bar{\rho}}.
\end{equation}
This equation can be confusing at first glance because it seems to be a function of both $k$ and $r$ -- which are themselves not independent variables. We shall resolve this confusion presently.
$P_{\rm hh}(k,r|m_1,m_2)$ is the power spectrum of halo centres for halos of mass $m_1$ and $m_2$. This is in general a complicated function of mass and scale, but the most common approach is to approximate it by a first-order linear bias of the matter power,
\begin{equation}
    \label{eq:halo_centre_power}
    P_{\rm hh}(k,r|m_1,m_2) \approx b(m_1,r)b(m_2,r)P_{\rm m}(k),
\end{equation}
where $b(m,r)$ is the first-order bias at $m$ (cf. \S\ref{sec:biastheory}), possibly with a scale-dependent correction, and $P_{\rm m}(k)$ is the matter power spectrum (cf. \S\ref{sec:powerspec})\footnote{Note that throughout, we use a subscript $m$ to denote either the linear matter power or the nonlinear matter power derived using analytic corrections, as opposed to the full halo model.}. Depending on the application, it is not uncommon \citep[eg.][]{Cooray2002, Smith2011} to adopt for $P_{\rm m}$ empirical nonlinear corrections on the linear matter power spectrum from \textsc{halofit} \citep{Smith2003}\footnote{This is slightly circular, as the \textsc{halofit} model itself is inspired by the halo model (and is tuned to emulate the clustering of dark matter within and between halos).}. 
The motivation for using the nonlinear matter power here is to obtain better agreement with simulations at quasi-linear scales, where the power is suppressed by a number of effects, including mutual exclusion of individual halos and the breakdown of linear theory (understandably) at the precise scales at which halos are forming \citep{Mead2021}.
Unfortunately, these somewhat circular corrections are not self-consistent -- adding contributions from the 1-halo term into the power for the 2-halo term.
Thus corrections made in the scale-dependent bias (which fundamentally should account for the halo exclusion, cf. \ref{sec:theory:exclusion}) are then forced to do two jobs: to model the true effect of halo exclusion, and to undo the fudged nonlinear corrections.
In principle, it is better to keep these effects separate, and deal self-consistently with the latter when modelling $P_{\rm m}$.
Thus, studies of pure matter correlations most often adopt either the linear matter power spectrum, or corrections from perturbation theory or similar \citep[eg.][]{Seljak2015,Mead2021}.
These extensions partially correct the quasi-linear regime in a self-consistent way. 
Still other studies use emulated halo-halo power spectra directly, and thus remove the need to model the one-to-two-halo transition entirely \citep[eg.][]{Nishimichi2021}.
We leave its definition free in our implementation (with a default of `linear'), and caution the user to pay close attention to its value for any particular application.

The general form for the 2-halo term then becomes
\begin{align}
\label{eq:dmpower2a}
    P_{\rm DM}^{\rm 2h}(k,r) =  \int\int & I_{\rm DM}(k,m_1) I_{\rm DM}(k,m_2) \nonumber \\
    &\times b(m_1,r)b(m_2,r)P_{\rm m}(k) {\rm d}m_2 {\rm d}m_1
\end{align}
This is separable, so long as the integration limits are not inter-dependent (this is not the case in general, cf. \S\ref{sec:theory:exclusion}), and in this simplest case (where we do not attempt to exclude correlations within halos) we have
\begin{equation}
    \label{eq:dmpower2}
    P_{\rm DM}^{\rm 2h}(k,r) = P_{\rm m}(k) \left[ \int  I_\text{DM}(k,m) b(m,r) {\rm d}m \right]^2.
\end{equation}
The right-most factor is often called the ``effective bias'' of matter, and under the halo model ansatz must by definition approach unity as $k\rightarrow0$ (since matter is not biased against itself at large scales). In practice, several effects can artificially inhibit this behaviour, and we provide mechanisms for dealing with these in \textsc{halomod} (cf. \S\ref{sec:halomod:frameworks:base}).

Note that all of these expressions denote the 2-halo term as a function of both $k$ and $r$ (as opposed to the 1-halo term which is just a function of $k$). 
This arises either due to the appropriate limits of the integration being dependent on the real-space size of halos, or because the halo bias may be scale-dependent. 
In cases when either of these is true, to recover the power spectrum as a function of $k$ only, we Hankel-transform to $\xi(r)$ and then back to $P(k)$.

\subsection{Matter Power Spectrum}
\label{sec:powerspec}
The matter power spectrum $P_{\rm m}(k)$ characterises the distribution of matter 
density perturbations as a function of wavenumber $k$.
When density fluctuations are small, so that $\delta \ll 1$, the statistical properties of the perturbations can be derived analytically via perturbation theory. 
Generally, it is assumed that the fluctuation spectrum was initially scale-free (i.e. a power-law), and the present day (linear) power spectrum is modified by a transfer function which captures the effects of the transition from a radiation-dominated to matter-dominated Universe \citep{Bond1984}:
\begin{equation}
	\label{eq:powerfromtransfer}
	P_\text{m,lin}(k) = Ak^n T^2(k),
\end{equation}
where $A$ is the normalisation constant and $n$ is the spectral index. 
We follow convention and use the
parameter $\sigma_8$, which measures the mass variance on a scale of $8 h^{-1}$Mpc at $z=0$, 
to calculate $A$. The transfer function is particularly sensitive to the nature of
the dark matter and the baryon density parameter $\Omega_\text{b}$.

A commonly used dimensionless equivalent can be defined as 
\begin{equation}
    \Delta^2(k) = \frac{k^3}{2\pi^2} P(k),
\end{equation}
which is the contribution to the variance in logarithmic wavenumber bins.

The CDM transfer function has been modelled extensively in the literature. 
A number of functional forms have been proposed \citep[][hereafter EH]{Bond1984,Bardeen1986,Eisenstein1998}, along with a series of \textit{Boltzmann} codes, which calculate the transfer function from first principles with perturbation theory techniques \citep{Zaldarriaga2000,Lewis2000,Blas2011}. 

\begin{figure}
  \centering
  \includegraphics[width=\linewidth]{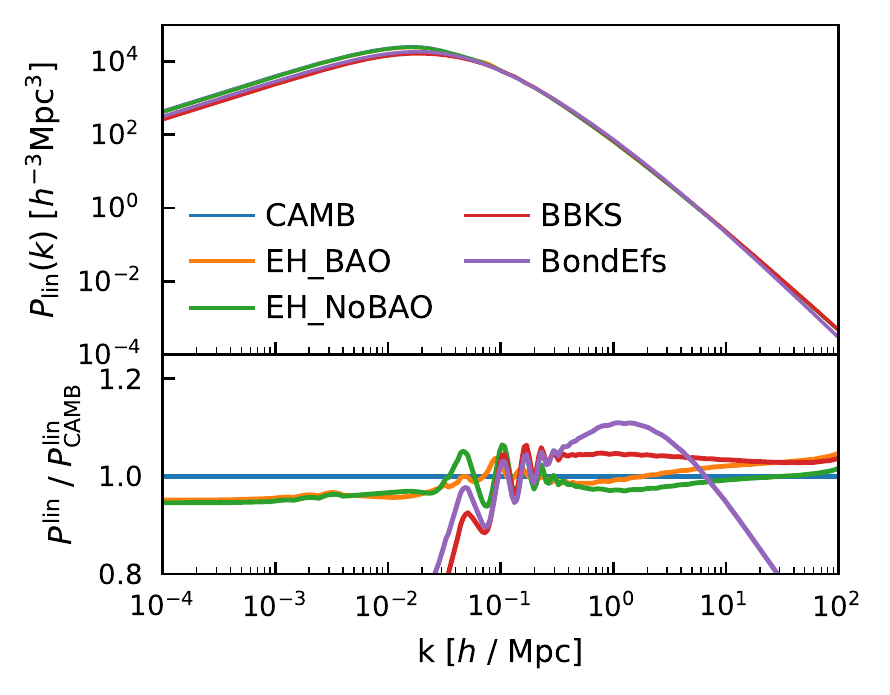}
  \caption[Comparison of transfer function models]{A selection of transfer function models, shown as linear power spectra, including analytic approximations \citep{Bond1984,Bardeen1986,Eisenstein1998} and Boltzmann solutions. Details for each model (referenced to the alias given in the legend) can be found in Table \ref{tab:models_transfer}. The general shape is reproduced in each case, with details of baryon acoustic oscillations differing. In the bottom panel, we show the relative fraction compared to the CAMB transfer function, which renders all models without baryons as having oscillations on the acoustic scales. Note that the models are normalized to the same value of $\sigma_8$ (though some models, like BBKS, appear to be low across most of the range, their brief excursion above CAMB in the centre is highly weighted).}
  \label{fig:power}
\end{figure}

\cref{fig:power} shows the typical shape of the CDM linear power spectrum, while highlighting differences between forms for $T(k)$ found in the literature. Both small- and large-scales have small fluctuations, with a characteristic turnover at scales of $\sim 0.02h/$Mpc. The low-$k$ form asymptotes to $P(k \ll 0.02 ) \propto k^{n_s}$, while the high-$k$ behaviour is more complicated, not obeying a power-law relation. Only the Boltzmann code is able to fully capture the wiggles introduced by the baryon acoustic oscillations (BAOs), though the EH form including baryons closely follows up to $k \sim 2 h^{-1}{\rm Mpc}$. 
In fact, all curves (except that of \citealt{Bond1984}) are within $\sim 10\%$ over the plotted range.

The calculation of the halo-centre power spectrum is commonly approximated by \cref{eq:halo_centre_power}, i.e.
\begin{equation*}
    P_{\rm hh}(k,r|m_1,m_2) \approx b(m_1,r)b(m_2,r)P_{\rm m}(k),
\end{equation*}
 in which an estimate of the matter power spectrum is linearly biased. As previously discussed, it is not uncommon to use the non-linear power spectrum from \textsc{halofit} to model $P_m$. We do not present the details here, except to mention that the modifications are applied on large scales as ratios of $P_\text{m,lin}(k)$, and on smaller scales via empirical fitting functions, which are obtained via arguments motivated by the halo model. These modifications are expressed as a series of equations with coefficients fitted to numerical simulations. Updated coefficients were introduced in \citet{Takahashi2012}. Note also that the normalisation of $P_\text{m,nl}(k)$ is ident\-ical to $P_\text{m,lin}(k)$, and is calculated using the linear form, since $\sigma_8$ refers to the linear power spectrum. The effect of introducing non-linearities is to increase small-scale power, with the recent coefficient updates slightly enhancing this effect.

\subsection{Mass Definition}
\label{sec:theory:massdef}
The halo model centres on the `halo' as its basic unit.
One of the most fundamental properties of the halo is its mass, which is required to define the density profile, concentration-mass relation and its bias with respect to background matter, but there is no unique way in which to define this property. 

The two dominant approaches to defining the halo mass are the \textit{Friends-of-Friends} (FOF) and \textit{Spherical Overdensity} (SO) definitions. 
The FOF definition is useful in the context of simulations, in which we have known particle locations. These particles are linked together into a halo if their separation is less than some chosen linking length, and the fully linked network provides a halo of arbitrary shape.

The SO algorithm finds peaks in the density field and grows spheres around them until the contained density drops below a threshold density.
All particles within the final sphere are considered part of the halo. 
This definition is more useful in the context of observations (of clusters, for example) as it can be applied to the measured (projected) density fields\footnote{Note that the 3D SO definition is not \textit{immediately} applicable to observations without forward-modelling and scaling relations, but they are far more applicable than FoF masses.}. 

Within each definition approach exists the potential for an infinite number of explicit definitions. The FOF approach has the linking length parameter (typically denoted $b$), whilst the SO approach has the overdensity threshold, $\Delta_{\rm h}$. Moreover, the conversion between FOF and SO masses is difficult because a linking length does not uniquely correspond to an overdensity \citep{More2011}.

Indeed, even conversion of relevant quantities (eg. mass functions, concentration-mass relations and bias functions) between different SO definitions is only ever approximate, due to practical difficulties in identifying halos in simulations. 
Accuracy of such conversions is discussed for concentration-mass relations in \citet{Diemer2015}, who give an uncertainty of 15-20\% at high mass, and for mass functions in \citet{Bocquet2016}, whose fits imply inaccuracies of 20-40\% below halo masses of $10^{15}M_\odot h^{-1}$.

While it is thus difficult to compare, even in principle, between FOF and SO definitions, comparison of halo model quantities between explicit SO definitions is approximately possible as long as one has a model for the density distribution within a halo (cf. \S\ref{sec:halomod:components:profile}).
Armed with an average spherical halo profile, one can compute the average mass of a halo under a different density threshold (clearly higher density thresholds resulting in lower masses) (cf. \S\ref{sec:halomod:components:mass-def}). 
We will use the standard symbols $m$ and $R$ for the halo mass and radius throughout this paper, typically without specifying which definition is being applied (most equations are agnostic of this definition).

\subsection{Variance and Spatial Filter}
\label{sec:theory:filter}
The halo model depends on knowing the overdensity (and variance) associated with a position $\vec{x}$. 
The mass variance is defined over some smoothing scale, $R_{\rm L}$:
\begin{equation}
	\label{eq:massvariance}
	\sigma^2(R_{\rm L}(m)) = \frac{1}{2\pi^2}\int_0^\infty{k^2P_{\rm m, lin}(k)W^2(kR){\rm d}k}.
\end{equation}
$W(x)$ here denotes a spatial filter.
In particular, halos arise from patches in the primordial density field with some shape, described by a function $w(\mathbf{x})$.
While the shape of the filter may be arbitrary (and should be matched to the typical shapes of the primordial patches that give rise to the halos \citep{Dalal2010,Chan2017,Diemer2019}, the most common choices are a spherical top-hat (either in real or Fourier space) or a Gaussian.
Under the assumption that the chosen filter is spherically symmetric (not the case in general, but assumed in \halomod), the Fourier-transform of the filter is given by the spherical-hankel transform:
\begin{equation}
    W(k) = 4\pi \int_0^\infty w(r) r^2 j_0(kr) {\rm d}r.
\end{equation}

Having defined a spatial filter on the primordial density field, one can proceed to define the Lagrangian radius, $R(m)$, which is the radius of a spherical region of homogeneous space containing the mass $m$ (or more generally, the scale of a particular filter function required to contain the mass, cf. \S\ref{sec:halomod:components:window}). For the standard top-hat, this is
\begin{equation}
    R_{\rm L}(m) = \left(\frac{3m}{4\pi\bar{\rho}_0}\right)^{1 / 3},
\end{equation}
with $\bar{\rho}_0$ the mean density of the Universe at the present epoch.

The ``mass variance'', $\sigma^2(R_{\rm L})$ is thus the variance on a scale from which the matter in halos is thought to have originated. Notably, the mass variance in spheres of Lagrangian radius 8 Mpc/$h$ yields the power spectrum normalisation, $\sigma_8$.

\subsection{Mass Function}
\label{sec:massfunction}
The mass function quantifies the number of dark matter halos per unit
mass per unit comoving volume of the Universe. 
The most commonly adopted theoretical formulation of the mass function is via the extended Press-Schechter (EPS) formalism \citep{Press1974,Bond1991}, in which density peaks form through gravitational instability and evolve into halos when they cross a density threshold, $\delta_c$. 
In this formalism, the mass function is a universal function of the peak-height, $\nu = \delta_c/\sigma$\footnote{Note that some authors use $\nu = \delta^2_c/\sigma^2(m)$, i.e. the square of our definition.}, which captures the variation from cosmological parameters and the redshift \citep[but note that several recent studies find significant non-universality with redshift, and have provided fits that explicitly account for this departure]{Jenkins2001}. 

The mass function can be defined as
\begin{equation}
\label{eq:hmf}
  n(m) \equiv \frac{{\rm d} n}{{\rm d}m} = \frac{\rho_0}{m^2} f(\sigma) \left|\frac{{\rm d}\ln\sigma}{{\rm d}\ln m}\right|,
\end{equation}
or using the peak-height, as
\begin{equation}
    \label{eq:theory:hmf_nu}
    \frac{{\rm d}n}{{\rm d}m} = \frac{\rho_0}{m} f(\nu) \frac{{\rm d} \nu}{{\rm d} m}.
\end{equation}
For an Einstein-de Sitter universe (in which the matter density equals the critical density and there is no dark energy), $\delta_c \approx 1.686$, and its value is very weakly dependent on cosmology\footnote{Note that even small changes in $\delta_c$ can affect the high-mass end of the HMF significantly. However, many authors assume $\delta_c = 1.686$ regardless of cosmology and thus many fits of the HMF are defined under this assumption.}
The function $f(\sigma) \equiv \nu f(\nu)$, denoting the fraction of matter collapsed into halos of peak-height in a logarithmic bin around $\nu$, has been modelled extensively in the literature, and is generally fit to numerical simulations\footnote{Note that some authors use $f(\nu)$ to describe what we have defined as $\nu f(\nu)$.}. 

It can be advantageous (but not strictly necessary) to choose a fitting function which is normalisable, ie.
\begin{equation}
    \label{eq:fnu1}
    \int f(\nu) {\rm d}\nu = 1.
\end{equation}
Several fits in the literature obey this relation \citep[eg.][]{Sheth2001,Peacock2007,Tinker2010},  which ensures that all mass is bound in halos. This is particularly attractive for HM calculations, in which it is consistent with the basic assumptions of the framework.

Note also that from \cref{eq:theory:hmf_nu} we can determine $d\nu$ as a function of $dm$, and integrals of the form $\int n(m) g(m) dm$ may be re-written
\begin{equation}
    \label{eq:integrate_over_nu}
    \int g(\nu(m)) \frac{\rho_0}{m} f(\nu) {\rm d}\nu.
\end{equation}
In particular, the sum of all mass in the universe requires setting $g(m) = m$, and the requirement that the mass density in halos equal $\rho_0$ is easily seen to be equivalent to the condition \cref{eq:fnu1}. 
Integrating \cref{eq:integrate_over_nu} can often be numerically advantageous, since relevant bounds on $f(\nu)$ can often be found independently of cosmology and redshift.

\begin{figure}
  \centering
  \includegraphics[width=\linewidth]{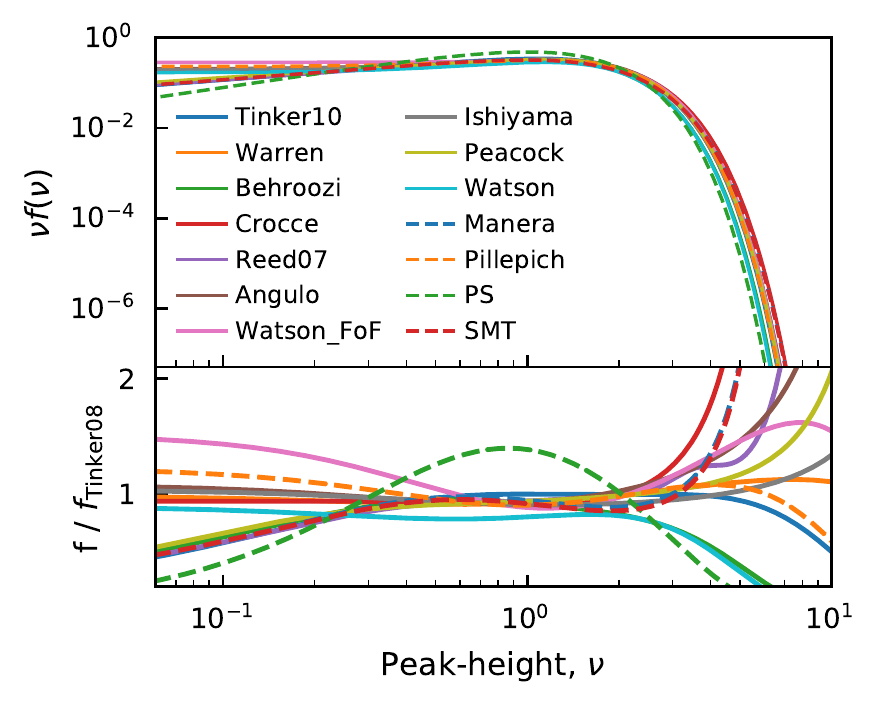}
  \caption[Several HMF fits]{Several HMFs from the literature, including the `original' form of \protect\citet[][`PS']{Press1974}, motivated by spherical collapse, and \textsc{SMT} motivated by ellipsoidal collapse from \protect\citet{Sheth2001}. Note that the mass definitions and cosmologies assumed by the fits are rather disparate -- this is not meant to be an apples-to-apples comparison. While $\nu f(\nu)$ is almost independent of cosmology \citep{Jenkins2001}, and in principle independent of mass definition (peak-height is based on the initial Lagrangian field), in practice $f(\nu)$ is measured from simulations which must employ some halo-finding algorithm and mass definition.}
  \label{fig:mfs}
\end{figure}

\cref{fig:mfs} shows the HMF for several of the fits available in \textsc{halomod} (via \textsc{hmf}). According to these fitting functions, the HMF behaves as a power-law at low masses, with an exponential cut-off above a typical mass $m_* \approx 5\times10^{14}$ (this mass changes with redshift). There is clearly significant variations between the fits, especially at high masses, for a range of reasons, including halo finder differences, alternate mass definitions, and simulation resolution \citep[cf.][]{Murray2013a,Knebe2011}. 

\subsection{Bias}
\label{sec:biastheory}
It is well-known that dark matter halos tend to trace the underlying density field in a biased manner \citep[eg.]{Cole1989}. In principal, the form of this bias can be of arbitrary order (as a function of the matter overdensity), non-local (i.e. dependent on the surrounding environment) and stochastic. However, it is most common within studies involving the halo model to consider only the first-order, local, deterministic bias as a function of the halo mass \citep{Mo1996}: 
\begin{equation}
    \Delta_{\rm h}(\bar{x},m) = b(m) \delta(\bar{x}).
\end{equation}
Peaks theory is able to make predictions of the form of $b(m)$, by considering the probability of collapse of a region of given smoothing scale $m$ in the context of an altered density threshold $\delta_c - \delta(\vec{x})$, often called the peak-background split method \citep[for more details, see eg.][]{Bond1991,Zentner2007,Tinker2010,Manera2010}. These predictions are specified in terms of the peak-height parameter $\nu$, and are therefore expected to exhibit universality with respect to cosmology and redshift.

In the spherical collapse paradigm, the bias is intimately related to the mass function \citep{Cole1989,Cooray2002}, 
\begin{equation}
    b(m) = 1 - \frac{\partial}{\partial \delta_c} \ln n(m),
\end{equation}
which when combined with \cref{eq:hmf}, gives
\begin{equation}
    b(m) = 1 - \frac{1}{\delta_c} - \frac{\partial}{\partial \delta_c} \ln f(\nu).
\end{equation}
In this formalism, if all mass is contained within halos, according to \cref{eq:fnu1}, then by construction, the following consistency relation holds,
\begin{equation}
\label{eq:consistency}
    \int b(\nu) f(\nu) {\rm d}\nu = 1, 
\end{equation}
which merely says that the average bias is unity, so that matter is not biased against itself\footnote{Non-unity average bias in the context of the halo model would be an inconsistency, as all halos contain all the mass, and the mass cannot be biased, on average, against itself.}. 
Note that if Eq. \ref{eq:consistency} holds, the large-scale ($k\rightarrow0$) matter power is gauranteed to be equivalent to $P_m$ (which just says that the effective bias at large scales is unity).

In the approximation of spherical collapse, under the original model of \citep{Press1974}, we have $f(\nu) \propto \exp(-\nu^2/2)$, which yields a bias of  \citep{Cole1989,Mo1996}
\begin{equation}
    b(\nu) = 1+ \frac{\nu -1}{\delta_c}.
\end{equation}
The general qualities of the spherical collapse form are (recall that $\nu = 1$ at $M_\star \sim 5\times10^{12}$ at $z=0$, and corresponds the mass of halos that are, on average, collapsing at a given redshift): 
\begin{itemize}
    \item  a transition from low-mass anti-bias to high-mass bias at $\nu = 1$
    \item a small-scale asymptotic limit of $b = 1 - 1/\delta_c \approx 0.4$
    \item a large scale asymptotic behaviour of $b \propto \nu$. 
\end{itemize}

Given the simplistic nature of the spherical collapse model, it is remarkable that this bias function roughly reproduces the results of $N$-body simulations. However, in detail, this form over-predicts the bias at high masses ($\nu >\sim1.5$), and consequently slightly under-predicts the bias for low-mass halos. 

The approximations induced by the spherical collapse assumption have been relaxed in several studies, most notably by \citet{Sheth2001}, who obtain a form motivated by ellipsoidal collapse, but calibrated with simulations. Forms motivated by stochastic barriers \citep{Corasaniti2011}, non-Markovian barriers \citep{Ma2011} and local density-maxima \citep{Paranjape2013a} are more recent developments.

Unfortunately, the spherical collapse paradigm leads to an artificial enhancement of the bias for halos with $\nu \lesssim 1$ \citep{Manera2010,Tinker2010}, due to the fact that low-mass halos are increasingly elliptical. This has resulted in a proliferation of direct empirical fits to high-resolution $N$-body simulations \citep{Jing1998,Jing1999,Seljak2004,Tinker2005,Mandelbaum2005,Pillepich2010}.
Motivated by the broad success of spherical collapse, these fits are usually \citep[but see][]{Jing1998,Seljak2004} still formulated in terms of the peak-height (indeed, often they utilise the forms yielded by the peak-background split, with updated parameters based on direct fits to simulations).

\begin{figure}
  \centering
  \includegraphics[width=\linewidth]{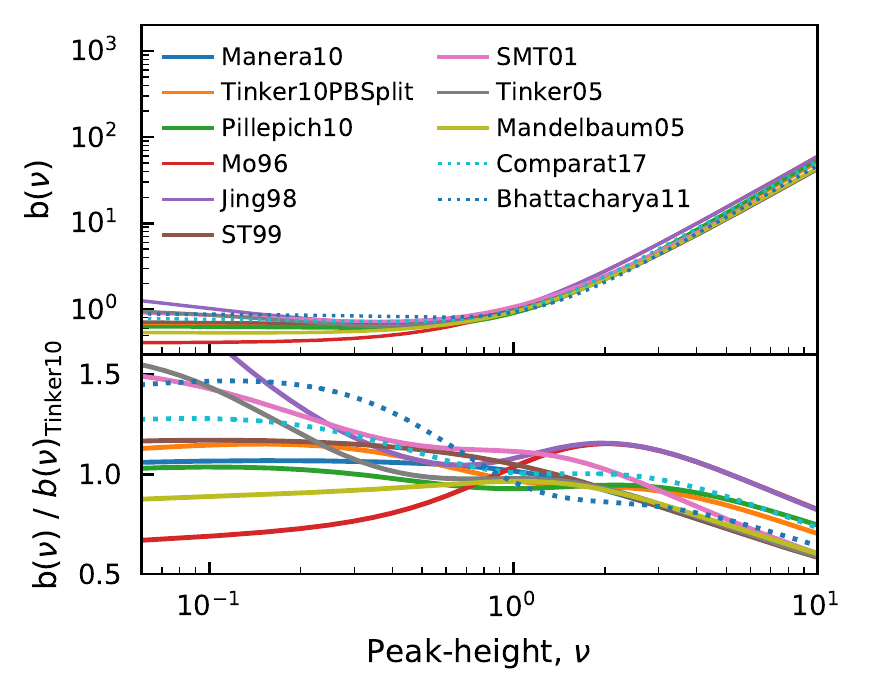}
  \caption[Several bias functions from the literature]{Several linear bias functions from the literature.Most functions, including \protect\citet{Mo1996}, \protect\citet{Sheth1999}, \protect\citet{Sheth2001} and those with the same form \protect\citep{Mandelbaum2005,Tinker2005} have high-mass asymptotic behaviour of $b \propto \nu$. The (default) form of \protect\citet{Tinker2010} is an empirical fit with an arbitrarily chosen form, and has a high-mass asymptotic limit of $b \propto  \nu^{1.2}$. The models plotted in dotted lines are from \textsc{colossus}, called by \halomod\ using its explicit interface}
  \label{fig:bias_functions}
\end{figure}

In \cref{fig:bias_functions} we show a sample of linear scale-independent bias relations found in the literature and included in \halomod. We note the similarity of high-mass behaviour, though the high-mass amplitude is substantially modified between several of the forms. 

\subsubsection{Scale-dependence}
\label{sec:theory:bias:sd}
In reality, the assumption that the bias is a function only of halo mass, to the exclusion of environment, is overly simplistic \citep{Sunayama2015}. 
The bias depends not only on the masses of the paired halos, but also the scale over which we consider the correlations. 
The scale-dependence of halo bias is a complex phenomenon, which has received a great deal of attention \citep[eg.][]{Paranjape2013,Lapi2014,Poole2015}. Foregoing these complexities, simple empirical models specify the scale-dependence as a function $S(\xi_{\rm m}(r))$,
\begin{equation}
    b^2(m,r) = b^2(m) S(\xi_{\rm m,nlin}(r)).
\end{equation}
An example form for $S$ is given by \citet{Tinker2005} as
\begin{equation}
	\label{eq:sdb}
	S(r) = \frac{\left[1+1.17\xi_{\rm m,nlin}(r)\right]^{1.49}}{\left[1+0.69\xi_{\rm m,nlin}(r)\right]^{2.09}}.
\end{equation}
At large scales, where $\xi \ll 1$, the scale-dependence is negligible, but smaller scales can receive significant ($\sim 50\%$) suppression. 
The correlation function used here is the non-linear matter correlation function from \textsc{halofit}.


\subsection{Halo Density Profiles}
\label{sec:profilestheory}
It is widely accepted that the density profiles of DM halos are more or less self-similar if they are rescaled by a suitable central density and radial scale. These two parameters can then be expressed more intuitively as a mass and a concentration. In this section, we define our nomenclature for density profiles. Our general, systematic formulation enables the user to elegantly insert specific functional forms. In particular, the average density profile follows the form
\begin{equation}
    \label{eq:universalprofile}
    \rho(r|m,z) = \rho_s(c, \bar{\rho}) \times f(r/r_s),
\end{equation}
where $\rho(r|m,z)$ has units of density, and $c(m,z) = r_{\Delta_{\rm h}} / r_s$ is the \textit{concentration parameter}, which we shall treat as an explicit function of the halo mass, redshift, and cosmology (although it exhibits appreciable scatter around any such relationship). Here, $\bar{\rho}$ is the mean matter density of the universe at redshift $z$.  The \textit{scale radius} $r_s$ is often defined as the radius where the logarithmic slope of the profile is $-2$, but other definitions are possible in principle. The factor $\rho_s$ controls the amplitude of the profile and is set by the known mass given the halo's radius (here we use halo radius generically as $r_{\Delta_{\rm h}}$, where $\Delta_{\rm h}$ is the halo overdensity with respect to the background). In particular, it is always true that the halo radius and mass are related by
\begin{equation}
    \label{eq:mass-radius-halo}
    m = \frac{4\pi r^3_{\Delta_{\rm h}}}{3} \bar{\rho} \Delta_{\rm h}.
\end{equation}
We may consider halos to have either a `hard' or `soft' edge; in the former case, the extent of the halo is precisely at $r_{\Delta_{\rm h}}$, and the density abruptly falls to zero beyond this. In the latter case, the halo extends to infinity. The latter case is only well-defined in the case that the total integrated mass converges. The HM generally considers halos with a hard boundary, but \halomod\ includes models for both, and we will present the equations for `non-truncated' halos as well.

While $\rho_s$ does not directly depend on mass in our formulation, the dependence on the mean density and the secondary dependence of $c$ on mass translate into a relation where more massive and earlier-forming halos have higher central densities. The $f(r/r_s)$ term controls the shape of the profile and is a function of the scaled radius $x=r/r_s$. Throughout the following discussion we shall often refer to a `reduced' form of the profile (or related quantities) in which simple normalisation factors are removed. An example of such a reduced form is $f$ in \cref{eq:universalprofile}.


\subsubsection{Truncated Halos}
For halos truncated at the halo radius, the amplitude of the profile is set by the total mass inside the halo radius, which we recover by integrating \cref{eq:universalprofile}:
\begin{equation}
\label{eq:minrv}
m = 4\pi \rho_s(c) r_s^3 \int_0^c x^2 f(x) {\rm d}x,
\end{equation}
where we have made the substitution $r = xr_s$ in which $x(r_{\Delta_{\rm h}}) = c$.
If we set
\begin{equation}
\label{eq:hofc}
h(c) = \int_0^c x^2 f(x) {\rm d}x
\end{equation}
then we obtain
\begin{equation}
\label{eq:hofm}
\rho_s(c) = \frac{m}{4 \pi r_s^3 h(c)} \equiv \frac{c^3 \Delta_{\rm h}}{3h(c)}\bar{\rho}
\end{equation}
Thus for all profiles, $h(c)$ uniquely defines the amplitude.

The calculation of 2-point statistics involves two more functions derived from the halo density profile: the normalised Fourier transform $u(k|m,z)$ and the self-convolution $\lambda(r|m,z)$. The normalised Fourier transform of the halo profile is defined as \citep{Cooray2002}
\begin{equation}
\label{eq:ugeneral}
u(\mathbf{k}|m,z) = \frac{1}{m}\int \rho(\mathbf{r}|m,z) e^{-i\mathbf{r}\cdot\mathbf{k}}{\rm d}^3\mathbf{r}
\end{equation}
in which $\mathbf{k},\mathbf{r}$ are the Cartesian wave-vector and position vector respectively.
In the case of a spherically symmetric distribution truncated at the halo radius (as we assume) this may be treated by a Hankel transform, in which we use $\kappa = kr_s$, and $x$ as previously defined:
\begin{equation}
u(\kappa|m,z) = \frac{4\pi r_s^3}{m}\rho_s(c)\int_0^c x^2 \frac{\sin(\kappa x)}{\kappa x}f(x) {\rm d}x.
\end{equation}
Using \cref{eq:hofm} we have
\begin{equation}
\label{eq:uspherical}
u(\kappa |m,z) = \frac{p(\kappa,c)}{h(c)}
\end{equation}
where
\begin{equation}
\label{eq:pofck}
p(\kappa,c) = \int_0^c x^2 \frac{\sin(\kappa x)}{\kappa x}f(x) {\rm d}x
\end{equation}
is the reduced form.

The self-convolution of the density profile is
\begin{equation}
\label{eq:simplifiedlam}
 \lambda(x|m,z) = \rho_s^2(c) r_s^3 l(x,c)
\end{equation}
and has units of $M^2/V$. The reduced form is given by
\begin{equation}
\label{eq:lofxc}
l(x,c) = \int f(\mathbf{y},c)f(|\mathbf{y}+\mathbf{x}|,c){\rm d}^3y,
\end{equation}
in which $x$ and $y$ are in units of $r_s$ as usual. It is generally difficult, if not impossible, to solve this integral analytically. 
However, there are several profiles for which do afford a solution, including the popular profile of \citet[][hereafter NFW]{Navarro1997}.

Thus, any profile is fully defined by its reduced profile shape, $f(x)$, which can be used to produce the dimensionless quantities $h(c)$, $p(\kappa,c)$, and $l(x,c)$. We utilise this set of variables in the \verb|halomod| implementation. 

\subsubsection{Non-truncated Profiles}
In the case that the profile is not truncated at $r_{\Delta_{\rm h}}$, we still define $c \equiv r_{\Delta_{\rm h}}/r_s$, and assume \cref{eq:mass-radius-halo} relates halo mass to $r_{\Delta_{\rm h}}$. 
The integrated mass of the halo is defined by \cref{eq:minrv}, but the upper integration limit is set to $+\infty$. This integral must converge for the given profile in order for it to be well-defined as a non-truncated profile.

For non-truncated profiles, neither $h$, $p$ or $l$ are dependent on the concentration parameter, and for each, the integration is performed to $+\infty$. Nevertheless, the overall profile is still scaled by the concentration and central density. 

\subsection{Concentration-Mass Relation}
\label{sec:theory:concentration}
Given the logic of self-similar fitting functions for the density profile laid out in the previous section, it is clear that the $c$--$m$ relation plays a critical role: if we can find a universally valid description of $c(m)$, the halo density profile can be determined solely based on the mass. However, concentration has long been known to depend on mass, redshift, and cosmology in a complex fashion, leading to a proliferation of models that fall into roughly three categories.

First, NFW suggested that concentration is intimately related to the assembly history of halos, which can be summarized as their age or formation time. Physically, this dependence can be understood by the transition from the fast to the slow accretion regime, after which the halo boundary (and its concentration) grow mostly via pseudo-evolution while the scale radius remains more or less constant \citep{Navarro1997, Bullock2001, Bosch2002, Wechsler2002, Zhao2003, Dalal2010, Diemer2013, Ludlow2013, Wang2020}. This logic has motivated numerous models based on some quantification of the age of halos \citep[e.g.,][]{Eke2001, Zhao2009, Giocoli2012, Ludlow2014, Ludlow2016, Correa2015b}. 

While age-based models follow physical insight, apply to individual halos, and are often universal (meaning they apply to all masses, redshifts, and cosmologies), they can be computationally complicated. Thus, a second popular way of modeling the $c$--$m$ relation is as a series of power laws 
\begin{equation}
    c(m_{\Delta_{\rm h}},a) = A\left(\frac{m}{M'}\right)^B a^C \,,
\end{equation}
which are typically valid only over the particular mass and redshift range where they were calibrated \citep[e.g.,][]{Dolag2004, Duffy2008, Maccio2008, Bhattacharya2013, Dutton2014, Klypin2016, Child2018}. Moreover, each calibration is valid only for a particular cosmology, which poses a significant barrier for purposes such as halo modeling.


Recently, a third avenue for modeling concentration has emerged. \citet{Prada2012} noticed that the redshift evolution of the $c$--$\nu$ relation is much smaller than that of the $c$--$m$ relation; they parameterized the differences with a fitting function. Subsequent work has shown that the residual evolution can be understood as additional, weaker dependencies on parameters that break the universality of structure formation in a $\Lambda$CDM universe, such as the power spectrum slope, $n_{\rm eff}$, and the evolution of the linear growth factor, $\alpha_{\rm eff}$ \citep{Diemer2015, Diemer2019, Ishiyama2020}. While similar parameters have been considered previously \citep[e.g.,][]{Bullock2001, Eke2001, Zhao2009}, the new models explicitly avoid reliance on any other variables. For example, when expressing concentration as $c(\nu, n_{\rm eff}, \alpha_{\rm eff})$, any dependence in cosmology is implicitly taken into account.

\subsection{Halo exclusion}
\label{sec:theory:exclusion}
On small scales, the 2-halo term as outlined in \S\ref{sec:theory:2halo} and integrated over all halo masses naturally counts pairs between theoretically overlapping haloes. In reality, correlations at these scales are much more likely to arise from within the one halo (or, alternatively put, particles within a halo's radius in a simulation are typically assumed to be a part of that halo).
Modelling halo exclusion in the 2-halo term attempts to account for this fact, so that these pairs aren't double-counted in both the 1-halo and 2-halo term. 

In practice, this affects the correlation function at the crossover between the 1-halo and 2-halo terms at the 30\% level \citep{Schneider2012}, by suppressing the 2-halo term.

\subsubsection{Empirical approaches}
There have been a number of proposed models to account for this effect. At a purely empirical level, one may consider modifying $P_{\rm hh}(k)$. While it is customary to use a (biased) nonlinear estimate of the matter power, \citet{Cooray2002} suggest that using the linear matter power  will decrease the power at small scales, therefore increasing the fidelity of the model. In a similar way, \citet{Schneider2013} use a smoothing of the power on roughly transition scales to eliminate the excess correlations:
\begin{equation}
    P_{\rm hh}(k,m_1,m_2) \approx b(m_1)b(m_2) W(kR_T) P_{\rm m, nl}(k),
\end{equation}
where $P_{m, nl}(k)$ is the matter power from \textsc{halofit}, $W$ is the Fourier transform of a top-hat in real space,
\begin{equation}
    W(x) = 3\frac{\sin x - x\cos x}{x^3},
\end{equation}
and $R_T\approx2 $Mpc$h^{-1}$.
This latter argument can be used to reduce errors to below 10\%.

\subsubsection{Physical approaches}
By far the more common approach is to set upper limits (say $m'_1$ and $m'_2$) on the 2-halo integral, \cref{eq:dmpower2a}, using physical arguments. This approach has been described in some detail in \citet{Tinker2005} (hereafter T05), 
and we briefly reproduce that discussion here. While we follow T05 and discuss the galaxy quantities, precisely the same arguments hold for DM quantities with the replacement $I_{\rm g} \rightarrow I_\text{DM}$ (and $\bar{n}_g \rightarrow \bar{\rho}$). 

The  underlying idea  is to set $m_1'$ and $m_2'$ such that the scale in question is not smaller than the sum of the halo radii, i.e. $r \ge R_\text{vir}(m_1') + R_\text{vir}(m_2')$. 
Note that in this approach, the density is also modified, so that
\begin{equation}
    \label{eq:ng2}
    \bar{n}'^{2}_g = \int_0^{m'_1} \int_0^{m'_2} \prod_{i=1}^2 n(m_i) N_{\rm t}(m_i) {\rm d}m_i.
\end{equation}
In this case, the final 2-halo correlation function needs to be re-scaled by
\begin{equation}
    \xi_{\rm gg}^{\rm 2h}(r) = \left(\frac{\bar{n}_g'}{\bar{n}_g}\right)^2\left[1+\xi_{\rm gg}^{\rm '2h}(r)\right] - 1.
\end{equation}

The simplest model is to assume spherical halos, and that the two halos in question are the same size, so that  
\citep{Zheng2005}:
\begin{equation}
m_1' = m_2' = \frac{4}{3}\pi\left(\frac{r}{2}\right)^3\Delta_{\rm h} \bar{\rho}.
\end{equation}
This is efficient since the limits are independent, and we can use the separated form of Eq.  \ref{eq:dmpower2}. However, it severely under-counts pairs due to edge effects. 

We can do better by relaxing the assumption that the two halos are the same size. In this case,  $m_1'$ and $m_2'$ are related by 
\begin{eqnarray}
    R_\mathrm{\Delta}(m_2') &=& r - R_\mathrm{\Delta}(m_1).
\end{eqnarray}
In this case, we cannot separate the integrals, and so double-integration must be performed, which is relatively inefficient.

It is well-known that halos are typically triaxial \citep{Bullock2001a,Taylor2011,Zemp2011}, and therefore one can improve the results by considering the probability that each pair is in different halos, given an empirical distribution of axis ratios. In this case, \cref{eq:dmpower2a} is augmented by the distribution:
\begin{equation}
    \frac{P_{g}^{\rm 2h}(k,r)}{ P_{\rm m}(k,r)}=   \int \int \prod_{i=1}^2  I_{\rm g}(k,m_i)b(m_i,r) \mathcal{P}(x) {\rm d}m_i.
\end{equation}
Here, $\mathcal{P}(x)$ is the probability that pairs at a given separation are in separate halos, where $x=r/(R_{\Delta}(m_1) + R_{\Delta}(m_2))$. T05 suggest using
\begin{eqnarray}
    \mathcal{P}(y) &=& 3y^2 - 2y^3 \\
    y &=& \frac{x-0.8}{0.29}
\end{eqnarray}
where $\mathcal{P}(y<0) = 0$ and $\mathcal{P}(y>1) = 1$. This is a significant improvement over the spherical case, however it still requires double-integration (at every value of $k$!) due to the appearance of $m_1$ and $m_2$ in $\mathcal{P}(x)$. 

T05 propose a solution to the efficiency problem by performing one double-integral to calculate $\bar{n}_g'$ for the full ellipsoidal case (where $\mathcal{P}(x)$ also augments Eq.\ref{eq:ng2}). From here, one can ``match" the number density numerically with the simpler single-integral:
\begin{equation}
\label{eq:theory:ng_dash}
 \bar{n}'_g = \int_0^{m'} n(m)N_{\rm t}(m){\rm d}m,
\end{equation}
cumulatively integrating until the integral matches $\bar{n}_g'$, thus acquiring $m'$. 
Setting $m_1' = m_2' = m'$, one can use the separated form,  \cref{eq:dmpower2}, to generate the final galaxy power spectrum.

\begin{figure}
  \centering
  \includegraphics[width=\linewidth]{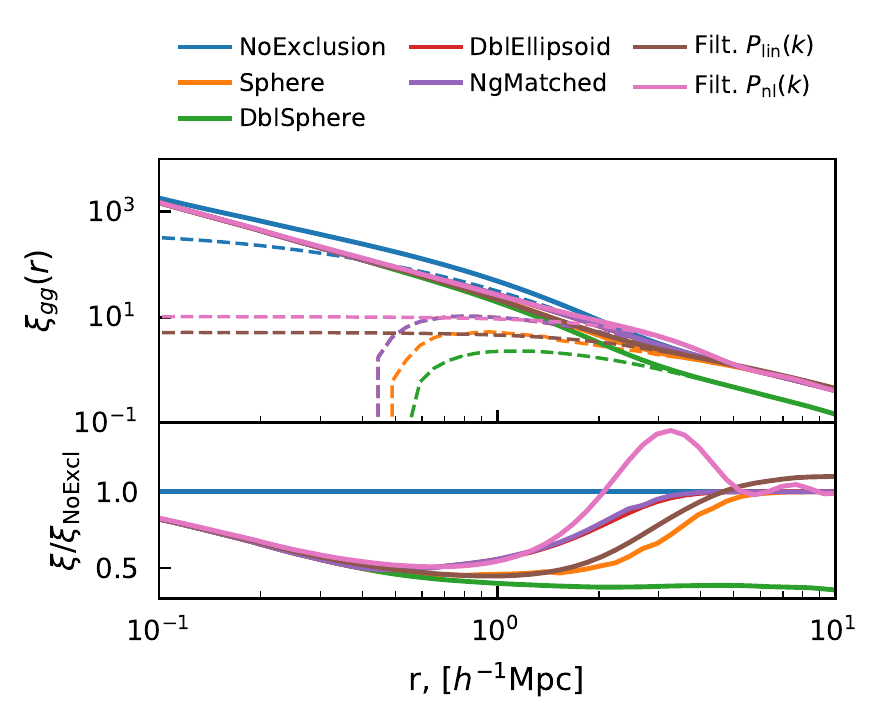}
  \caption[Effect of halo exclusion]{Effects of various halo exclusion models around the transition region. Solid lines are the full $\xi_{\rm gg}(r)$, while the dashed lines show the 2-halo term. In both cases where the halo-centre power spectrum is modified (brown and pink curves), the halo exclusion is not performed. }
  \label{fig:halo_exclusion}
\end{figure}

We show the effect of several of these models in  \cref{fig:halo_exclusion}. The spherical exclusion is the most effective at reducing pairs, with the $n_g$-matched method exhibiting a more gentle suppression. Interestingly, merely using the linear power without any exclusion is the closest to the $n_g$-matched case at the transition scales. 

Note that the recent model of \cite{Garcia2020}, in which halo exclusion is included explicitly via a soft halo radius, is not yet included in \halomod.

\section{Extending the Halo Model to Galaxies}
\label{sec:TracerHaloModel}

\subsection{HOD models}
\label{sec:theory-gal:hod}
Before we can extend the previous theory to a treatment of tracers (eg. optical galaxies), we must have a model for the expected occupation of the tracer within a given halo, $P(T|m)$, termed the halo occupation distribution (for discrete tracers such as galaxy counts, this is typically written $P(N|m)$ and must be a discrete distribution). Note that this model is assumed to depend solely on the halo mass, which perhaps surprisingly captures the majority of the behaviour of interest. There is a known second-order dependence on environment, the so-called `assembly bias' \citep{Sunayama2015}, but we do not consider it in this work.

Any HOD model is principally composed of two parts -- a parametrisation of the mean occupation of halos of a given mass, $\Nt_m$\footnote{Note that we use notation consistent with discrete tracers throughout, but these functions are generalizable to non-discrete distributions.}, and a distribution about this mean $P(N_{\rm t}|\Nt_m)$ (note that throughout this work, we imply that any values $\langle N \rangle$ are functions of mass, and drop the notation from here on). The mean occupation plays the major role in providing a form for the mass dependence of typical galaxy abundances, while the distribution is crucial in calculating pair (and triplet etc.) count statistics. 

Historically, the parametrisation of the mean occupation began as a single function, but early work \citep{Kauffmann2004,Zheng2005,Zehavi2005} showed an advantage in treating `central' and `satellite' galaxies independently. In hydrodynamical simulations, these are empirically identifiable classes of galaxies, and semi-analytic methods also treat them separately. Central galaxies (in optical) tend to be large and bright, and occupy the region near the halo's central potential well. Satellite galaxies are less bright, and tend to follow the halo's density profile\footnote{This is not strictly true -- satellites may follow an alternate profile to the underlying dark matter (eg. flattening towards the center). The point is that they follow a profile. \halomod\ allows both the halo and tracer profiles to be specified independently.}.  Note that these qualitative descriptions are not necessarily true for all samples of tracers, but it typically remains useful to separate occupation of the tracer into such central and satellite classes.

Finally then, we have a mean occupation function for each class, and the total occupation is their sum:
\begin{equation}
    \Nt = \Nc + \Ns.
\end{equation}
Identifying these classes as independent is helpful in determining the distributions about the means. Halos may only contain a single central galaxy (or none at all if the sample selection does not allow it). Thus their distribution is simply Bernoulli (or Binomial with a single trial):
\begin{equation}
    N_{\rm c} \sim \text{Bern}(\Nc).
\end{equation}
Satellite galaxies have been found to approximately follow Poisson distributions \citep{Kauffmann2004,Zheng2005}:
\begin{equation}
    N_{\rm s} \sim \text{Pois}(\Ns).
\end{equation}

Most studies also impose what we will term the `central condition'. Explicitly, the condition states that $N_{\rm c} = 0 \Rightarrow N_{\rm s} = 0$ (i.e. no central galaxies in a halo implies no satellites).
It arises from the fact that central galaxies are more luminous than their satellite counterparts, and therefore for any sample selection based on luminosity (or similar) thresholds, the central galaxy will be the first to be included.  Our implementation does not enforce this condition (as there are conceivably sample selections which circumvent it), but does enable it, which simplifies some calculations.

For 2-point statistics, important quantities arising from the HOD are the mean pair counts within a halo, which we address separately for centrals and satellites.
We wish to derive these quantities in forms explicitly dependent only on the mean occupation functions $\Nc$ and $\Ns$, for which we have specific parametrisations. 
Furthermore, we treat self-pairs separately from cross-pairs, as self-pairs have different correlation properties.
Noting that ultimately we will be comparing these pair counts to the mean number density squared (which counts each pair twice), we have:
\begin{align}
    N_{\rm cross-pair}^{\rm cen-cen} &= 0 \\
    N_{\rm cross-pair}^{\rm cen-sat} &= 2 \langle N_c N_s \rangle \\
    N_{\rm cross-pair}^{\rm sat-sat} &= \langle N_s (N_s -1) \rangle \langle N_s \rangle^2
\end{align}
where the last equality is made under the assumption of Poisson-distributed satellite counts.
The cen-sat term is problematic (as it is not a function purely of mean counts of either central or satellite), however if we impose the central condition, then $\langle N_{\rm c} N_{\rm s} \rangle = \Ns$, since we need only count halos in which $N_{\rm c} = 1$. 
Conversely, if the parameterisations for $\Nc$ and $\Ns$ do not impose the central condition then we can approximate, following \citet{Zehavi2005}, with $\langle N_{\rm c} N_{\rm s} \rangle = \Nc \Ns$, given that $\Ns \ll 1$ when $\Nc < 1$. For $\Nc = 0$ or 1, it is no longer an approximation, and thus for step-function parameterisations of $\Nc$, it is completely accurate (for arbitrary parameterisations of $\Ns$).

Finally, if the central condition is required, but the parameterisations of choice do not impose it (i.e. $\Ns > 0$ when $\Nc<1$, or ``satellites may exist in a particular halo where centrals do not''), one can manipulate the equations to enforce it using a common technique \citep[eg.][]{Beutler2013}, in which we set $\Ns = \Nc \Ns'$, where $\Ns'$ is the original parameterisation. In this case, the truncation of $\Nc$ forces the appropriate truncation of $\Ns$. One must be careful in this method, however,  to ensure that the HOD parameters retain the same meaning as the original. For a step-function $\Nc$ this is not a problem, but there may be subtle differences induced for smoothed truncations (these are likely to be negligible for reasonable models).

Clearly, the number of self-pairs is simply $\langle N_t \rangle$.

There are several parameterisations for the HOD in the literature. In this work, and correspondingly in the \verb|halomod| framework, we focus on those that utilise the separation of
central and satellite galaxies. We present a summary of those included in \textsc{halomod} in Table \ref{tab:models_hod}. 

 Most formulations contain the same essential features, so the most important features of the HOD are present in the simplest case. For the central term: a step function characterised by a step at the parameter $M_\text{min}$, which is heavily dependent on the sample selection. The deeper the survey, the lower this parameter will be. The satellite term is primarily characterised by a power law which quantifies the growth of galaxies with halo mass by the index $\alpha$, and a scaling mass $M_1$ which is the mass at which an average halo in the sample contains a single satellite galaxy.

\begin{figure}
\centering
\includegraphics[width=\linewidth]{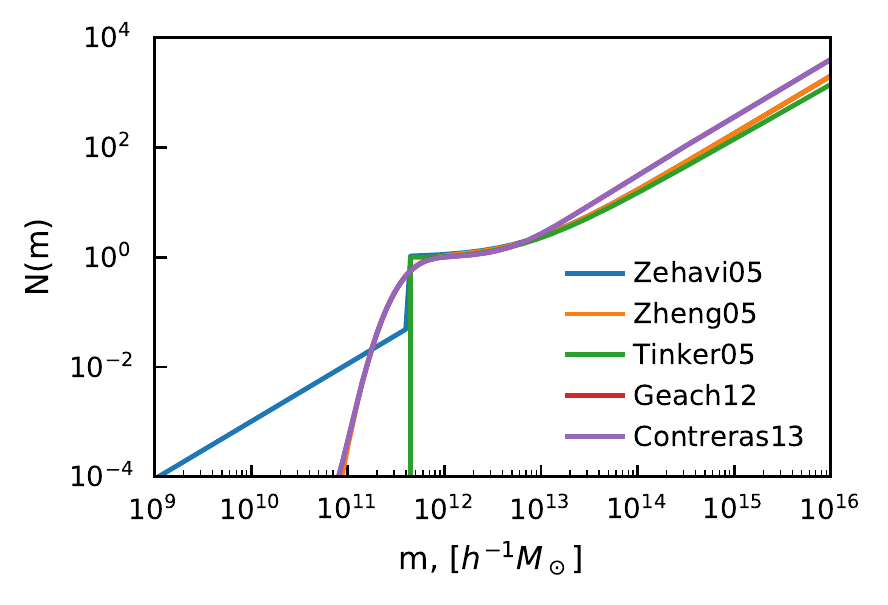} 
\caption[Halo occupation for several parameterisations]{Total halo occupation for several parameterisations, highlighting the core similarities and differences between the models. Note that the central condition is false for the models shown.}
\label{fig:hod}	
\end{figure}

The various refinements to this general structure include smoothing the truncation of central galaxies assuming a log-normal distribution of halo masses at this threshold, and truncating the satellite power law at an arbitrary mass scale $M_0$. Further refinements can be specific to a certain kind of sampling, for example samples gained at different wavelengths. An illustrative plot of the core similarities and detailed differences can be found in \cref{fig:hod}.

\subsection{Extension of framework to galaxies}
\label{sec:theory-gal:framework}
The extension of equations \ref{eq:dmcorr1}, \ref{eq:dmpower1} and \ref{eq:dmpower2} to include galaxies is relatively simple under the assumption that all galaxies reside in halos, and their intra-halo abundance follows that of the density profile for satellites. 

For the 2-halo term, we make a simple modification to the DM equations, replacing factors of $m/\bar{\rho}$ with $N_{\rm t}(m)/\bar{n}_g$, where $N_{\rm t}(m)$ is the mean occupation function from the HOD (where for simplicity we omit angle brackets for the rest of this work), and $\bar{n}_g$ is the mean number density of galaxies,
\begin{equation}
    \label{eq:meandens}
 \bar{n}_g = \int n(m)N_{\rm t}(m) {\rm d}m.
\end{equation}
With this modification the 2-halo term is simply described by
\begin{equation}
    \label{eq:Ig}
    I_{\rm g}(k,m) = n(m) u(k|m) N_{\rm t}(m)/\bar{n}_g,
\end{equation}
which serves as a direct replacement for $I_\text{DM}$ in equations \ref{eq:dmpower2a} and \ref{eq:dmpower2}.
Note that this is not \textit{quite} accurate given the separation of galaxies into centrals and satellites, because in this case the central galaxy is constrained to be at the center instead of following the profile (whereas Eq. \ref{eq:Ig} assumes all of the galaxies---including the central---are located following the density profile). Nevertheless, the error induced by this approximation is extremely small.
Note further that no ``self-pairs'' are included in the 2-halo term, since it always considers two separate halos (thus it contains no ``shot-noise'').

The 1-halo term needs a little more care. 
Using the formalism of \S\ref{sec:theory-gal:hod}, we separate the pair counts into contributions from central-satellite ($c-s$) and satellite-satellite ($s-s$) cross-pairs (we shall deal with self-pairs shortly). 
The first of these is a one-point quantity dependent on the density profile, while the latter is given by the self-convolution of the profile (similar to DM). 
Thus we have
\begin{equation}
\label{eq:1-halo-galaxy}
1 + \xi_{\rm gg}^{\rm 1h}(r) = \frac{1}{\bar{n}_g^2}\int n(m) \left[2\langle N_{\rm c} N_{\rm s} \rangle \frac{\rho(r|m)}{m} + N_{\rm s}^2 \frac{\lambda(r|m)}{m^2}\right] {\rm d}m,
\end{equation}
where the quantity $\langle N_{\rm c} N_{\rm s} \rangle$ is treated differently depending on whether the central condition is imposed (cf. \S\ref{sec:theory-gal:hod}).

The behaviour of this expression as the separation approaches zero is of some interest. We firstly note that as $r$ falls below some (very small) threshold, we should no longer trust the HOD model -- some level of ``galaxy exclusion'' must apply \citep{Ginzburg2017}, rendering the the distribution of $N_s$ non-Poisson and non-independent of the surrounding locations. Thus Eq. \ref{eq:1-halo-galaxy} will tend to over-estimate the correlation function below this threshold. This is most evident at $r=0$, where Eq. \ref{eq:1-halo-galaxy} predicts a non-zero correlation, even though we have explicitly avoided self-pairs. 

On the other hand, self-pairs are really there. They form a delta-function contribution at $r=0$, with a magnitude given by the mean galaxy density\footnote{Note that for non-discrete tracers, which \halomod\ supports, the `shot-noise' contribution is not simply given by the tracer number density, but requires knowledge of the scatter in the tracer HOD population (which is no longer Poisson, cf. \citealt{Chen2021}).}. 
This translates into a constant with magnitude $1/\bar{n}_g$ in the power spectrum, and this is properly the `shot-noise'. 
In \textsc{halomod}, the power spectra are defined without shot-noise (i.e. purely as the FT of Eq. \ref{eq:1-halo-galaxy}), which can be added in by the user if desired simply as a constant. The power spectrum is thus
\begin{equation}
    P_{\rm gg}^{\rm 1h}(k) = \int {\rm d}m\ n(m)u(k|m) \left[N_s^2 u(k|m) + 2 \langle N_s N_c \rangle \right].
\end{equation}
Note again that the \texttt{force\_1halo\_turnover} option forces the 1-halo term to turnover on scales larger than the size of halos (cf. \S\ref{sec:halomod:frameworks:base}), to reduce the spurious `shot-noise', and the true shot-noise can be manually added back in to the power spectrum.

\subsection{Derived Quantities}
There are several quantities of interest which can be derived from the HM/HOD framework.

Firstly, on large scales, where $u\rightarrow1$, and the 1-halo term is negligible, the 2-halo term is approximated by
\begin{equation}
 P_{\rm gg}(k) \approx b^2_\text{eff}P_{\rm m}(k),
\end{equation}
where
\begin{equation}
\label{eq:beff}
 b_\text{eff} = \frac{1}{\bar{n}_g}\int n(m)b(m)N_{\rm t}(m) {\rm d}m
\end{equation}
is called the ``effective bias".

We may likewise calculate an ``effective concentration" which is the mean halo concentration within the sample:
\begin{equation}
     c_\text{eff} = \frac{1}{\bar{n}_g}\int n(m)c(m)N_{\rm t}(m) {\rm d}m,
\end{equation}
and an ``effective mass":
\begin{equation}
     M_\text{eff} = \frac{1}{\bar{n}_g}\int n(m)mN_{\rm t}(m) {\rm d}m.
\end{equation}

An important quantity in studies of galaxy formation and evolution is the fraction of galaxies that are satellites. The evolution of this quantity through time can trace the effects of various physical processes. It is simply defined as
\begin{equation}
    f_\text{sat} = \frac{\int n(m) N_{\rm s}(m) {\rm d}m}{\int n(m) N_{\rm t}(m) {\rm d}m}.
\end{equation}

\section{The \textsc{halomod} library}
\label{sec:halomod}

In this section we present our new implementation of the HM, \textsc{halomod}. Our intention is first to give an overview of the philosophy behind, and general characteristics and features of the code. From there, we describe in detail various parts of the code, beginning in \S\ref{sec:halomod:frameworks} with the ``frameworks" which tie together the various components, and then the individual components themselves in \S\ref{sec:halomod:components}. These sections have a pragmatic focus, introducing any pertinent numerical techniques, and summarising the various models included. Finally, we comment on some of the extra functionality present in the package in \S\ref{sec:halomod:extra}.

\subsection{Philosophy and Architecture}
\label{sec:halomod:overview}
\textsc{Halomod} is a \python package, built on the \textsc{hmf} package\footnote{\url{https://github.com/halomod/hmf}} \citep{Murray2013a}, which contains the necessary components and algorithms to calculate HM quantities, and also tracer statistics via a HOD\footnote{For now, it only implements the 2-point statistics, but this will be expanded in future versions.}. 
\textsc{hmf} provides modules for cosmology, transfer functions, spatial filters, mass definitions and mass functions, along with defining a coherent and flexible software framework. 
\halomod\ inherits that framework and provides halo profiles, concentration relations, bias functions, halo exclusion models and HOD models and combines them into the full halo model.

The principles outlined in the introduction, i.e.  ``Intuitive'', ``Simple'', ``Efficient'', ``Extendible/Flexible'', ``Comprehensive'' and ``Open'', have been the guiding principles in the development of \textsc{halomod} (and also \textsc{hmf}). Our code provides a well-defined architecture for defining components and frameworks for halo modelling, including many basic and derived quantities of interest, and a large fraction of the available models in the literature. Each component is completely flexible, either by supplying custom parameters or entirely new parameterisations \textit{without touching the source code}. We now expand upon some of these aspects.

Note that many of the features that we discuss in this section reflect updates to the \textsc{hmf} package (compared to its presentation in \citet{Murray2013}, which are inherited in \textsc{halomod}. Furthermore, clearly this paper represents a snapshot of the package in time, with updates likely to occur. 

\subsubsection{Intuitive Usage}
\label{sec:halomod:overview:usage}
To demonstrate the intuitive usage of \halomod, it is perhaps most useful to give a quick example of how \verb|halomod| can be invoked. 
Necessarily, the following example is both simple and rather arbitrary, but it should give a flavour of the usage of \halomod.

To create a new halo model class (this one will support galaxy power spectra and correlation functions) does not require passing any parameters, as all parameters have reasonable defaults:
	\begin{lstlisting}[language=Python]
	import halomod as hm
	tr = hm.TracerHaloModel()
	\end{lstlisting}
After creating the model, all of the relevant quantities may be accessed as attributes (in fact, they are not simple attributes that have been pre-computed, they are calculated lazily on first access and then cached).
Thus, we can plot the auto-correlation function of galaxies, for example; 
\begin{lstlisting}[language=Python]
plt.loglog(tr.r, tr.corr_auto_tracer, label='z=0')
\end{lstlisting}
To update a parameter, one can simply set it on the instance, so for example to change the redshift and plot the new correlation function:
%
\begin{lstlisting}[language=Python]
	tr.z = 1
	plt.loglog(tr.r, tr.corr_auto_tracer, label='z=1')
\end{lstlisting}
Updating any parameter immediately invalidates the cache of quantities that depend on it, and thus that quantity is recalculated (but not quantities on which it depends that don't themselves depend on the changed parameter -- like the cosmological transfer function in this example). 

\begin{figure}
    \centering
    \includegraphics[width=\linewidth]{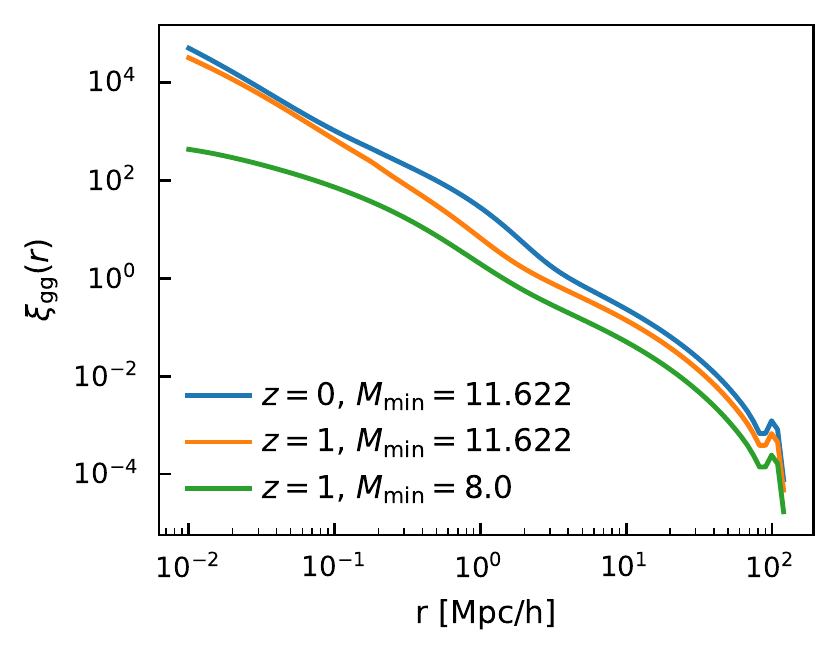}
    \caption{Simple example of galaxy auto-correlation functions produced with three inputs (two different redshifts and a different HOD model).
    Notice the BAO peak at $\sim100 h^{-1}{\rm Mpc}$. Code to produce the figure is presented in \S\ref{sec:halomod:overview:usage}.}
    \label{fig:toy_example}
\end{figure}

Every sub-component of the halo model (see the following sections) allows its parameters to be changed directly from this top-level interface. So, if we wanted to update the $M_{\rm min}$ parameter of the halo model, we could simply do
%
\begin{lstlisting}[language=Python]
tr.hod_params = {"M_min": 8.0}
plt.loglog(tr.r, tr.corr_auto_tracer, label='Mmin=8')
\end{lstlisting}
After applying relevant axis labels, we then produce Fig. \ref{fig:toy_example}.

\subsubsection{Simple Conceptual Framework}
\label{sec:halomod:overview:concept}
The backbone of \textsc{halomod}, directly inherited from \textsc{hmf}, is split into two categories: `frameworks' and `components', which are well-defined by a base class, \framework\ and \component\ respectively. Individual elements of the calculation, such as a halo profile, HMF fitting function, or any other unit which may be calculated in more than one way (e.g., growth factors) are represented by a \component. \texttt{Framework}s are the structures which tie each \component\ together to calculate a set of quantities (e.g., correlation functions, power spectra or mass functions). Frameworks may also contain other frameworks (for example the \texttt{Cross\-Correlation} framework contains two \texttt{TracerHaloModel} frameworks -- one for each tracer to be cross-correlated).

This structure makes it simple to identify a component of interest for extended modelling, and provides a common interface and set of methods to each type. For example, if one is developing a new improved model of halo bias, it is clearly identified as a \component, with a particular structure. Its place in the overall framework is well-specified, it by default receives any defining parameters it requires, and specific examples of implementations of other models of the component are easily found in the \verb|bias| module. 

It is expected that the most common usage will be to create a \framework\ instance, and access available quantities through it. Our general approach has been to treat frameworks as a self-contained unit. This means that all parameters are passed to the initial constructor, after which every relevant quantity is ready for immediate access, without method-specific parameters. Indeed, to the user, once a \framework\ instance has been created (with all its options), any derived quantity (such as mass functions, correlation functions, mass variance etc.) is accessed just like a stored variable. 

This approach is not typical of much software -- usually there is benefit in segregating object-wide variables from method-specific ones. However, we find in our case that most variables are used across many methods, and are more naturally interpreted as being a part of the framework itself. 

In fact, our adherence to this approach is bolstered by our generic distinction (within \texttt{Framework}s) between a \parameter\ and a \cached. Every input variable is treated as a class-wide \parameter, which receives its own characteristics (this is very useful for our caching mechanism, see \S\ref{sec:halomod:overview:efficiency}). Every evaluated quantity is treated as a \cached, which likewise has its own properties. The set of all \texttt{parameter}s of a given \framework\ is then a unique description of the entire object (and can be used to test equality between framework instances). 

Though the state of a \framework\ is defined by its parameters, we provide a simple mechanism of modifying the parameters for dynamic calculations, through an \verb|update| method, which can receive any \parameter\ (or, as noted in the previous subsection, simply by setting the attribute)\footnote{This method is defined in the base \texttt{Framework} class, and is therefore simply inherited by any specific framework.}.

Our approach to the passing of parameters has been to provide defaults for \textit{every} variable in a \framework\footnote{These defaults are currently majoritarily static, but we plan on dynamifying them in future versions.}. These defaults are chosen to be broadly in line with the most common usage, which greatly simplifies common tasks. However, it must be noted that this has the drawback that if user code relies on defaults, and code updates modify them, then results can vary without being noticed. 

In summary, the architecture of \hmf\ and \halomod\ consists of
\begin{itemize}
    \item A flexible \component\ for each independent element of the HM.
    \item A range of specific \textit{models} for each \component.
    \item Powerful \framework\ classes that synthesise multiple \verb|Components| to provide plug-and-play functionality.
    \item \texttt{Frameworks} consist of \texttt{parameters} which can be input and modified by the user, and \cached\ objects, which are calculated from the \verb|parameters| and are read-only and always consistent with the current settings.
\end{itemize}

\subsubsection{Efficiency}
\label{sec:halomod:overview:efficiency}

While \halomod\ generally prioritizes ease-of-use over efficiency, it is optimized in a number of ways:
\begin{itemize}
    \item Heavy use of \textsc{numpy}\footnote{\url{http://www.numpy.org/}} to enable vectorized calculations, close to compiled code performance in most cases.
    \item Calculation of all quantities at a fixed set of user-defined vectors (as opposed to providing methods that ingest evaluated variables such as mass or wavenumber)\footnote{An exception to this rule applies in the case of the 2-halo term, which can require two sequential Hankel transforms. For this, we evaluate $\xi_{2h}(r)$ on a background grid of $r$ which adequately captures the behaviour required for the inverse transform, and we interpolate this grid for the user's input vector of $r$.}.
    \item Use of ``just-in-time'' (JIT) compiled functions (via \textsc{numba}\footnote{\url{http://numba.pydata.org/}}) where vectorization is impossible or disadvantageous.
    \item A novel and robust caching system with extremely accurate and efficient cache invalidation (i.e. cached quantities always stay up-to-date with parameter changes, but nothing updates that doesn't have to). See \ref{app:caching} for details on this system, which utilises the core architecture of the \framework\ to achieve its goals.
\end{itemize}

\subsubsection{Flexibility}
\label{sec:halomod:overview:flexibility}
We have referred to our imposed dichotomy between \framework\ and \component\ objects, and in the previous subsection we outlined a performance benefit derived from the internal structure of the \framework. Likewise, we have implemented a uniform convention for \component\ specification which enhances flexibility.

Any component that is implemented as a \component\ is a variable within a \framework. As such, it is determined by passing a \parameter, in the form of either a string representing the name of the model, or the class itself. 

Every \component\ has a top-level dictionary specifying the model-specific parameters and their default values (these are not fundamental quantities used in the calculation, eg. $\nu$ in the $f(\nu)$ for the HMF, but rather free parameters of the model, if any). These are processed by the base class so that any newly defined model automatically has them available. In the \framework, the model parameters are passed, by convention, as \verb|<cmpnt>_params|. Furthermore, the \verb|update| method in each \framework\ intelligently updates parameter dictionaries so that previous updates are persistent unless explicitly overwritten.

This general system offers the highest level of flexibility and extendibility. In a single session, it is extremely simple to switch between several models for a given component, thereby making useful comparisons (and this without needing to recalculate basic quantities that do not depend on the variable component). On the other hand, using entirely new custom models can be achieved without modifying the source code, by inheriting from the base \component\ and passing the new model directly to a \framework. We reiterate this point, as we expect this to be particularly advantageous in the current climate of rapid development: \texttt{Components} are \textit{pluggable}. A user can write their own new HMF fit, or bias function, or HOD model, in their own notebook (or wherever they write code), and use it right away in the context of the full halo model. 

Furthermore, to a greater or lesser degree, each \component\ has built-in methods which either remove the burden on the user, or supply extra functionality for free when the user develops a new model. This is perhaps most clear in the \verb|Profiles| component (cf. \S\ref{sec:halomod:components:profile}), though some components have very little to add to the basic model (eg. the \verb|Bias| component, cf. \S\ref{sec:halomod:components:bias}). 

\texttt{Components} are primarily designed to integrate with the \texttt{Frameworks}, however, they are not limited to this usage. They are also intended to be useful in their own right, for more specific applications. 


\subsubsection{Interoperability}
\textsc{halomod} is intended to be a package that is adopted and maintained by the community. To this end, it uses well-known and well-maintained libraries to enable much of its functionality. 
For example, cosmologies are defined via \textsc{astropy}\footnote{\url{https://www.astropy.org/}} and we use the standard open-source \textsc{numpy} and \textsc{scipy} packages for numeric and scientific routines. 
We also adopt the open-source \textsc{hankel}\footnote{\url{https://github.com/steven-murray/hankel}} \citep{Murray2019} package for computing some of the Hankel transforms.

We also provide extensive interoperability with the \textsc{colossus} code\footnote{\url{http://www.benediktdiemer.com/code/colossus/}} \citep{Diemer2018}, by providing component constructors for models contained in \textsc{colossus}. 
In practice, this means that the user can invoke something like the following
%
\begin{lstlisting}[language=Python]
	bias = make_colossus_bias("comparat17")
	model = TracerHaloModel(bias_model=bias)
\end{lstlisting}
and get the full range of functionality from the \textsc{colossus}-defined bias model within the \textsc{halomod} framework (even setting of model-dependent parameters).
At the time of publishing, these constructors are defined for bias and halo concentration models, as these models are part of the core functionality of \textsc{colossus} and thus there is a real chance that the models may exist in \textsc{colossus} and not \textsc{halomod}. 

Furthermore, versions of \textsc{colossus} from v1.2.16 onwards have implemented a simple pair of functions enabling conversion of \textsc{astropy} cosmologies to and from the native \textsc{colossus} cosmologies. 
Since \halomod\ cosmologies are essentially derived from \textsc{astropy}, this enables a greater degree of interoperability between \textsc{colossus}
 and \halomod.
 
\subsection{Frameworks}
\label{sec:halomod:frameworks}
In this section we present several of the frameworks that \textsc{halomod} provides. We note that several other frameworks are found in the \textsc{hmf} package, from which those presented here inherit. Furthermore, \textsc{halomod} also provides a \texttt{DMHaloModel} framework for calculating non-linear matter power spectra and correlation functions (the same are available within the \texttt{TracerHaloModel}, but the \texttt{DMHaloModel} requires fewer inputs). 

\subsubsection{Base Halo Model}
\label{sec:halomod:frameworks:base}
The basic framework in \textsc{halomod} is \texttt{TracerHaloModel}, which provides general access to most derived quantities relating to 2-point statistics. Table \ref{tab:halomodel_parameters} contains a useful summary of parameters for \texttt{TracerHaloModel}, giving an indication of its flexibility. Likewise, table \ref{tab:halomodel_properties} lists the available properties. We note that these are expected to be updated and expanded in future versions, and thus are not intended to provide a strict API, but rather an illustrative summary.

\begin{table*}
\centering
 {\tabulinesep=1mm
\begin{tabu} to \linewidth{X[1.6l]X[1.4l]X[5l]} 
\toprule[0.05cm]
\textsc{Parameter} & \textsc{Default}$^\dagger$ & \textsc{Description} \\
\toprule[0.05cm]

\multicolumn{2}{l}{\textsc{Components/Parameters}*} \\
\texttt{cosmo\_<>} & \texttt{Planck13} & Underlying cosmology. Must be an \texttt{FLRW} object from \textsc{astropy}. This model differs from other \texttt{Components}, being passed as an \textit{instance}, not a \textit{class}.\\
\texttt{transfer\_<>} & \texttt{CAMB} & A model to use for the transfer function. \\
\texttt{growth\_<>} & \texttt{GrowthFactor} & A model for the growth function. The default is to perform the explicit integration.\\
\texttt{mdef\_<>} & \texttt{None} & A mass definition (default is the definition under which the chosen \texttt{hmf} was measured). \\
\texttt{hmf\_<>} & \texttt{Tinker08} & A model specifying the HMF fitting function, $\nu f(\nu)$.\\
\texttt{filter\_<>}& \texttt{TopHat} & A filter function model. \\   
\texttt{hod\_<>} & \texttt{Zehavi05} & A HOD model. \\   
 \texttt{profile\_<>} & \texttt{NFW} & A halo profile. \\
 \texttt{concentration\_<>}  & \texttt{Duffy08} & A concentration-mass-redshift relation \\
 \texttt{bias\_<>} & \texttt{Tinker10} & A scale-independent linear bias model. \\
 \texttt{sd\_bias\_<>}& \texttt{Tinker\_SD05} & A model for the scale-dependence of the linear bias. \\
  \texttt{exclusion\_<>} & \texttt{NgMatched} & A halo exclusion model. \\

\midrule
\multicolumn{2}{l}{\textsc{Resolution/Location}} \\
\texttt{lnk\_min}, \texttt{lnk\_max}, \texttt{dlnk} & -8,8, 0.05  & Min/Max/$\Delta$ logarithmic wavenumber for linear power spectra and tabulated halo model spectra\\
\texttt{hm\_logk\_min}, \texttt{hm\_logk\_max}, \texttt{hm\_dlogk} & -2.5,1.5, 0.05  & Min/Max/$\Delta$ logarithmic wavenumber for output halo model spectra \\
\texttt{dlog10m} & 0.01 & Logarithmic mass interval \\
\texttt{rmin}, \texttt{rmax}, \texttt{rnum}, \texttt{rlog} & 0.1, 50.0, 20, True & Min/Max/\# for pair-separation vector, and whether its vector should be equi-log-spaced. \\
\texttt{r\_table\_num} & 100 & Length of background table of $r$ \\

\midrule
\multicolumn{2}{l}{\textsc{Physical}} \\
\texttt{sigma\_8} & 0.8344 & Normalisation of power spectrum, $\sigma_8$ \\
\texttt{n} & 0.9624 & Spectral index, $n_s$ \\
\texttt{z} & 0.0 & Redshift \\
\texttt{delta\_c} & 1.686 & Critical overdensity for collapse, $\delta_c$. \\
\texttt{ng} & None & Optional specification of mean galaxy density. If present, an HOD parameter (generally $M_\text{min}$) is fixed. \\

\midrule
\multicolumn{2}{l}{\textsc{Options}} \\
\texttt{takahashi} & True & Apply updated \textsc{halofit} coefficients from \citet{Takahashi2012}, otherwise use original coefficiencts from \citet{Smith2003}. \\
\texttt{hc\_spectrum} & Linear & Model for $P_{\rm hh}^c(k)$. Either ``linear", ``nonlinear" (\textsc{halofit}), or ``filtered-nl" (smoothed on a scale of 2$h^{-1}$Mpc). \\
\texttt{force\_1halo\_turnover} & True & Whether to cut off the 1-halo power on extremely large scales.  \\
\texttt{force\_unity\_dm\_bias} & True & Whether to renormalize the 2-halo term such that the DM bias is unity for $k\rightarrow0$.  \\

\bottomrule[0.05cm]
\end{tabu}}
\caption[Summary of parameters included in \texttt{TracerHaloModel}]{All parameters of \texttt{TracerHaloModel}. Parameters in the ``Components/Parameters" section are each specified by two parameters, ending in \texttt{model} and \texttt{params} (i.e. \texttt{hmf\_<>} represents two parameters: \texttt{hmf\_model} and \texttt{hmf\_params}). Using the HMF as an example, \texttt{hmf\_model} is the actual model specification, and this is listed in the \textsc{Default} column. Conversely, \texttt{hmf\_params} is a dictionary of parameters for the model, which is not listed in this table, but can be found in the online API documentation. The parameters available for each \textit{model} within a particular component may be different. Notice also that the physical parameters $\sigma_8$ and $n_s$ are directly set within the framework, rather than being a part of the cosmology parameters. The reason for this is outlined in \S\ref{sec:halomod:components:cosmology}.}
\label{tab:halomodel_parameters}
\end{table*}

\begin{table*}
\centering
 {\tabulinesep=1mm
\begin{tabu} to \linewidth{X[1l]X[4.5l]} 
\toprule[0.05cm]
\textsc{Type} & \textsc{Quantities}*  \\
\toprule[0.05cm]
Scalar & $b_\text{eff}$, $f_\text{cen}$, $f_\text{sat}$, $D^+(z)$, $M_\text{eff}$, $M_\star$, $\bar{\rho}(z)$, $\bar{n}_g$, $n_\text{eff}$ \\
Length-$m$ & $R$, $b$, $\bar{c}$, $\nu f(\nu)$, $\frac{dn}{d\ln m}$, $\frac{dn}{d\log_{10} m}$, $\frac{dn}{d m}$, $n(>m)$, $\Nc$, $\Ns$, $\Nt$, $L(n=1)$, $\ln \sigma^{-1}$, $\nu$,  $\rho(>m)$, $\rho(<m)$ \\
Length-$k$ & $T$, $P_{\rm m}$, $P_{{\rm m}, \text{halofit}}$, $\Delta^2$, $\Delta^2_\text{halofit}$, $P_{\rm gg}^{ss}$, $P_{\rm gg}^{\rm 2h}$, $P_{mm}$, $P_{mm}^{\rm 1h}$, $P_{mm}^{\rm 2h}$, $P_{mg}^{\rm 1h}$, $P_{mg}^{\rm 2h}$, $P_{mg}$ \\
Length-$r$ & $\xi_{\rm gg}^{\rm 1h}$, $\xi_{\rm gg}^{cs}$, $\xi_{\rm gg}^{ss}$, $\xi_{\rm gg}^{\rm 2h}$, $\xi_{\rm gg}$, $\xi_{\rm DM}^{\rm 1h}$, $\xi_{\rm DM}^{\rm 2h}$, $\xi_{\rm DM}$, $\xi_{\rm m}$, $\xi_{m, \text{halofit}}$, $\xi_{mg}^{\rm 1h}$, $\xi_{mg}^{\rm 2h}$, $\xi_{mg}$, $S(\xi_{\rm m}(r))$ \\
Function of $r,m$ & $\rho$, $\lambda$ \\
Function of $k,m$ & $u$ \\
 
\bottomrule[0.05cm]
\end{tabu}}
\caption[All included properties of \texttt{TracerHaloModel}]{All properties of \texttt{TracerHaloModel}. Listed are those quantities that are directly accessible as explicit properties. Other quantities, such as the window function, are accessible indirectly through a model instance variable.}
\label{tab:halomodel_properties}
\end{table*}

Some of the features of the \texttt{HaloModel} framework, ignoring contributions of individual components, are as follows.

\paragraph*{Mean density matching} 
Most HOD parameterisations have a parameter $M_\text{min}$, which either explicitly or loosely defines the minimum halo mass expected to host galaxies in the sample. This parameter is tightly correlated with the mean galaxy density, as it determines the lower limit of the integral in \cref{eq:meandens}. It is common (though not ubiquitous) in survey analyses to let the known value of $\bar{n}_g$ determine $M_\text{min}$ (given other parameters of the HOD) \citep[eg][]{Beutler2013}. \texttt{TracerHaloModel} supports this technique for all HOD models, either by direct cumulative integration, or by numerical minimization. This is caught at every update of parameters.

\paragraph*{Mass range setting}
The mass range in \texttt{TracerHaloModel} is user-definable, but sets intelligent limits. An upper limit of $M_\text{max} = 10^{18}h^{-1}M_\odot$ is set to ensure convergence of mass integral, and the lower limit is set to $M_\text{min} = 10^{0}h^{-1}M_\odot$ for matter calculations, and is based on the value of $M_\text{min}$ in the HOD for tracer calculations. For step-function models, the lower limit is exactly $M_\text{min}$, which increases accuracy in an important regime. For smooth-cutoff models, the lower limit is determined by the \verb|mmin| property of the HOD (see \S\ref{sec:halomod:components:hod}).

\paragraph*{Halo-centre power spectrum}
A key quantity in the large-scale clustering is the power spectrum of halo centres, $P_{\rm hh}^c(k)$ (cf. \S\ref{sec:theory:2halo} and \S\ref{sec:theory:exclusion}). This can be modelled via the linear matter power, nonlinear matter power, or even some smoothed version of either. We leave this choice free for the user, through the parameter \verb|hc_spectrum|. 

\paragraph*{Convolutions}
The satellite-satellite term can be calculated directly in real-space if and only if there is an analytic solution for the self-convolution of the halo profile. \texttt{TracerHaloModel} uses this form directly if possible, saving integration time, but resorts to a calculation in Fourier-space (and subsequent Hankel transform) if necessary. 

\paragraph*{Hankel transforms} 
The Hankel transform is a delicate operation, involving the integration of a highly oscillatory function. We use a novel method for performing this transform, which we detail in \ref{app:hankel}.

\paragraph*{Tabulation}
The Hankel transform of a power spectrum to a correlation function is typically relatively simple and accurate(using the framework for Hankel transforms defined in \ref{app:hankel}).
This is especially so since the power spectrum is naturally defined over a large range and at quite high resolution (as is required for the calculation of the mass variance, for example). 
However, the converse is not true -- transforming a correlation function to a power spectrum is typically quite limited in its range of accuracy. 

It is thus tempting to calculate power spectra at the $k$-vector at which the linear power spectrum is defined, and consider only ever transforming into real space. If this is the case, the vector of $r$ at which the correlation functions are defined can be low-resolution and limited in range, since it would be used only as an output, not as an input to some other function that requires it to be high-resolution.

However, this turns out not to be realistic in practice. 
In particular, some models of halo exclusion mix a scale $r$ -- related to the halo radius -- with the 2-halo power spectrum. Scale-dependent bias does likewise. To obtain a standard $P(k)$ thus requires performing a Hankel transform of the power \textit{at each} $r$, then an inverse Hankel transform to get back to $P(k)$. 

This requires using a higher-resolution, and importantly, large-range vector of $r$ for the intermediate correlation function. Furthermore, the calculation must be confident that the vector of $r$ is suitably well-chosen. We thus use a set tabulated background $r$ vector, and interpolate the results of $\xi(r)$ onto the user-provided $r$-vector. 

We also provide an option for the user to define a more limited range of $k$ at which to output the halo model power spectra, and these are also interpolated from their base table, which is defined on the $k$-vector used for the linear spectra. 

Problematically, the Hankel transform algorithm we employ often requires \textit{extrapolated} values of the integrand in order for the integral to converge for some scales\footnote{This is partly because the default $k$ range is quite small, as seen in Table \ref{tab:halomodel_parameters}. However, for numerical transfer functions such as those from CAMB, the intrinsically calculated range of $k$ is rather small and requires extrapolation even onto the set of $k$ input by the user.}, particularly when transforming from $\xi(r)$ to $P(k)$.
Typical cubic spline interpolation performs very poorly on extrapolation in general, often blowing up so much that the transform can be completely ruined.
We implement a robust \verb|ExtendedSpline| object to circumvent these issues. 
The \verb|ExtendedSpline| interpolates (typically with cubic interpolation) interior to the given co-ordinates, but is supplemented by settable conditions at both ends. 
The conditions at either end can be to return zeros, to continue to extrapolate the spline, to return some user-defined asymptotic function (which is normalised to match at the boundary point) or to extrapolate linearly in log-space (i.e. assuming the asymptotic behaviour on either end of the function is roughly power-law-like). 
We typically set correlation functions to return zero above the point at which they cross zero (typically about 125 Mpc/$h$) and exhibit power-law behaviour on small scales (as long as the slope of the function is shallower than an index of -2 the smallest scales do not contribute significantly in any case). 
Power spectra are considered to be power-laws at high $k$ and power laws explicitly with an index of $n_s$ at low $k$ (except when they are cut-off for some reason, like the 1-halo term, in which case they return zero below the cutoff).

We provide callable functions for each of the power spectra and correlation functions of the halo model, based on these extended splines evaluated on the high-resolution background tabulated data. 
Nevertheless, we still provide the default evaluation of these splines on the user-defined $r$ and $k$ vectors.

\paragraph*{1-halo turnover}
At large scales, the 1-halo power spectrum converges to a non-zero constant (this is clear from \cref{eq:dmpower1}, since $u$ asymptotes to unity as $k\rightarrow0$). This is generally considered to be the ``shot-noise'' term, and is clearly unphysical when considering a smooth distribution. 
\texttt{TracerHaloModel} offers to force the 1-halo term to turnover on scales larger than about 10 halo radii. 

Explicitly, it changes the lower mass limit of the integration in \cref{eq:dmpower1} to be
\begin{equation}
    m_{\rm lim}(k) = \frac{4\pi}{3} \left(\frac{\pi}{10k}\right)^3 \bar{\rho}_0 \Delta_{\rm h}.
\end{equation}

\paragraph*{Large-scale matter-matter bias}
The large-scale matter power under the halo model ansatz (i.e. ``all mass is inside halos'') is by definition equivalent to the linear power. 
Another way of saying this is that the large-scale effective bias, Eq. \ref{eq:beff}, for matter is unity. This, in principle, is true if the consistency relation, Eq. \ref{eq:consistency}, holds. However, there are several ways in which this condition will fail -- either in principle or in practice. 

The first is that in some halo models, the ansatz itself is specifically broken. 
For example, halo models of warm dark matter may invoke a ``smooth'' dark matter 
component that is not collapsed into halos \citep[eg.][]{Smith2011a}. In this case,
the HMF model is specifically tuned to not integrate to the mean density.

The second is when a bias or HMF relation is chosen such that Eq. \ref{eq:consistency} is not satisfied. Sometimes these relations may provide better
accuracy at large masses, at the expense of being incompatible with the halo model ansatz. Typically, in these cases, one still \textit{desires} the behaviour that the matter power converge to the linear power at large scales. The best way to avoid this situation is to avoid choosing incompatible relations when computing matter statistics (such relations are usually fine for tracer statistics, as these generally trace only halos above some minimum mass).

The third is simply due to the inherent limitation of the numerical integration. Even with compatible HMF and bias relations, the numerical integration to compute $b(k, r)$ is over a finite range of mass, and thus will not necessarily converge to unity at large scales. This is always undesirable. 

To deal with these issues, \textsc{halomod} includes an option \texttt{force\_unity\_dm\_bias}, which if set, manually sets the effective bias of matter to unity, and then renormalizes the 2-halo term such that it approaches these effective bias at large scales. It does this irrespective of the HMF and bias relation chosen, which is generally the desired behaviour.

\subsubsection{Projected correlation function}
\label{sec:projcorr}
In analyses of galaxy surveys, it is more common to measure \textit{projected}, rather than real-space, correlation functions (another alternative is the \textit{angular} CF, which we outline in the next section).
We provide an extended framework, inheriting from \texttt{TracerHaloModel}, for this calculation. The primary reason it is separated from the basic class is that it requires extra parameters specific to its calculation. 





The primary addition with respect to the base \texttt{Tracer\-Halo\-Model} class is the conversion of the real-space 2PCF to projected space. The transformation from real-space to projected is defined as
\begin{equation}
	\label{eq:wprp}
	w_p(r_p) = 2\int_{r_p}^\infty \frac{r\xi(r)}{\sqrt{r^2-r_p^2}}{\rm d}r.
\end{equation}
The implementation of this integral is rather delicate due to the singularity at the lower bound. 
This renders both limits of the integration non-physical, and convergence requirements must be met by both. We discuss our prescription for these limits in appendix \ref{app:proj}.

\subsubsection{Angular correlation functions}
\label{sec:halomod:frameworks:angular}
Angular correlation functions can be computed using the \verb|AngularCF| class (which again inherits from \texttt{TracerHaloModel}). 

The angular correlation function between two density fields (potentially the same field in an auto-correlation) is defined as \citep{Simon2007}
\begin{equation}
    w(\theta) \approx \int_0^\infty {\rm d}r_1\int_0^\infty {\rm d}r_2\ p_1(r_1) p_2(r_2) \xi(R,\bar{r}),
\end{equation} 
where $p_1$ and $p_2$ are probability distributions of finding tracers from population 1 and 2 respectively at line-of-sight comoving distance $r$ (i.e. the redshift distribution of sources), and
\begin{equation}
   R \equiv \sqrt{r_1^2 + r_2^2 - 2r_1r_2 \cos \theta}, 
\end{equation}
is a projected separation, and 
\begin{equation}
   \bar{r} = (r_1+r_2)/2. 
\end{equation}

In \halomod, we exclusively use the Limber approximation \citep{Limber1953},
which is a good approximation if $p(r)$ is a wide distribution, and $\theta$ is relatively small. In this approximation we have
\begin{equation}
   w(\theta) = 2\int_0^\infty {\rm d}\bar{r} p_1(\bar{r}) p_2(\bar{r}) \int_0^{\infty} {\rm d}\Delta r \xi(R,\bar{r}), 
\end{equation}
where 
\begin{equation}
   R \equiv \sqrt{\bar{r}^2 \theta^2 + \Delta r^2}.
\end{equation}

\subsubsection{Tracer-Tracer Cross-Correlations}
\label{sec:halomod:frameworks:crosscorr}
While the \texttt{TracerHaloModel} class is able to compute cross-correlations between matter and a tracer field, it is often useful to be able to compute cross-correlations between two tracer fields (eg. optical galaxies and neutral hydrogen, or two classes of optical galaxies). This functionality is supported via the \texttt{CrossCorrelations} class, to which the user provides two \texttt{TracerHaloModel} instances. Note that currently, halo exclusion is not supported in cross-correlations.

In addition to defining the properties of the individual tracer populations, one must define the way the two populations correlate within halos (this is immediately clear when considering two overlapping populations of galaxies, for example). Within \halomod, we use the following relations:
\begin{equation}
    \langle T^{c,s}_{1} T^{c,s}_2 \rangle = \langle T^{c,s}_1 \rangle \langle T^{c,s}_2 \rangle + \sigma^{c,s}_1 \sigma^{c,s}_1 R^{c-s, s-c, s-s}_{12} - \delta_{ss}Q_{ss},
\end{equation}
where $c$ and $s$ stand for central and satellite, and may be the same or different for each of the populations (i.e. we have central-satellite, satellite-central and satellite-satellite terms, but not a central-central term since the central tracer, if it exists, must be identical between the populations).
The final term, which is only non-zero in the satellite-satellite case, is in that case the average occupation of tracers shared between the population per halo of mass $m$. 
This relation is nothing more than the standard definition of the variance of a product of random variables, in which the $\sigma$ is the standard deviation in the occupation for each population, and $R$ is a correlation coefficient between -1 and 1. 
For independent/uncorrelated populations, $R=0$.
For an autocorrelation, $R=1$ and $Q=N_s$, and we obtain the same expressions presented in \S\ref{sec:theory-gal:hod}.
More details and a worked example in the context of cross-correlations of optical galaxies and \textsc{Hi} is provided in \citet{Wolz2019}.

In \halomod, we allow the user to specify the variance of the occupation (for both centrals and satellites) within the \texttt{HOD} component (cf. \S\ref{sec:halomod:components:hod}). We additionally provide a \texttt{HODCross} component that allows combining two HOD models, and allows specifying $R$ for the three available combinations of centrals and satellites.

\subsubsection{Warm dark matter models}
\label{sec:halomod:frameworks:wdm}
For each of the frameworks, we also implement a version suited for warm dark matter (WDM) models. The WDM frameworks are designed to set relevant default models and perform any framework-level WDM-specific modifications.

In particular, we implement the WDM transfer function from \cite{Viel2005}, the WDM-compatible concentration relation of \cite{Ludlow2016}, along with the Sharp-$k$ filter proposed by \cite{Schneider2012} and the general large-scale structure framework for WDM outlined in \cite{Smith2011a}.

\subsection{Components}
\label{sec:halomod:components}
This section provides a brief summary of each of the components within \textsc{halomod}. 
For each component, we provide a table or otherwise concise summary of the models already implemented within \textsc{halomod}, and a comparison plot if relevant.

\subsubsection{Cosmology}
\label{sec:halomod:components:cosmology}
Instead of defining our own basic cosmology models, we use the mature \textsc{astropy} package\footnote{\url{http://www.astropy.org}} \citep{Robitaille2013}. In brief, this package implements a connected series of cosmogaphic models based on the Friedmann-Lemaitre-Robertson-Walker (FLRW) metric, with several standard instances based on results from CMB missions. 

We do not present the summary details here, but note that basic cosmographic quantities such as comoving distance, and density parameters such as the mean density as a function of redshift, are implemented in these models. 

One commonly misunderstood aspect of the cosmology component is that it \textit{does not} include the large-scale-structure (LSS) parameters $\sigma_8$ and the spectral index $n_s$. This is perhaps counter-intuitive, since CMB constraints such as those from WMAP and Planck typically provide constraints on these parameters as part of the fundamental cosmology, and furthermore these parameters are covariant with cosmographic parameters (especially $\sigma_8$ and $\Omega_m$). However, \textsc{astropy} cosmologies are purely cosmographic, and \hmf\ inherits this behaviour. These LSS parameters are instead provided as inputs to the \texttt{Transfer} framework for which they are necessary to determine the normalisation and slope of the power spectrum.

\subsubsection{Transfer functions}
\label{sec:halomod:components:transfer}
We provide a common interface for calculating transfer functions using a variety of methods: from analytic fits, to Boltzmann code, to importing data from file. In each case, the required function returns the logarithmic transfer as a function of logarithmic wavenumber, as lower-order splines can be accurately fit in log space. 

We use the \textsc{pycamb}\footnote{\url{https://github.com/cmbant/CAMB}} wrapper to provide on-the-fly access to \textsc{camb} routines. We also note that we include a standalone version of \textsc{halofit} that may be used to apply nonlinear corrections to any of the included power spectra.

The various transfer models included in \textsc{hmf} are summarized in Table \ref{tab:models_transfer}. A comparison of the models with default parameters is shown in \cref{fig:power}.





\begin{table*}
\centering
 {\tabulinesep=1.3mm
\begin{tabu}{X[1.5l]X[1l]X[3l]X[3l]} 
\toprule[0.05cm]
\textsc{Ref} & \textsc{Name} & \textsc{Formula} & \textsc{Params} \\
\midrule[0.05cm]
\citet{Bond1984} & \texttt{BondEfs} & $\left[1 + \left(\sum_{i=2}^4 (ck)^i \right)^\nu\right]^{-1/\nu}$ & $\displaystyle \frac{\Omega_m h^2}{0.3\times 0.7^2} c = (37.1,21.1,10.8)$, $\nu=1.12$ \\
\citet{Bardeen1986} & \texttt{BBKS} & $\displaystyle \frac{\ln(1+aq)}{aq} \left[\sum_{i=0}^4 (c_i q)^i\right]^{-1/4}$ & $\displaystyle q=\frac{k\exp[\Omega_b(1 + \sqrt{2h}/\Omega_m)]}{\Gamma h {\rm Mpc}^{-1} }$, $\Gamma = \Omega_mh$,  $a=2.34$, $c=$(1, 3.89, 16.1, 5.46, 6.71) \\
\citet{Eisenstein1998} (No BAO) & \texttt{EH\_NoBAO} & Original paper,Eq. [26] and [28]-[31] & \\
\citet{Eisenstein1998} (w/ BAO) & \texttt{EH\_BAO} & Original paper, \S 2 and Eq. [16]-[24] & \\
\citet{Lewis2000} & \texttt{CAMB} & Boltzmann Code & Many options \\
 & \texttt{FromFile} & Read from file & Filename \\
 & \texttt{FromArray} & Use pre-computed array & Array \\
 
\bottomrule[0.05cm]
\end{tabu}}
\caption{Summary of included \texttt{Transfer} models. Note that the $\Omega_b$-dependent exponential factor in the BBKS model of $q$ is drawn from \citet{Sugiyama1995} and \citet{Liddle2000}, and is optional within \halomod. Note also that these equations assume $k$ to be in units of $h/{\rm Mpc}$, which explains why $q$ is divided by $h$ instead of $h^2$.}
\label{tab:models_transfer}
\end{table*}

\subsubsection{Mass Definitions}
\label{sec:halomod:components:mass-def}
\textsc{halomod} contains a flexible system for defining halo masses. 
Both FOF and SO halos are supported, and three general SO subclasses in which the overdensity is with respect to the mean background or the critical density, or is defined as the virial overdensity as a function of redshift \citep{Bryan1998}. 

All mass definitions have a free parameter -- either the linking length or the overdensity criterion. All of the definitions are able to calculate the halo density (or overdensity with respect to either mean or critical). For the FOF definition we use the approximate formula from \cite{White2001};
\begin{equation}
    \rho_h^{\rm FOF} \approx \bar{\rho} \frac{9}{2\pi b^3},
\end{equation}
with $b$ the linking length. Note that this formula is very approximate \citep{More2011}, as FOF halos can't readily be identified with SO halos. 

\begin{figure}
  \centering
  \includegraphics[width=\linewidth]{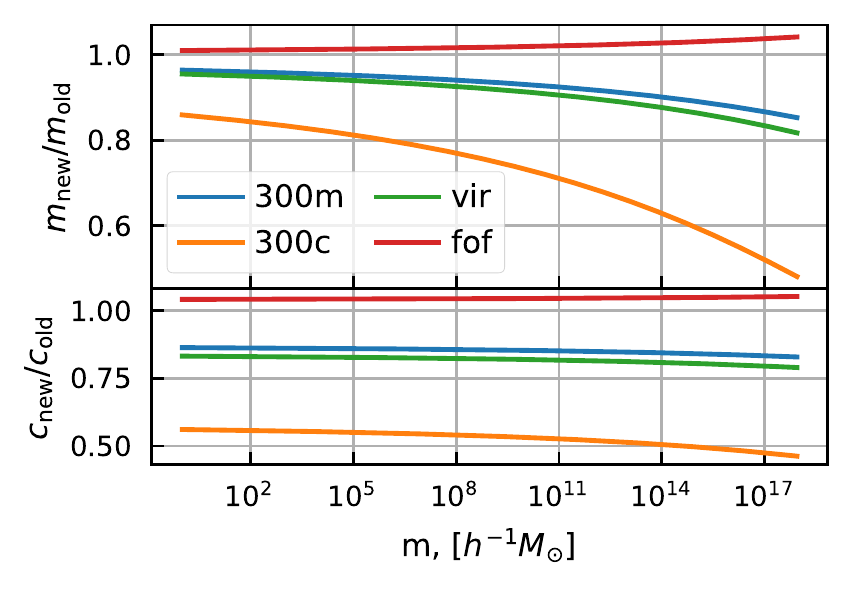}
  \caption[Mass conversion]{Conversion of mass and concentration between different halo definitions. Top-panel shows ratio of converted mass to input mass, as a function of input mass. Different colors represent different halo definitions. Bottom panel shows ratio of converted
  concentration to input concentration, as a function of input mass.
  Input mass definition is 200m. Note that the trend with increasing overdensity is to decrease both the mass and concentration. 
  The halo profile used in the conversion is the NFW profile, and the concentration-mass relation is the simple power law of \cite{Duffy2008}.}
  \label{fig:mass_conversion}
\end{figure}

To change mass definition requires solving for a new halo concentration, according to
\begin{equation}
    \label{eq:convert_mass_c}
    \frac{c_{\rm old}^3}{c_{\rm new}^3} \frac{h(c_{\rm new})}{h(c_{\rm old})} = \frac{\Delta_{\rm new}}{\Delta_{\rm old}},
\end{equation}
which is clearly profile-dependent. Then the new mass is just
\begin{equation}
    \label{eq:convert_mass_m}
    m_{\rm new} = m_{\rm old} \frac{c_{\rm new}^3}{c_{\rm old}^3} \frac{\Delta_{\rm new}}{\Delta_{\rm old}}.
\end{equation}
\halomod\  supports changing masses and concentrations (partially) between mass definitions in this way. 
See Fig. \ref{fig:mass_conversion} for a comparison of mass and concentration for different halo mass definitions.

\subsubsection{Window functions/Filters}
\label{sec:halomod:components:window}
The window/filter function contributes to the basic unit of the spherical collapse formalism, the peak-height $\nu$. It is also responsible for the conversion of distance scales to mass scales (cf. \S\ref{sec:theory:filter}).

Our general approach in this component is to specify all the quantities in terms of the Fourier co-ordinate, $k$, and the smoothing scale $R$. The class itself defines the relationship between mass $m$ and the smoothing scale. The primary quantity of interest is the window function itself, specified in Fourier space, $W(x)$ (one can optionally specify the real-space version as well, though it is not involved in typical HM calculations). With this function, the mass variance can be calculated using \cref{eq:massvariance}. 

To calculate $dn/dm$ requires also the logarithmic derivative of the variance with mass. To remain as general as possible, we use the following identity
\begin{equation}
    \label{eq:dlnssdlnm}
     \frac{{\rm d} \ln \sigma}{{\rm d} \ln m} \equiv \frac{1}{2}\frac{{\rm d} \ln \sigma^2}{{\rm d} \ln R} \frac{{\rm d} \ln R}{{\rm d} \ln m},
\end{equation}
where 
\begin{equation}
    \label{eq:dlnssdlnr}
    \frac{{\rm d}\ln \sigma^2}{{\rm d}\ln R} = \frac{1}{\pi^2 \sigma^2}\int W(kR) \frac{{\rm d}W(kR)}{{\rm d}\ln (kR)} P(k) k^2 {\rm d}k.
\end{equation}
The factor $d\ln R/d\ln m$ is typically 1/3, though this is not necessarily the case for window functions of arbitrary shape\footnote{The mass referenced here is the mass within a shaped volume defined by the window function.} \citep{Schneider2013}. Thus, the defining quantity is the derivative of the window function, $dW(x)/dx$. 

We also note that we generalise the calculation of the mass variance to all moments of the smoothed density field:
\begin{equation}
    \label{eq:moments}
    \sigma_n^2(r) = \frac{1}{2\pi^2}\int k^{2n} P(k) W^2(kR) {\rm d}k,
\end{equation}
for which $\sigma_0^2$ is the usual mass variance.

\begin{figure}
  \centering
  \includegraphics[width=\linewidth]{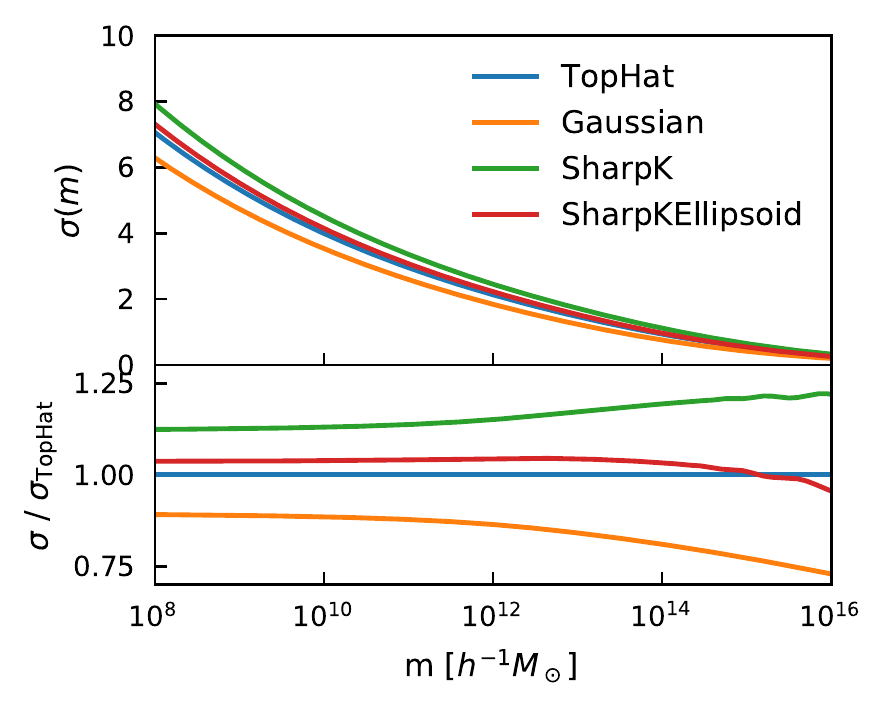}
  \caption[Mass variance using three different filters]{The mass variance determined by the three most common window functions. The normalisation of the power spectrum in each case uses the \texttt{TopHat} filter, for which $\sigma_8$ is defined.}
  \label{fig:filter_sigma}
\end{figure}

Table \ref{tab:models_window} summarises the window functions included in \textsc{halomod}, while \cref{fig:filter_sigma} displays the value of $\sigma(m)$ for each. Note that there is a complex interplay between the role of $W(kR)$ and the mass assignment which drives changes between the models. Note also that details for \texttt{SharpKEllipsoid} can be found in Appendix A of \cite{Schneider2013}.

\begin{table}
\centering
 {\tabulinesep=1.3mm
\begin{tabu}{X[1.1l]X[2.4l]} 
\toprule[0.05cm]
\textsc{Name} & \textsc{Formula}  \\
\midrule[0.05cm]

\multirow{4}{*}{\texttt{TopHat}}  & $\displaystyle W(r,R) = H(R-r)$ \\
 & $\displaystyle W(x=kR) = \frac{3}{x^3}\left(\sin x - x \cos x\right)$  \\
 & $\displaystyle m(R) = \frac{4\pi}{3}R^3 \bar{\rho}$ \\
 & $\displaystyle \frac{{\rm d} W}{{\rm d} \ln x} = \frac{9  x \cos x + 3 (x^2 - 3)  \sin x) }{ x ^3}$  \\
\midrule

\multirow{5}{*}{\texttt{SharpK}}  & $\displaystyle W(r,R) = \frac{\sin(r/R) - (r/R)\cos(r/R)}{2\pi^2 r^3}$   \\
 & $\displaystyle W(x=kR) = H(kR -1)$ \\
 & $\displaystyle m(R) = \frac{4\pi}{3}\left[cR\right]^3 \bar{\rho}$   \\
 & $\displaystyle \frac{{\rm d} W}{{\rm d} \ln x} = \delta_D(x-1)$  \\
 & $\displaystyle \frac{{\rm d} \ln \sigma^2(r)}{{\rm d}\ln r} = -\frac{P(1/r)}{2\pi^2 \sigma^2(r) r^3} $  \\
 \midrule
 
\multirow{1}{*}{\texttt{SharpKEllipsoid}}  & Same as \texttt{SharpK} except $m(R)$.  \\
\midrule

\multirow{4}{*}{\texttt{Gaussian}}  & $\displaystyle W(r,R) = \frac{\exp(-r^2/2R^2)}{(2\pi)^{3/2} R^3}$  \\
 & $\displaystyle W(x=kR) = \exp(-x^2/2)$ \\
 & $\displaystyle m(R) = (2\pi)^{3/2}R^3 \bar{\rho}$  \\
 & $\displaystyle \frac{{\rm d} W}{{\rm d} \ln x} = -xW(x)$  \\

\bottomrule[0.05cm]
\end{tabu}}
\caption[Summary of included \texttt{Filter} models]{Summary of included \texttt{Filter} models. Note: $H$ is the Heaviside step-function, $\delta_D$ is the Dirac-delta function and $P$ is the power spectrum. Also, in \texttt{SharpK}, the mass-assignment is not well-defined, but we use the given formula with $c\approx2.5$ a fit to simulations \protect\citep{Benson2012,Schneider2013}.}
\label{tab:models_window}
\end{table}

\subsubsection{HMF}
\label{sec:halomod:components:hmf}
The diversity among fitting functions for the HMF, not only in functional forms, but in general approach and explicit parameter dependence, requires that quite a number of variables be available in the general case. However, the typical fit uses just $\nu$ (or sometimes $\sigma$) and its own model parameters, reflecting the expected universality of the fit. 

Each HMF model has a class attribute \texttt{normalized}, which specifies whether the particular HMF integrates to the mean density (i.e. all mass is in halos). Note that this is an \textit{in principle} attribute -- when actually integrating the HMF, a limited mass range is used, and there is no guarantee that any particular model (even if normalized in principle) will integrate to the mean density.
%

The mass function is dependent on the halo mass definition.
For a given halo mass, a higher overdensity criterion means that fewer peaks will exist. 
This effectively changes $f(\nu)$, since all other factors in Eq. \ref{eq:theory:hmf_nu} remain constant with mass definition (as they depend on the Lagrangian mass). 
The precise form of this dependence is extremely complicated; 
for one, changing mass definition does not necessarily evolve the masses of individual halos smoothly -- there is considerable scatter in the distribution of profiles for halos of a given mass, and for high masses where the mass function is extremely sensitive to mass, this can result in Eddington bias.
Furthermore, the under the halo model ansatz, all mass is contained in halos, so by contracting a halo to a new overdensity, one does not merely change the mass of the given halo, but also necessarily creates several new lower-mass halos that were previously subsumed in the larger halo. 
Thus, converting a mass function from one definition to another is fraught with peril.

Nevertheless, there is a reasonably simple approximate conversion that can be made. For example, \citet{Bocquet2016} uses the chain rule to give the relation
\begin{equation}
    \frac{dn}{dm'} = \frac{dn}{dm} \frac{dm}{dm'} = f(\nu) \frac{\rho_0}{m'} \frac{{\rm d}\nu}{{\rm d}m'}\times \frac{m'}{m},
\end{equation}
where the modification to the mass function is contained in the final $m'/m$, which can be computed (approximately) using Eqs. \ref{eq:convert_mass_c} and \ref{eq:convert_mass_m}.
However, it is important to notice the assumptions in this conversion: here we are preserving the total number density of halos (over all mass), but not the total mass density. That is, for each existing halo at some mass definition, we are reducing/extending that halo to encompass less/more mass, but not creating/destroying more halos from the remains. 
Importantly, this means that normalized HMFs will no longer obey this important halo model consistency relation for other mass definitions. 
Furthermore, there are significant errors introduced by such `translations' in comparison to properly re-fitting the mass function measured at the alternative definition (eg. fits from \citet{Bocquet2016} suggest a 20\% error at low mass which diverges exponentially above the nonlinear mass). 
\textsc{hmf} allows automatically converting mass functions from their native mass definition to other definitions using this formalism, but this feature must be switched on manually by turning off the setting \texttt{disable\_mass\_conversion} (otherwise an inconsistent halo definition will cause an error to be raised).
Note that most fits that use spherical overdensity halos are directly calibrated to varying overdensity criteria \citep[eg.][]{Tinker2008,Watson2013,Bocquet2016}, and \textsc{halomod} does not use the above conversion in these cases.

We present a selection of the 21 available fitting functions available in \textsc{hmf} in \cref{fig:mfs}.
We refer the reader to \cite{Murray2013a} for a table of included fitting functions, noting that five more fits have been added to the collection, specifically those of \citet{Tinker2010,Behroozi2013a,Manera2010,Pillepich2010} and \citet{Ishiyama2015}. 


\subsubsection{Bias}
\label{sec:halomod:components:bias}
Though the majority of bias functions are specified in `universal' form with respect to the peak-height, there are some older functions that explicitly require cosmology and are specified in terms of a scaled mass, $m/M_\star$. We thus provide access to these quantities, which may also serve to enable more fine-tuned models than possible with pure peak-background split arguments in the future. 

Table \ref{tab:models_bias} compiles the range of included models, and \cref{fig:bias_functions} displays a selection of them.
Note that \halomod\ defines an explicit interface with \textsc{colossus}, enabling all of its bias models to be used in \halomod, including those of \cite{Bhattacharya2011} and \cite{Comparat2017}, which are not included directly in \halomod.

Importantly, Table \ref{tab:models_bias} lists the paired HMF for each bias model, for which the consistency relation, Eq. \ref{eq:consistency}, holds. 
By default, \textsc{halomod} will choose the paired HMF once a bias function is chosen.
The paired HMF of any bias model can be determined by accessing the \texttt{pair\_hmf} attribute of the model.

Note also that in the case that the HMF and bias function chosen support the consistency relation, by \textit{default} the large-scale 2-halo term of matter will be normalized to ensure it converges to $P_m(k)$ (i.e. a unity effective bias of matter). This needs to be manually enforced, because the numerical integration of the biased HMF does not necessarily converge to the mean matter density over the restricted range of mass included in the calculation. This normalization is implemented as a scalar multiplication of the entire 2-halo term, so as to avoid sharp features in the spectrum. This behaviour can be turned \textit{off} by setting \texttt{force\_unity\_dm\_bias} to false. 

\begingroup
 \small
\begin{table*} 
\centering

\begin{tabular}{>{\raggedright}m{3cm} >{\raggedright}m{2.6cm} >{\raggedright}m{5.2cm} >{\raggedright\arraybackslash}m{4.2cm} >{\raggedright\arraybackslash}m{1.2cm}}
\toprule
 \textsc{Ref} & \textsc{Name} & \textsc{Formula} & \textsc{Params} & \textsc{hmf}\\
\toprule
 
 \citet{Mo1996} & \texttt{Mo96} & $\displaystyle 1 + \frac{\nu - 1}{\delta_c}$ &  & \texttt{PS} \\
 \midrule
 \citet{Jing1998} & \texttt{Jing98} & $\displaystyle (a/x^2 +1)^{b-cn_s} \left(1 + \frac{x-1}{\delta_c}\right)$ & $x = (m/M_\star)^2$, $a=0.5$, $b=0.06$, $c=0.02$ & \\
 \midrule
 
 \citet{Sheth1999} & \texttt{ST99} & \multirow{2}{*}{$\displaystyle 1 + \frac{q\nu -1}{\delta_c} + \frac{2p/\delta_c}{1+(q\nu)^p}$} & $q=0.707$, $p=0.3$ & \multirow{2}{*}{\texttt{SMT} }\\
\citet{Mandelbaum2005} & \texttt{Mandelbaum05} & & $q=0.73$, $p=0.15$ & \\
\citet{Manera2010} & \texttt{Manera10} & & $q=0.709$, $p=0.248$ & \\
\midrule

\citet{Sheth2001} &\texttt{SMT01} & \multirow{2}{*} { \pbox{20cm}{$\displaystyle  1 +\frac{1}{\sqrt{a}\delta_c}\bigg[s\sqrt{a}(a\nu)+\sqrt{a}b(a\nu)^{1-c}$ \\
$\displaystyle \hphantom{1 +}-\left.\frac{(a\nu)^c}{(a\nu)^c + b(1-c)(1-c/2)}\right]$}} & $a=0.707$, $b=0.5$, $c=0.6$ & \multirow{2}{*}{\texttt{SMT}} \\

 \citet{Tinker2005} & \texttt{Tinker05} \phantom{hey there} &  & $a=0.707$, $b=0.35$, $c=0.8$  &\\
 & & & &\\
\midrule

\citet{Seljak2004} & \texttt{Seljak04} & $\displaystyle a + bx^c + \frac{d}{ex+1} + fx^g$ & $x=m/M_\star$, $a=0.53$, $b=0.39$, $c=0.45$, $d=0.13$, $e=40$, $f=5\times10^{-4}$, $g=1.5$ & \\
\midrule
\citet{Seljak2004} & \texttt{Seljak04Cosmo} & $\displaystyle b_\text{Sel} + \log_{10}x \left[a_1(\Omega'_m + n'_s) + a_2(\sigma'_8 + h')\right]$ & $a_1 = 0.4$, $a_2=0.3$ $\Omega'_m=\Omega_m-0.3$, $n'_s = n_s-1$, $\sigma'_8 = \sigma_8-0.9$, $h' = h-0.7$& \\
\midrule
\citet{Pillepich2010} & \texttt{Pillepich10} & $\displaystyle B_0 + B_1\sqrt{\nu} + B_2\nu$ & $B_0=0.647$, $B_1=-0.320$, $B_2=0.568$ & \\
\midrule
 \citet{Tinker2010} & \texttt{Tinker10} & $\displaystyle 1 - A\frac{\nu^{a/2}}{\nu^{a/2} + \delta_c^a} + B\nu^{b/2} + C\nu^{c/2}$ & $A=1+0.24y\exp[-(4/y)^4]$, $a = 0.44 y - 0.88$, $B = 0.183$, $b = 1.5$, $C = 0.019+0.107y+0.19\exp[-(4/y)^4]$, $c = 2.4$, $y=\log_{10}\Delta_{\rm h}$. &  \\
 \midrule
 \citet{Tinker2010} & \texttt{Tinker10PBsplit} & $\displaystyle 1 + \frac{\gamma\nu-(1+2\eta)}{\delta_c} + \frac{2\phi/\delta_c}{1+(\beta^2\nu)^\phi}$ & Same as $f_\text{Tinker10}$ & \texttt{Tinker10} \\
 \midrule
 & \texttt{UnityBias} & $b(\nu) = 1$ & & \\
 \bottomrule
 \end{tabular}
 \caption[Summary of included \texttt{Bias} models]{Summary of included \texttt{Bias} models. Note that the fit of \texttt{Manera10} is dependent on Friends-of-Friends linking length and redshift, and we give the result for $l = 0.2$ at $z=0$.}
 
\label{tab:models_bias}

\end{table*}
\endgroup

\subsubsection{Halo Profiles}
\label{sec:halomod:components:profile}
We developed a systematic representation of two-param\-eter universal halo profiles in \S\ref{sec:profilestheory}. Our implementation follows this development closely. There is a richness of predictions available given just the basic profile shape, which makes the class structure of the \verb|Profile| component extremely valuable. We have not yet implemented several generic predictions that may be added in future versions, such as the gravitational potential profile.

The \verb|Profile| component provides a wide range of derived quantities, accessible with a simple definition of the profile shape. To make this more explicit, suppose that one wished to implement a ``cored" NFW, such that the inner density was a constant $\rho_s$. The minimum code that the user must write (remember, this is user-side code, not touching the \halomod\ source) would be:
\begin{lstlisting}[language=Python]
from halomod.profiles import Profile
class CoredNFW(Profile):
	def f(self,x):
		return 1./(1 + x*(1+x)**2)
\end{lstlisting}
The class itself can then be used directly or passed to any relevant \framework, and all of the necessary quantities will be available. For increased efficiency, it may also be beneficial to analytically define $p(\kappa,c)$, $h(c)$ and even $l(x,c)$ if available. If not specified, these quantities are evaluated via numerical integration (except for $l(x,c)$ -- it is usually more efficient to use an analytic Fourier-transform $u^2(k|m)$ and then use a Hankel transform on the result).

We define our implemented profiles in Table \ref{tab:models_profile}, and show them in and we note that for each, analytic forms have been given wherever possible to improve efficiency. In general it is non-trivial to produce analytic forms for the self-convolution. Fortunately, this can be achieved with the popular NFW profile \citep{Sheth2001a}, as we show in Table \ref{tab:models_profile}. The associated forms for the $T_i$ are
\begin{subequations}
    \label{eq:nfw_t}
    \begin{align}
        T_1 &= \frac{-4(1+a)+2ax(1+2a)+a^2x^2}{2x^2(1+a)^2(2+x)},\\
        T_2 &= \frac{1}{x^2}\ln\left[\frac{(1+a-ax)(1+x)}{1+a}\right], \\
        T_3 &= \frac{\ln(1+x)}{x(2+x)^2}, \\
        T_4 &= \frac{\ln[(1+a)/(ax+a-1)]}{x(2+x)^2}, \\
        T_5 &= \frac{a^2x-2a}{2x(1+a)^2(2+x)}.
    \end{align}
\end{subequations}

\begin{figure*}
    \centering
    \includegraphics{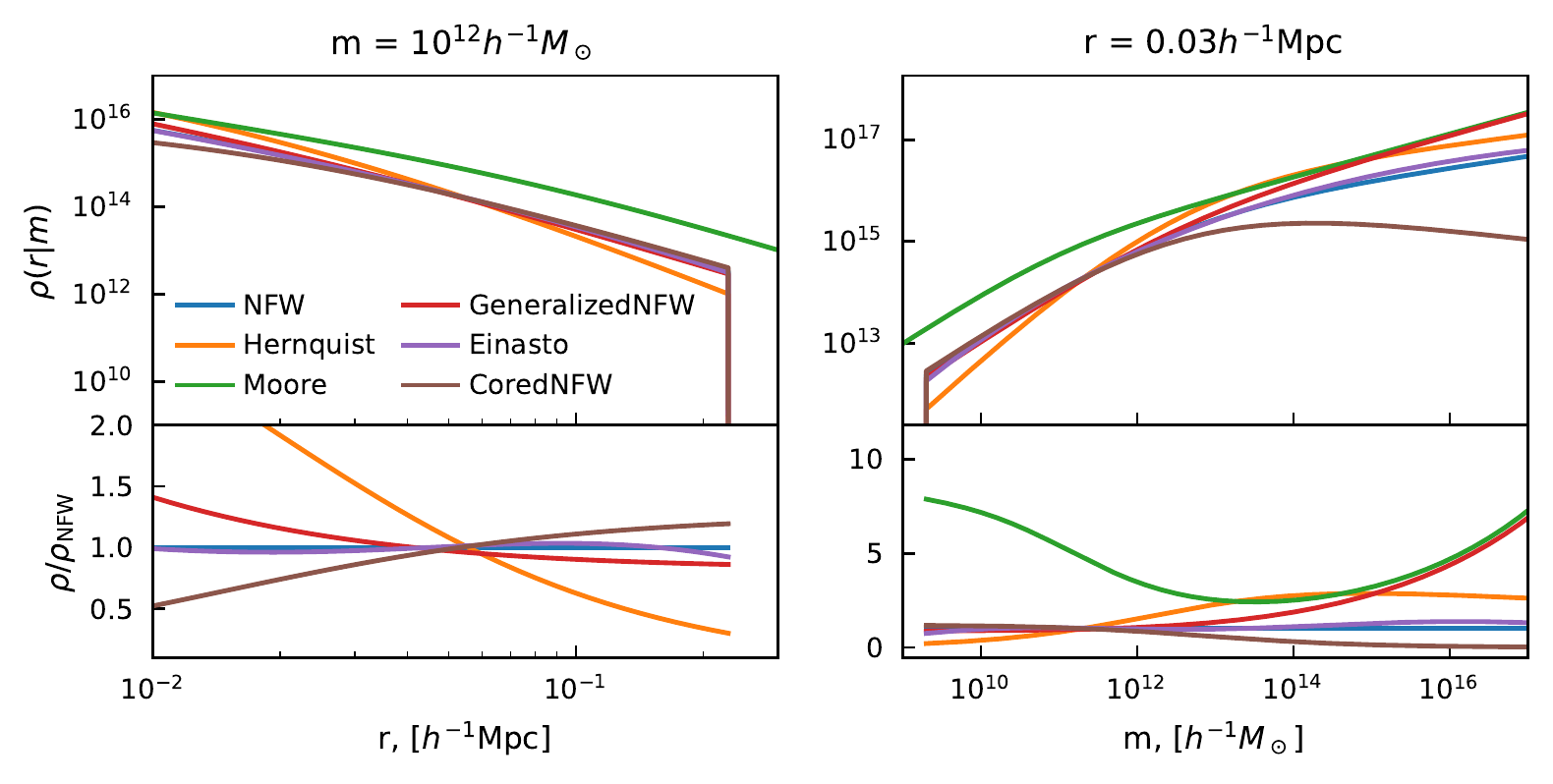}
    \caption{Illustration of the halo profile models implemented in \texttt{halomod}. Left panel shows the profile as a function of radius at $m=10^{12} h^{-1}M_\odot$, and the right panel shows the profile as a function of mass at $r = 0.03 h^{-1}$Mpc. The general trends are similar, though the inner and outer slopes vary between models. The generalized NFW shown here has $\alpha=1.5$.}
    \label{fig:profiles}
\end{figure*}

For the profile of \citet{Moore1998} and the generalised NFW and Moore profiles, the non-truncated self-convolution can be reduced to a single integral \citep{Ma2000},
\begin{equation}
    l(x) = \frac{2\pi}{x}\int_0^\infty yf(y) F^X_\alpha(x,y) {\rm d}y,
\end{equation}
where $F$ describes the angular part of the 3D integral. In the case of the generalised NFW, this is fully analytic:
\begin{equation}
\begin{split}
    F^N_\alpha(x,y) = \frac{1}{2-\alpha}\Bigg[ &\frac{(x+y)^{2-\alpha}}{(1+x+y)^{2-\alpha}} \\& \red{-} \frac{|x-y|^{2-\alpha}}{(1+|x-y|)^{2-\alpha}}\Bigg],
\end{split}
\end{equation}
while in the generalised Moore profile, we have
\begin{equation}
    F^M_\alpha(x,y) = \int_{|x-y|}^{x+y} \frac{z {\rm dz}}{z^p(1+z^{3-\alpha})},
\end{equation}
which for the Moore profile with $\alpha=3/2$ yields
\begin{equation}
\begin{split}
    F^M_{3/2} = \frac{1}{3}\Big[ &2\sqrt{3} \tan^{-1}\left(\frac{-1+2\sqrt{z}}{\sqrt{3}}\right) \\&+ \ln\left(\frac{1+2\sqrt{z} + z}{1-\sqrt{z}+z}\right)\Big]\Bigg|_{|x-y|}^{x+y}.
\end{split}
\end{equation}

\begingroup
\small
\begin{table*} 
\begin{tabular}{ m{4cm} m{12cm}} 
\toprule
\textsc{Name \& Refs.} & \textsc{Formulae} \\
\toprule
\texttt{NFW} & $\displaystyle f(x) = x^{-1}(1+x)^{-2}$ \\
\citet{Navarro1997} & $\displaystyle  h(c) = \frac{c}{1+c} + \ln(1+c)$ \\
\citet{Sheth2001a} & $\displaystyle    p(\kappa,c) = \cos (\kappa) \left[\text{Ci}(c \kappa+\kappa)-\text{Ci}(\kappa)\right] 
                   +\sin (\kappa) \left[\text{Si}(c \kappa+\kappa)-\text{Si}(\kappa)\right]-\frac{\sin (c \kappa)}{c \kappa+\kappa}$ \\
\citet{Ma2000} & $\displaystyle    p(\kappa) =  \frac{1}{2} \left[(\pi -2 \text{Si}|\kappa|) \sin|\kappa|-2 \cos (\kappa) \text{Ci}| \kappa| \right]$ \\
& $\displaystyle  l(x,c) = 4\pi \begin{cases} (T_1 + T_2 +T_3) & 0\leq x \leq c \\
 	(T_4 + T_5) & c\leq x \leq 2c \end{cases}$ \\
 & $\displaystyle l(x) = \frac{8\pi}{x^2(x+2)}\left[\frac{(x^2+2x+2)\ln(1+x)}{x(x+2)} -1 \right]$ \\
 \midrule
 
\texttt{Moore}   &   $\displaystyle f(x) = \frac{1}{x^{3/2}(1+x^{3/2})}$ \\
   \citet{Moore1998}  & $\displaystyle h(c) = \frac{2}{3}\ln \left(c^{3/2}+1\right) $ \\
\citet{Ma2000} & $\displaystyle    p(\kappa) = \frac{G^{7,3}_{3,9}\left(\frac{\kappa^6}{46656}\middle|\frac{1}{12}\left(
\begin{array}{c}
 2,5,11 \\
 2,2,5,6,8,10,11,0,4 \\ 
\end{array}
\right) \right)}{4 \sqrt{3} \pi^{5/2} | \kappa |}$ \\
& $\displaystyle    l(x) = \frac{2\pi}{x}\int_0^\infty  yf(y) F^M_{3/2}(x,y) {\rm d}y$ \\
\midrule

\texttt{Hernquist}   & $\displaystyle f(x) = \frac{1}{x(1+x)^3}$ \\
  \citet{Hernquist1990} & $\displaystyle  h(c) = \frac{c^2}{2(1+c)^2}$ \\
\citet{Sheth2001a} & $\displaystyle   p(\kappa) = \frac{1}{4} \left[-|\kappa| (2 \text{Ci}|\kappa| \sin|\kappa|+\pi\cos(\kappa)) \right.
    \left.+2\kappa \text{Si}(\kappa) \cos (\kappa)+2\right]$ \\ 
  & $\displaystyle p(\kappa,c) =  1/2 - \frac{\sin (c \kappa)+(c+1)
   \kappa \cos (c \kappa)}{2\kappa(c+1)^2} $ \\
& $\displaystyle \ \ \ \ \ \ \ \ \ 
   + \frac{k}{2} \left[\sin (\kappa) (\text{Ci}(c \kappa+\kappa)-\text{Ci}(\kappa)) \right.
   \left. +\cos (\kappa) (\text{Si}(\kappa)-\text{Si}(c \kappa+\kappa))\right] $\\
  & $\displaystyle  l(x) = \frac{16\pi(h_1 - h_2)}{x^4 * (2 + x)^4} $\\
  & $\displaystyle  h_1(x) = \frac{24 + 60 x + 56 x ^2 + 24x^3 + 6x^4 + x^5}{1 + x} $ \\
  & $\displaystyle  h_2(x) = 12(1 + x)(2 + 2x + x^2)\ln(1 + x) / x$ \\
 \midrule
 
\texttt{GeneralizedNFW} & $\displaystyle    f(x) = \frac{1}{x^\alpha (1+x)^{3-\alpha}}, \ \ \ \  h(c) = -(-c)^\alpha c^{-\alpha}B_{-c}(3-\alpha,\alpha-2)$\\
\citet{Ma2000} & $\displaystyle   p(\kappa) = \frac{2^{-\alpha}}{\sqrt{\pi}\Gamma(3-\alpha)}G^{3,2}_{2,4}\left(\frac{\kappa^2}{4}\middle| {\begin{array}{c}
                      (\alpha-2)/2,(\alpha-1)/2 \\
                        0,0,1/2,-1/2 \\
                    \end{array}}\right) $ \\
& $\displaystyle l(x) = \frac{2\pi}{(2-\alpha)x} \int_0^\infty  yf(y) F^N_\alpha(x,y) {\rm d}y$ \\
\midrule

\texttt{GeneralizedMoore} & $\displaystyle   f(x) = \frac{1}{x^\alpha (1+x^{3-\alpha})} $\\
\citet{Ma2000} & $\displaystyle    l(x)  = \frac{2\pi}{x}\int_0^\infty  yf(y) F^M_\alpha(x,y) {\rm d}y $ \\
\midrule

\texttt{Uniform} & $ \displaystyle f(x) = 1, \ \ \ \    h(c) = c^3/3, \ \ \ \  p(\kappa,c) = \frac{\sin(c\kappa) - c\kappa \cos(c\kappa)}{\kappa^3}$\\
\midrule

\texttt{Einasto} & $\displaystyle f(x|\alpha) = \exp\left(-\frac{2(x^\alpha - 1)}{\alpha}\right)$\\
\citet{Einasto1965} & $\displaystyle  h(c|\alpha) = \frac{e^{2 / \alpha}}{\alpha} \left(\frac{2}{\alpha}\right)^{-3/\alpha} \Gamma\left(\frac{3}{\alpha}, \frac{2 c^\alpha}{\alpha}\right)$\\

\bottomrule
\end{tabular}
\caption[Summary of included \texttt{Profile} models]{Summary of included \texttt{Profile} models. Note that for $p$ and $l$, functions explicitly including $c$ are the usual profiles truncated at $r_\Delta$, whereas those that do not explicitly include $c$ are not truncated (these are rather unhelpful in terms of the HM). $G$ is the Meijer-$G$ function. The functions $T_i$ in the NFW definition are found in the text, \cref{eq:nfw_t}.}
\label{tab:models_profile}
\end{table*}

\endgroup

\subsubsection{Concentration-mass relations}
\label{sec:halomod:components:concentration}
In section \ref{sec:theory:concentration}, we described different classes of models of the $c$--$m$ relation. We have implemented a number of models in \halomod; we summarize them in table \ref{tab:models_concentration} but refer the reader to the respective papers for details.

In particular, we have implemented two age-based models, namely those of \citet{Bullock2001} and \citet{Ludlow2016}. We use interpolating splines to efficiently perform the numerical inversions involved. The results of both models can be approximated with simpler fitting functions, which are also included. Moreover, we add the power-law fits of \citet{Duffy2008} and \citet{Zehavi2011}. 
Note that the power-law fits provided are based on a particular cosmology, and will be inaccurate if used for a model with a different cosmology (the sensitivity to the cosmology is not high, and \textsc{halomod} does not warn the user if the cosmology does not match). Thus, for general purpose applications we recommend an analytic model. 

Concentration-mass relations, as we have seen in \S\ref{sec:halomod:components:mass-def}, are sensitive to the halo definition. 
In principle, the concentrations can be `translated' from one definition to another \citep{Diemer2015} given a halo profile. However, there is a significant scatter in this relationship, and it is better to work within the halo definition native to the measured concentration-mass relation.
For included concentration models in \textsc{halomod}, if the halo definition is not compatible with the measurement, a warning is issued -- but there is no attempt to translate the relation. This behaviour may be updated in the future to automatically translate the relation between halo definitions.

\begin{figure}
  \centering
  \includegraphics[width=\linewidth]{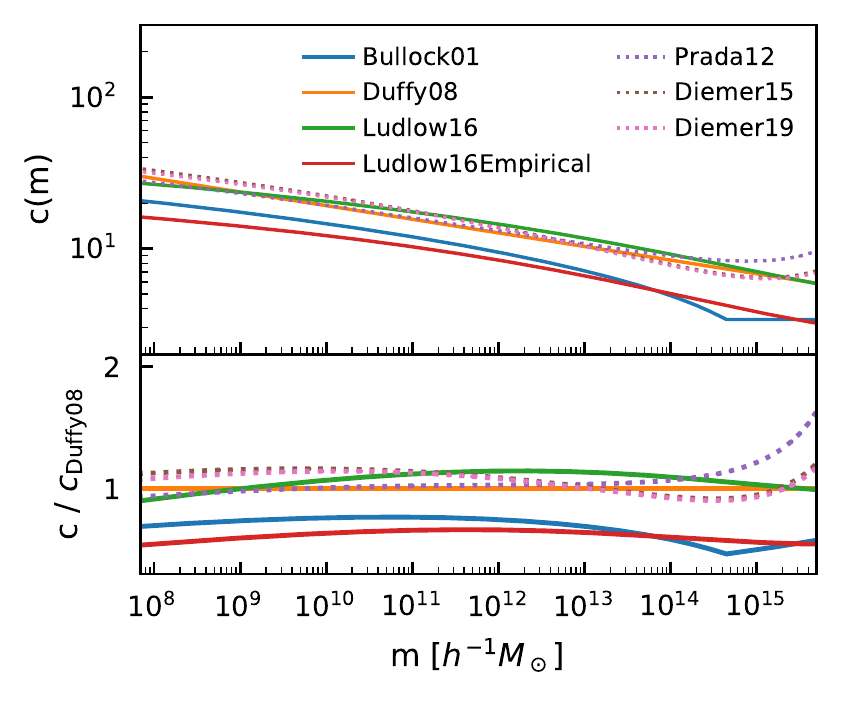}
  \caption[Concentration-mass-redshift relation for various models in the literature]{Illustration of the concentration-mass relations implemented in \halomod, focusing on analytic models (with the exception of the class power-law fit of \citealt{Duffy2008}). Models plotted with dotted curves are obtained via the explicit interface with \textsc{colossus}, and are able to be used within the entire halo model framework. }
  \label{fig:concentration}
\end{figure}

\cref{fig:concentration} illustrates the variations between the models at $z=0$ for the default cosmology of \cite{PlanckCollaboration2015}. 

Note that \halomod\ includes an explicit interface for using the excellent suite of concentration-mass relations from \textsc{colossus}, which provides access to extra models such as the recent analytical models of \cite{Prada2012} and \cite{Diemer2019}. Importantly, models from \textsc{colossus} are able to translate between halo mass definitions. These models are plotted for comparison in \cref{fig:concentration}, but we do not include them in Table \ref{tab:models_concentration} and refer the reader to their respective papers.

\begin{table*}
\centering
\begin{tabu}{X[1.7l]X[1.5l]X[2l]X[3.9l]}
\toprule[0.05cm] 
Ref. & Name & Formula & Params. \\
\toprule[0.05cm]
      
        \citet{Bullock2001} & \texttt{Bullock01} & $\displaystyle K\frac{1+z_c}{1+z}$ & $F=0.001$, $K=3.4$, $z_c$ given in Ref. \\ 
        
        \citet{Ludlow2016} & \texttt{Ludlow16} & See ref. & \\
        \citet{Ludlow2016} & \texttt{Ludlow16Empirical} & $\displaystyle \frac{c_0 (\nu/\nu_0)^{-\gamma_1}}{ \left(1 + \sqrt[\beta]{\nu/\nu_0}\right)^{\beta(\gamma_2 - \gamma_1)}} $ & $c_0 = 3.395(1+z)^{0.215}, \ \ \beta = 0.307(1+z)^{0.54},$ \\
        & & & $\displaystyle \gamma_1 = 0.628(1+z)^{-0.047}, \ \ \gamma_2 = 0.317(1+z)^{-0.893}$\\ 
        \midrule

        \citet{Bullock2001} & \texttt{Bullock01\_Power} & \multirow{3}{*}{$\displaystyle \frac{a}{(1+z)^c}\left(\frac{m}{M_\star}\right)^b$} & $a=9$, $b=-0.13$, $c=1$ \\

        \citet{Duffy2008} & \texttt{Duffy08} & & $a=6.71$, $b=-0.091$, $c=0.44$, $M_\star=2\times10^{12}$ \\

        \citet{Zehavi2011} & \texttt{Zehavi11} & & $a=11$, $b=-0.13$, $c=1$, $M_\star=2.26\times10^{12}$ \\
        
         \bottomrule[0.05cm]
\end{tabu}
\caption{Summary of concentration-mass-redshift relations implemented in \textsc{halomod}. Additional models can be imported from the \textsc{colossus} code. Note that the relation from \citet{Duffy2008} has multiple sets of parameters for different mass definitions and different halo samples. \textsc{halomod} includes all of these, and they can be manually specified and/or intelligently chosen.}
\label{tab:models_concentration}
\end{table*}

\subsubsection{HOD}
\label{sec:halomod:components:hod}
We keep the implementation of the HOD simple, requiring only the two mean occupation functions for centrals and satellites. Furthermore, there is an optional definition of $m_\text{min}$, which can be used to define a precise lower bound on halo mass. Generally this will be predominantly determined by the parameter $M_\text{min}$, and it defaults to this value if not specified. In general it may be some function of the parameters.

As alluded to in \S\ref{sec:theory-gal:hod}, there is a subtlety in the HOD definition -- the so-called `central condition' -- which is the statement that if no central galaxy exists in a halo in a given sample, then no satellites will be present. Note that the central condition \textit{implies} the statement $\Nc=0 \Rightarrow \Ns=0$, but the reverse is not true in general (one could then still have $\Nc = 0.5$ and $\Ns > 0$, and then a specific halo could still have a satellite but not central). However, if $\Nc$ is a step function from zero to one at some mass (and $\Ns=0$ when $\Nc=0$) then the reverse is true.

The following points outline our approach to this condition:
\begin{enumerate}
    \item The average satellite occupancy is taken to be the average over \textit{all} haloes, with and without centrals. This has subtle implications for how to mock up the galaxy population, because if one requires a central before placing a satellite, then the average number of satellites placed into \textit{available} haloes is increased if the central occupation is less than 1.
    
    \item If the central condition is enforced, then for all HOD classes (except see point 4), the mean satellite occupancy is modified. If the defined occupancy is $\Ns'$, then the returned occupancy is $\Ns = \Nc\Ns'$. This merely ensures that $\Ns=0$ when $\Nc=0$. Note that this will change the interpretation of parameters in the $\Ns$ model, unless $\Nc$ is a simple step function.
    
    \item The pair-wise counts involve a term $\langle N_{\rm c} N_{\rm s}\rangle$. When the central condition is enforced, this reduces trivially to $\Ns$. However, if the central condition is not enforced we \textit{assume} that the variates $N_{\rm c}$ and $N_{\rm s}$ are uncorrelated, and use $\langle N_{\rm c} N_{\rm s}\rangle = \Nc\Ns$.
    
    \item For A HOD class that is defined with the central condition intrinsically \textit{satisfiable} (i.e. $\Ns=0 \Rightarrow \Nc=0$), a variable can be set in the class definition, which will avoid the extra modification of point 2. Note that due to our above observation that being \textit{satisfiable} does not in general imply that it has been \textit{satisfied}, the pairwise counts still depend on whether the user asserts that the central condition is enforced or not (note that in the one case in which \textit{satisfiable} implies \textit{satisfied}, namely a step function $\Nc$, the pairwise counts are the same with or without the central condition).
\end{enumerate}

We present the implemented models in Table \ref{tab:models_hod}. An illustration of the various effects introduced by each HOD is presented in Fig. \ref{fig:hod}. Note that many of the HOD models are extensions of the more simple models, and are equivalent when setting the extra parameters to zero or one. Typically, \halomod\ will by default render these models the same as the more simple models, with the option of changing the extra parameters.

In addition to the galaxy-focused HOD models presented in Table \ref{tab:models_hod}, \halomod\ provides base classes for diffuse tracers of the underlying DM distribution, such as H\textsc{i} gas. 
In these distributions, it is assumed that there is no distinction between central and satellite, and the only relevant quantity is the mean value of the tracer squared (as a function of mass), which enters the power spectrum integral. 
Furthermore, non-unitless HODs are available in which the occupation can be defined in terms of a \textit{quantity} (eg. H\textsc{i} brightness temperature), but in which that quantity is assumed to be confined to discrete locations within the halos (like galaxies). In these distributions, the central-satellite distinction is maintained.


\begin{table*}
\small
\centering
\begin{tabu}{X[1.7l]X[4l]X[4l]}
\toprule[0.05cm] 
Ref. & $\Nc_m$ & $\Ns_m$ \\
\toprule[0.05cm]
      
        \citet{Zehavi2005} & $H(m-M_\text{min})$ & $\left(\frac{m}{M_1}\right)^\alpha$ \\ \midrule

        \citet{Zheng2005} & $\frac{1}{2}\left[1+\text{erf}\left(\frac{\log m-\log M_{min}}{\sigma_{\log M}}\right)\right]$ & $\left(\frac{m-M_0}{M'_1}\right)^\alpha$ \\ \midrule

        \citet{Tinker2005} & $H(m-M_\text{min})$ & $\exp\left(-\frac{M_\text{cut}}{m-M_\text{min}}\right)\left(\frac{m}{M'_1}\right)H(m-M_\text{min})$ \\ \midrule
                                        
        \citet{Geach2012} & $\begin{aligned}&F_c^B(1-F_c^A)\exp \left[\frac{\log_{10}(m/M_c)^2}{2\sigma^2_{\log M}}\right] \\ &\times F_c^A \left[1+ \text{erf} \left(\frac{\log_{10}(m/M_c)}{\sigma_{\log M}}\right)\right]\end{aligned}$ & $F_s \left[1 + \text{erf} \left(\frac{\log_{10}(m/M_{min})}{\delta_{\log M}}\right)\right]\left(\frac{m}{M_{min}}\right)^\alpha$ \\ \midrule
          
         \citet{Contreras2013} & $N_{\rm c}^\text{Geach}$, $\sigma_{\log M} \rightarrow x\sigma_{\log M}$ & $N_{\rm s}^\text{Geach}$ \\
         \bottomrule[0.05cm]
\end{tabu}
\caption[Summary of included HOD parameterisations]{Summary of included HOD parameterisations. Here $H$ is the Heaviside step-function. Note that this table does not include HOD models for diffuse tracers.}
\label{tab:models_hod}
\end{table*}

\subsubsection{Halo Exclusion}
\label{sec:halomod:components:exclusion}
The primary task of the \verb|Exclusion| component is to evaluate \cref{eq:dmpower2a} (or \cref{eq:dmpower2} if appropriate) and the associated (potentially modified) mean density, \cref{eq:theory:ng_dash}. Each model, as discussed in \S\ref{sec:theory:exclusion}, is quite unique, and therefore there is not a great deal of redundancy that can be mitigate by class inheritance.
Nevertheless, for the sake of consistency, we implement exclusion using the standard \component\ architecture.

Performance becomes a definite issue with these calculations, especially for those requiring many double-integrations. Thus we introduce alternative methods, where necessary, accelerated by just-in-time compilation using \textsc{numba}. 

We show the effects of the halo exclusion for seven of the implemented models (two of them being modifications to $P_{\rm hh}^c(k)$ rather than physical halo exclusion models) in Fig. \ref{fig:halo_exclusion}. 
Of these models, it is expected that \verb|DblEllipsoid| and \verb|NgMatched| are the most accurate, and the latter is far more performant.

\subsection{Extra functionality}
\label{sec:halomod:extra}
In addition to the core \framework\ and \component\ elements, we have implemented several pieces of functionality aimed at performing the most commonly required tasks.

\subsubsection{Command-line interface}
\label{sec:halomod:extra:cli}
Though we expect that the vast majority of usage will be interactive or via custom scripts, for additional flexibility we also provide a command-line interface (CLI), which provides a convenient way to compute specified quantities for a range of parameters. This may be useful for batch scripts which only require computation of basic quantities over a large parameter space, or for creating simple interfaces with other programming languages. 

The underlying machinery for the CLI is contained in the \verb|functional| module, which provides a top-level interface for calculating given quantities within a given \framework, for all combinations of a set of given parameters. Due to the homogeneity of the \framework\ definitions, this interface needs only be defined once for all frameworks\footnote{Indeed, due to \halomod\  being inherited from \textsc{hmf}, \halomod\  requires only a thin wrapper around the function in \textsc{hmf} to update some defaults.}. 

The most challenging aspect of this function is the ordering of the implicit loops. As an example, consider a case where the user wishes to range over both redshift and $\Omega_m$ to calculate $n(m)$. It is much more efficient to use redshift as the inner loop, since fewer quantities depend on it, and therefore fewer re-calculations will be performed. This implicit ordering of parameters is accounted for in the routine (if necessary) by performing a very fast low-resolution calculation and determining the number of child quantities for each parameter. This is made possible through the caching system which we have already described (cf. \S\ref{sec:halomod:overview:efficiency}.

Along with the calculated quantities, the routine optionally returns unique labels for each of the iterations, since the return order is not pre-specified.

The CLI wraps this routine, allowing any parameter to be specified by name, either as a scalar or list of values to be iterated over. The CLI ingests a configuration file in \textsc{TOML}\footnote{\url{https://toml.io/en/}} format, but may be over-ridden via command-line arguments. The CLI outputs a complete configuration file (with all default parameters set explicitly) which can be used to re-compute the quantities later. It also outputs each desired quantity in its own file.

\subsubsection{HOD population}
\label{sec:halomod:extra:pophod}
Quite apart from the analytic formalism outlined in this paper, the HOD allows for creating galaxy catalogues by directly populating halos from $N$-body simulations. Though less efficient than its analytic counterpart, this method is more robust, since it does not depend on approximations for the several halo-based components. Indeed, several authors have used this method, either alone or in conjunction with the analytic calculation \citep{Skibba2015,Zheng2015}. It is also useful as a sanity check.

We provide a very basic set of tools for populating halo catalogues with galaxies using the HOD models included in \textsc{halomod}, synthesised into a CLI. Our implementation assumes a Bernoulli (Poisson) distribution for centrals (satellites), and a spherically symmetric profile for each halo, corresponding precisely to the analytic calculations. It is possible that this functionality will be extended in future versions to account for triaxial profiles \citep{Jing2002} and a range of galaxy classes such as colour, cf. \citet{Skibba2009}.

\section{\textsc{TheHaloMod} web-application}
\label{sec:thehalomod}
\citet{Murray2013a} presented \textsc{HMFcalc}, an online HMF calculator that has since been widely used in the community.
Web-applications can be useful, as they circumvent the overhead of installing custom software and using a command-line or interpreter interface. 
As such, they are more readily employed by researchers seeking to obtain a single model to corroborate the output of a simulation or observation.
Perhaps more interestingly, the intuitive and graphical nature of web-apps make them extremely useful for educational purposes -- whether in a classroom setting, or for self-learning.
It is with these motivations that we present a successor to \textsc{HMFcalc}\footnote{\textsc{HMFcalc} was located at \url{https://hmf.icrar.org}. This address now automatically forwards to the address of \textsc{TheHaloMod}, at \url{https://thehalomod.app}. All the functionality of \textsc{HMFcalc} is available in \textsc{TheHaloMod}.} that computes the full range of halo model quantities we have outlined in this paper -- called \textsc{TheHaloMod}\footnote{The \textit{code} for \textsc{TheHaloMod} is open-source and available at \url{https://github.com/halomod/TheHaloMod-SPA}.}.

\subsection{Interface}
The interface of \thm\ is expansive but simple.
It is a single-page application (SPA) that uses the \textsc{vue.js}\footnote{\url{https://www.vuejs.org/}} framework to render input forms dynamically.
The primary view presented to the user consists of a left-hand column containing an overview of the models the user has created and a drop-down menu for downloading various data products, and a right-hand panel for viewing plots of arbitrary halo model quantities.

The Models panel is initially empty until the user creates a model. 
Creating a model opens the input form view (technically this is on the same page, and overlays the primary view). We will discuss this view momentarily, but once a model has been defined, and a unique label is designated for it, the model appears in its own row in the Models panel on the main view. From here, the model may be renamed, edited (taking the user back to the input form view), cloned or deleted. 
Cloning the model makes a copy of model, but also takes the user to the input form view, to edit its parameters. 
In addition, there is an option for the user to report a bug specific to a particular model, which---along with a useful message from the user---collects and sends detailed information about the model to the developers. 

The Download panel contains a single drop-down menu, offering choices of file downloads. Presently, the available options are (i) an image of the plot (in SVG format), (ii) configuration files containing all of the parameters specified for each model (these come as ZIP archives, with one file per user-specified model, and each file can be in TOML or JSON format, where the TOML files are precisely in the format required to be input to the CLI, cf. \S\ref{sec:halomod:extra:cli}), and (iii) a ZIP archive containing ASCII data files with all possible halo model quantities stored in them (i.e. all quantities defined within the \texttt{TracerHaloModel}). The latter makes it simple for the user to produce their own plots with their preferred plotting tools. 

The Plot panel contains a number of tools to customize what is being plotted.
Most importantly, there are two drop-downs that specify the quantities that comprise the $x$- and $y$-axes. While typically the $x$-axis will either be halo mass, comoving scale or comoving wavenumber, more general quantities can be chosen (for example the peak-height $\nu$ instead of mass). The options available for the $y$-axis are dynamically updated based on the value of the $x$-axis, and include every possible quantity in the \texttt{TracerHaloModel}. Beyond this, there are options for setting the scale on either axis to be logarithmic (the default is intelligently chosen for each quantity).  Finally, the plot itself contains the chosen quantity for all models defined by the user. A legend to the right of the plot indicates corresponding models. 

The interface is further available in both light and dark modes.
It is also fluid -- adjusting to fit on phone, tablet and monitor screens. 

The input form view (cf. left-hand panel of Fig. \ref{fig:webpage-image}) presents all possible input parameters to a model. 
This is quite a few possible parameters -- including the choice of each kind of component model as well as all possible parameters to each of those models (eg. HOD models, bias models and HMF models). 
To make the interface concise and simple, the view consists of a left-hand panel that contains a dynamic index of parameter groups (essentially each of the components) so that particular parameters are easier to locate.
Each component has a similar layout for its inputs: a drop-down menu to choose the model, and a dynamically-rendered set of numerical inputs to change the unique parameters of the model (i.e. if \texttt{Tinker08} is chosen as the HMF model, only its parameters are shown to the user, and not those of any other mass function). 
The user is able to create and edit as many models as they wish, and these models persist between sessions. They are able to refresh and remove all their models via the Model panel on the main page.

\subsection{Architecture}
The overall architecture of \textsc{TheHaloMod} is broken up into two parts: server and client. 
From a high-level perspective, the server handles the API portion of the website, and the client holds the front-end. 
This way, the front-end can benefit from modern JavaScript UI libraries and the backend interfaces with the Python libraries \textsc{hmf} and \textsc{halomod}.

Automatic testing on both server and client are performed via Github Actions, which also handle automatic continuous deployment on every commit to the main branch.

The server is built with the \textsc{flask}\footnote{\url{https://flask.palletsprojects.com/en/2.0.x/}} Python web framework. 
This framework handles the (very simple) API for the single-page app. 
Since it is a Python framework, it is able to interface directly with \textsc{halomod}. 
Models defined by the user (on the input form view) are serialized as a set of parameters, and used to construct \texttt{TracerHaloModel} objects on the server. These objects are serializable by construction, and are stored on the server (uniquely for each user) using \textsc{redis}\footnote{\url{https://redis.io/}}. 
Each time a quantity is requested by the user (to create the plots on the main view), each model object is deserialized, and the relevant quantity is computed (or obtained from cache if it has already been computed, cf. \S\ref{sec:halomod:overview:efficiency}). 
Cloning a model (via the Model panel on the main view) literally copies the instance of the model on the server, and uses the inherent caching and updating ability of \textsc{hmf} \texttt{Framework} objects to improve the speed of calculation of the new model.
In this way, the web-app essentially provides a thin, visual, wrapper around the full capability of a \texttt{TracerHaloModel} object.

The frontend client is built with the \textsc{vue.js} web framework, and uses the \textsc{Material}\footnote{\url{https://vuematerial.io/}} CSS framework.
This provides a simple yet modern and flexible framework for both visual and interface design. 
The plots are created using \texttt{d3.js}\footnote{\url{https://d3js.org/}}.

To enhance performance, an initial high-resolution object is pre-computed on the server, which is cloned and updated upon the first model definition by the user. In particular, this is very effective at producing high-resolution transfer functions from \textsc{camb}, which otherwise are computationally intensive.

The \thm\ server is hosted by the Low-Frequency Cosmology (LOCO) Lab at Arizona State University. 
The client is separately hosted by \textsc{netlify}\footnote{\url{https://d3js.org/}}, which aids load speeds for users worldwide.
Deployment is aided by containerizing the entire application using Docker\footnote{\url{https://www.docker.com/}}.

\section{Example Application}
\label{sec:applications}

To illustrate the utility of \halomod\ in semi-realistic applications, we here present a worked example of fitting parameters to mock galaxy data\footnote{The full example notebook can be found at \url{https://halomod.readthedocs.io/en/latest/examples/fitting.html}}. 
We do not use real data, but instead create some simplistic mock data using \halomod\ itself.

We note that we do \textit{not} include native fitting capabilities in \halomod. The reason for this is that applications in constraining parameters, and their relevant likelihoods, are so diverse. It is impossible to capture the range of realistic likelihoods, and it is more useful to let the user adopt one of the many excellent Bayesian (or more general fitting) frameworks in the \python ecosystem\footnote{Examples include \textsc{emcee}, \textsc{PyMC3}, \textsc{PyStan} and \textsc{cobaya}.}.
However, \halomod\ was partially motivated by the desire to perform HOD fits on galaxy data, and it is thus written in a manner conducive to this end.
For instance, each \framework\ is pickle-able by the \verb|dill.pickle| library. 
Furthermore, due to the fact that all quantities of interest are simply obtained by accessing attributes, quite general likelihoods can be written that are able to dynamically switch between different quantities. We will utilise this feature in our example here.

In our example, we will simply create mock data to which we fit a matching HOD model. 
The mock data is generated as $\xi_{\rm gg}(r)$, with diagonal covariance matrix and Gaussian noise at an amplitude of 10\% of the amplitude of the correlation function itself. In addition, we use the constraint of galaxy number density (with a Gaussian uncertainty of $10^-4 h^3{\rm Mpc}^{-3}$) to break a degeneracy in the HOD parameters.
The appropriate likelihood is thus a simple $\chi^2$ between the mock and model $\xi_{\rm gg}(r)$, along with an extra Gaussian term for the mean galaxy number density.

We use \textsc{emcee} to perform the actual Bayesian inference in this example, though it should be clear how to adapt the likelihood to other frameworks. We can write a fairly generic likelihood function:

%
%
%
%
%
\begin{lstlisting}[language=Python]
def log_prob(
	param_values: np.ndarray, 
	param_names: List[str], 
	data: np.ndarray,
	model: TracerHaloModel, 
	derived: Optional[Tuple[str]]=()
):
	# Pack parameters into a dict
	params = dict(zip(param_names, param_values))

	# Allow for simple bounded flat priors.
	bounds = bounds or {}
	for key, val in params.items():
		bound = bounds.get(key, (-np.inf, np.inf))
		if not bound[0] < val < bound[1]:
			return -np.inf, []

	# Update model with all fitting parameters 
	params = flat_to_nested_dict(params)
	model.update(**params)
	
	# Get arbitrary derived data
	derived = [getattr(model, d) for d in derived]
	
	# Calculate chi^2 likelihood
	logl = chi_square(
		model=model.corr_auto_tracer, 
		data=data, 
		sigma=0.2 * np.abs(model.corr_auto_tracer)
	)
	
	# Return derived to make emcee "blobs"
	return logl, derived
\end{lstlisting}

This function contains a reference to another custom function, \verb|flat_to_nested_dict|, whose purpose is to take a dictionary with 
dot-path keys and convert it to a nested dictionary, so that eg. \texttt{\{'nested.key': 3\}} is converted to \texttt{\{'nested': \{'key': 3\}\}}. This makes it possible set arbitrary parameters of components within the \texttt{Tracer\-Halo\-Model}, as we shall see. 

Furthermore, notice that arbitrary derived data is able to be extracted due to the consistent API of the \framework, in which every quantity is accessed as an attribute. 
Due to the efficient caching system (cf. \S\ref{sec:halomod:overview:efficiency} and \ref{app:caching}), the order in which quantities are accessed is irrelevant, and only the quantities that are required are actually calculated.
We return these arbitrary derived data as a list of \verb|emcee| ``blobs'', so that we can plot posteriors of the derived parameters.

This likelihood function is more than generic enough for this example, and many others, and is a good starting place for most likelihoods that the user might wish to write.
One clear deficiency in the log probability function is the absence of sophisticated priors on the parameters. 
Most sophisticated Bayesian frameworks have abstract mechanisms for defining priors, but \verb|emcee| assumes that the priors will be added to the probability inside the \verb|log_prob| function. 
Due to the non-general nature of such a code, we omit it for these examples, implicitly assuming flat priors on all parameters. 
We do however include a potential bound on each parameter, which should serve as a simple starting point for extensions to more complex priors.

We start the MCMC chains in a small Gaussian ball around the true parameters (or close to the true parameters in the case of the mismatched model) and let them expand to fill the posterior\footnote{This is recommended in the \textsc{emcee} documentation.}. 
The chains have 100 walkers, and comprise 10000 iterations, of which the first 1000 are removed as burnin. 
As this is a toy examples, explicit convergence tests are not performed as they should be in real applications. 


Our mock data is created with a default \texttt{Tracer\-Halo\-Model} set at a redshift of $z=0.2$, and using the transfer model of \citet{Eisenstein1998}. The HOD model is \verb|Zehavi05| (cf. Table \ref{tab:models_hod}), with $M_{\rm min} = 10^{12}$, $M_1 = 10^{12.8}$ and $\alpha = 1.05$.

Constraining the parameters is simple. We construct the sampler like so:
\begin{lstlisting}[language=Python]
sampler = emcee.EnsembleSampler(
	nwalkers = 100,
	ndim = 3,
	log_prob_fn = log_prob,
	kwargs = {
		'param_names': [
			'hod_params.M_min', 
			'hod_params.M_1', 
			'hod_params.alpha'
		],  
		'data': (mock_data, mock_ngal), 
		'model': model,  
		'derived': [
			'satellite_fraction', 
			'mean_tracer_den',
			'bias_effective_tracer',
			'corr_auto_tracer'
		],
	},
	blobs_dtype=[
		("sat_frac", float), 
		("tracer_den", float), 
		("bias_effective_tracer", float),
		("corr_auto_tracer", (float, len(mock_data)))
	]
	)
\end{lstlisting}

and then create the initial positions and sample for 10000 iterations:
%
\begin{lstlisting}[language=Python]
initialpos = np.array([
	fiducial_model.hod.params['M_min'], 
	fiducial_model.hod.params['M_1'],
	fiducial_model.hod.params['alpha']
]) + 1e-4 * np.random.normal(
	size=(sampler.nwalkers, sampler.ndim)
)
	
sampler.run_mcmc(initialpos, nsteps=10000)
\end{lstlisting}
Note that in the setup of the \verb|EnsembleSampler| we have specified the parameters we wish to constrain simply by passing a list of dot-pathed names.
The first part of the name refers to the \parameter\ name in the \framework\ that we would like to update. In this case, \verb|hod_params| is itself a dictionary of parameters for the HOD, and so the second part of the dot-path is the name of the parameter within that dictionary. This makes it possible to constrain any continuous parameter defined at any level of the framework.

Note also that the \verb|model| variable is itself a \texttt{Tracer\-Halo\-Model} instance, cloned from the original fiducial model from which we created the \verb|mock_data| using the simple command \texttt{model = fid\_model.clone()}. 
This \verb|model| is mutable and is internally updated in the \verb|log_prob| function, so it is useful to keep the original fiducial model untouched.

\begin{figure*}
    \centering
    \includegraphics[width=\textwidth]{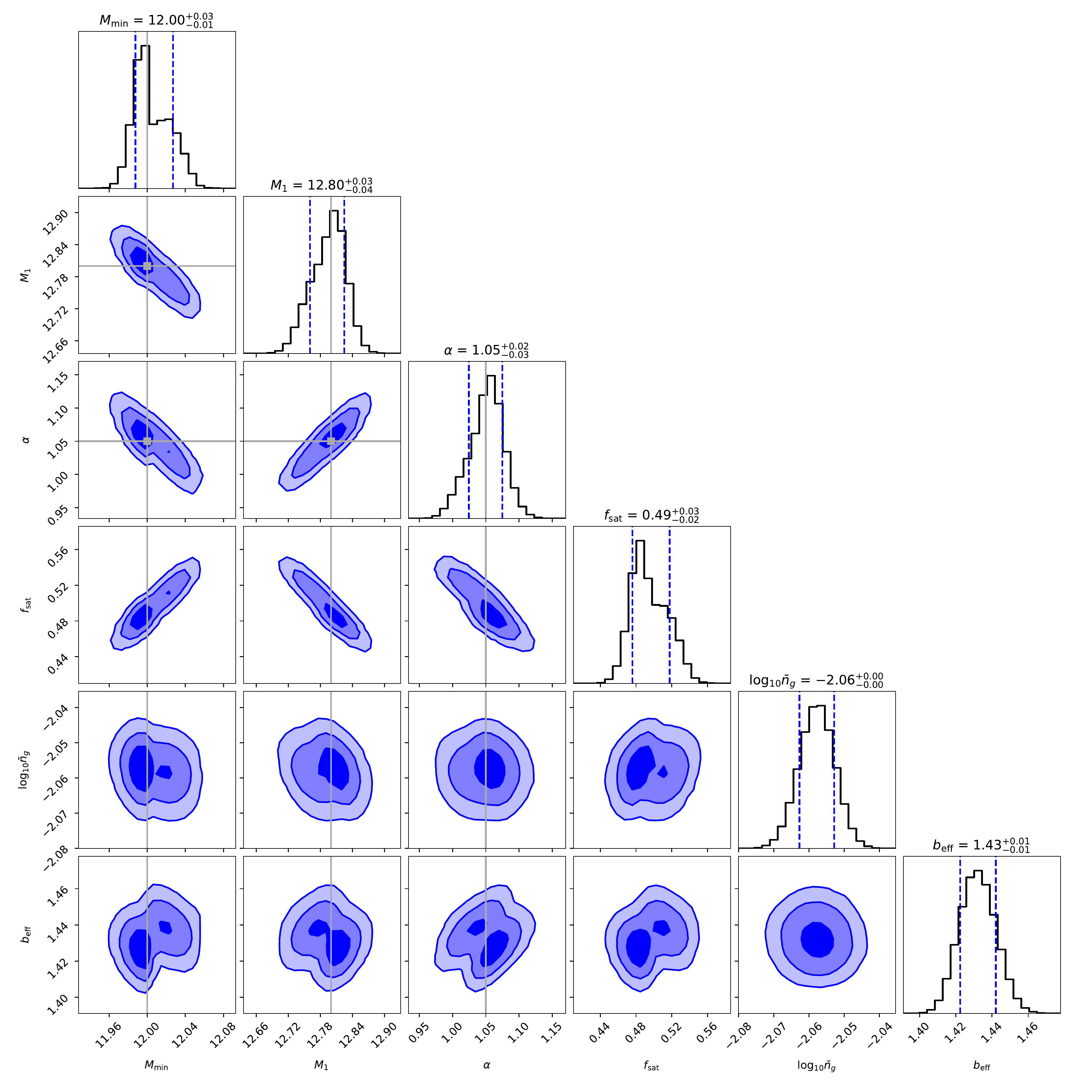}
    \caption{Corner-plot showing joint posteriors for each of the three HOD parameters, as well as the three derived parameters (satellite fraction, mean galaxy number density and linear galaxy bias). Example code to reproduce this plot is found throughout \S\ref{sec:applications}. }
    \label{fig:example1_corner}
\end{figure*}

After removing burnin and thinning by every $5^{th}$ sample to reduce correlations, we can make Fig. \ref{fig:example1_corner}, which shows that that parameters were recovered with moderate accuracy. 
Note that in this ``corner plot'', we have also shown the joint posteriors of the three derived scalars -- the satellite fraction, the mean galaxy density and the linear galaxy bias.  

\begin{figure}
    \centering
    \includegraphics[width=\linewidth]{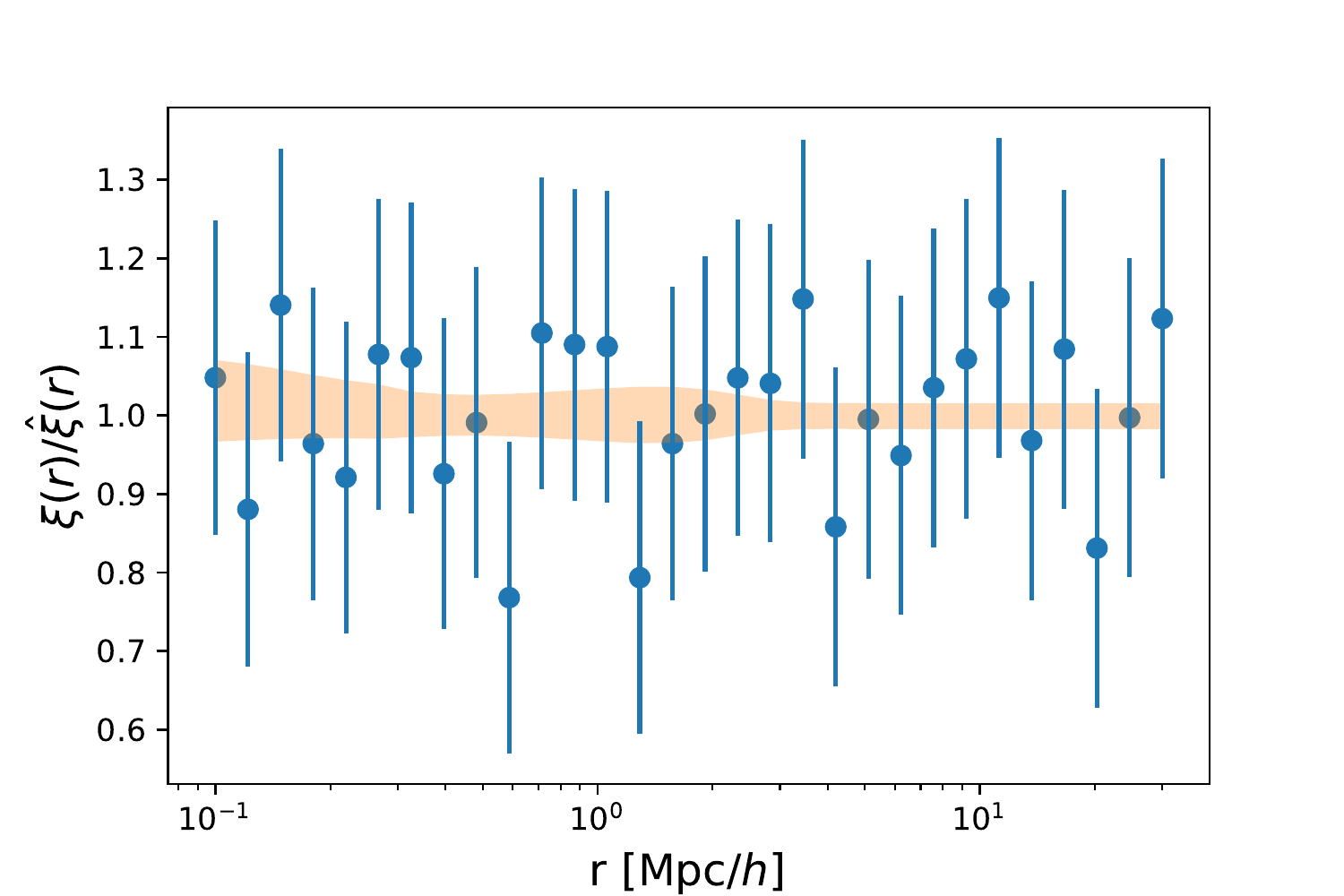}
    \caption{Ratio of mock data to the median posterior of $\xi(r)$. Here, the median posterior is calculated as the median (per $r$-bin) of the posterior \textit{samples} (not the model at the median of the posterior of the parameters). Error-bars shown are 10\% uncertainties as input to the mock data. The orange region is the 68\% central quantile of the posterior. The residuals are noise-like.}
    \label{fig:example_resids}
\end{figure}

Furthermore, Fig \ref{fig:example_resids} shows the posterior 68\%  percentile range of the correlation function, along with the ratio of the mock data to the median. The posterior of the model $\xi(r)$ is here easily obtained from the output sampler: \texttt{sampler.get\_blobs(flat=True)['corr\_auto\_tracer']}. \\
Notice that the residuals are noise-like, as is expected from this toy model.

It is easy to imagine extensions to this example -- using projected or angular correlation functions, more sophisticated HOD models, allowing cosmology to be free, or using multiple datasets simultaneously. 
In most cases, the necessary modifications to the example code presented here will be small, illustrating the simplicity and flexibility of \halomod.

\section{Future Development}
\label{sec:future}
The current implementation of \textsc{halomod} (likewise \textsc{hmf}) forms a solid framework for evaluating halo model quantities. Its flexible and extendable architecture enables user-side development with updated components and paradigms (eg. warm dark matter, or alternate dark energy scenarios). However, several features considered for future versions are worth mentioning:

\paragraph*{Additional and higher-order statistics}
Though the most popular usage of the halo model has been to analyse the 2-point clustering of galaxies, there has also been a range of studies with other quantities. In particular, galaxy-galaxy weak lensing, as an independent observable, has been shown to be able to break HOD degeneracies \citep{Leauthaud2012}. Future versions of \halomod\  will include these quantities in addition to higher-order statistics such as the bispectrum and 3-point correlation function. 

\paragraph*{Alternative Cosmologies}
As already mentioned, \hmf\ already has the infrastructure to support alternative cosmological scenarios. However, only one scenario is actually fleshed out -- WDM. It will be interesting to implement more exotic models such as Early Dark Energy, or $f(R)$ gravity. Ideally, such implementations will arise from the community.

\paragraph*{Support for \textsc{class}}
\textsc{class}\footnote{\url{https://lesgourg.github.io/class_public/class.html}} \citep{Lesgourgues2011} is a popular Boltzmann-solver for the transfer function. As such, it is a prime candidate as an extra model for the \verb|Transfer| \component.

\paragraph*{Performance and Accuracy Improvements}
There are known bottlenecks in the code -- especially as related to Hankel transforms and halo exclusion -- both in memory and CPU. 
These do not affect the user experience greatly over a single model, but make a definite impact when running many models. Further improvements in the interpolation and integration required for Hankel transforms will be highly useful. Reducing the number of nodes required for these integrals will also lower the burden of halo exclusion.

\paragraph*{Improved Web Interface}
Though \thm\ has a clean and usable interface, it was designed and implemented by a definite non-expert in the field of web development. 
Many advanced features that would enhance the user experience are possible: from fitting more useful information into a one-page app, to making the figure(s) interactive, to having the ability to store ``sessions'' for later use.

Most importantly, \hmf, \halomod\ and \thm\ are all open-source and intended to be projects nurtured and maintained by the community. 
This will make them of more impact and usefulness for the community.
Up-to-date planned improvements for all of these codes can be found by browsing their issues on GitHub\footnote{Eg. \url{https://github.com/halomod/halomod/issues}}. We highly encourage the reader to suggest new features via this mechanism!

\section{Summary}
\label{sec:summary}
We have presented a new code, \halomod, accompanied by a web-application, \thm, for the calculation of quantities within the halo model framework. This code aims to satisfy six principles: simple, intuitive, flexible/extendible, comprehensive, efficient and open; we have shown how these criteria have shaped the architecture of the library.

\halomod\  is designed to be a valuable resource for the community. 
To this end, we have presented an illustrative application which helps to get a flavour of how \halomod\ may be used `in the wild' for constraining physical parameters using galaxy surveys.

We expect \thm\ to be useful for both theorists and observers, but are especially excited about the prospects for using it educationally. 
\thm\ also serves as an example and motivation for other broadly-applicable codes to make use of scientific web-applications in order to reach a wider user-base.

\section*{Acknowledgements}
SGM acknowledges many useful discussions with colleagues that formed the basis of this paper, including Chris Power, Aaron Robotham, Chris Blake, Florian Beutler and David Palomara. 
Many thanks to the Low-frequency Cosmology (LoCo) lab at Arizona State University whose cluster, \textit{pleiades}, hosts the server on which \thm\ runs. In particular thanks goes to Matt Kolopanis, who greatly helped with setting up the server infrastructure.
ZC acknowledges support from the Overseas Research Scholar Awards from the School of Physics and Astronomy, The University of Manchester.
All figures in this paper were produced with \textsc{matplotlib} \citep{Hunter2007}. We acknowledge the extremely helpful \textsc{numpy} \citep{vanderWalt2011}, \textsc{astropy} \citep{Robitaille2013,AstropyCollaboration2018} and \textsc{SciPy} \citep{Virtanen2020} codes.

\bibliography{library}

\begin{thebibliography}{162}
\expandafter\ifx\csname natexlab\endcsname\relax\def\natexlab#1{#1}\fi
\providecommand{\url}[1]{\texttt{#1}}
\providecommand{\href}[2]{#2}
\providecommand{\path}[1]{#1}
\providecommand{\DOIprefix}{doi:}
\providecommand{\ArXivprefix}{arXiv:}
\providecommand{\URLprefix}{URL: }
\providecommand{\Pubmedprefix}{pmid:}
\providecommand{\doi}[1]{\href{http://dx.doi.org/#1}{\path{#1}}}
\providecommand{\Pubmed}[1]{\href{pmid:#1}{\path{#1}}}
\providecommand{\bibinfo}[2]{#2}
\ifx\xfnm\relax \def\xfnm[#1]{\unskip,\space#1}\fi
\bibitem[{{Astropy Collaboration} et~al.(2018){Astropy Collaboration},
  {Price-Whelan}, Sip{\H o}cz, G{\"u}nther, Lim, Crawford, Conseil, Shupe,
  Craig, Dencheva, Ginsburg, VanderPlas, Bradley, {P{\'e}rez-Su{\'a}rez}, {de
  Val-Borro}, Aldcroft, Cruz, Robitaille, Tollerud, Ardelean, Babej, Bach,
  Bachetti, Bakanov, Bamford, Barentsen, Barmby, Baumbach, Berry, Biscani,
  Boquien, Bostroem, Bouma, Brammer, Bray, Breytenbach, Buddelmeijer, Burke,
  Calderone, Cano~Rodr{\'i}guez, Cara, Cardoso, Cheedella, Copin, Corrales,
  Crichton, D'Avella, Deil, Depagne, Dietrich, Donath, Droettboom, Earl, Erben,
  Fabbro, Ferreira, Finethy, Fox, Garrison, Gibbons, Goldstein, Gommers, Greco,
  Greenfield, Groener, Grollier, Hagen, Hirst, Homeier, Horton, Hosseinzadeh,
  Hu, Hunkeler, Ivezi{\'c}, Jain, Jenness, Kanarek, Kendrew, Kern, Kerzendorf,
  Khvalko, King, Kirkby, Kulkarni, Kumar, Lee, Lenz, Littlefair, Ma, Macleod,
  Mastropietro, McCully, Montagnac, Morris, Mueller, Mumford, Muna, Murphy,
  Nelson, Nguyen, Ninan, N{\"o}the, Ogaz, Oh, Parejko, Parley, Pascual, Patil,
  Patil, Plunkett, Prochaska, Rastogi, Reddy~Janga, Sabater, Sakurikar,
  Seifert, Sherbert, {Sherwood-Taylor}, Shih, Sick, Silbiger, Singanamalla,
  Singer, Sladen, Sooley, Sornarajah, Streicher, Teuben, Thomas, Tremblay,
  Turner, Terr{\'o}n, {van Kerkwijk}, {de la Vega}, Watkins, Weaver, Whitmore,
  Woillez, Zabalza and {Astropy Contributors}}]{AstropyCollaboration2018}
\bibinfo{author}{{Astropy Collaboration}}, \bibinfo{author}{{Price-Whelan},
  A.M.}, \bibinfo{author}{Sip{\H o}cz, B.M.}, \bibinfo{author}{G{\"u}nther,
  H.M.}, \bibinfo{author}{Lim, P.L.}, \bibinfo{author}{Crawford, S.M.},
  \bibinfo{author}{Conseil, S.}, \bibinfo{author}{Shupe, D.L.},
  \bibinfo{author}{Craig, M.W.}, \bibinfo{author}{Dencheva, N.},
  \bibinfo{author}{Ginsburg, A.}, \bibinfo{author}{VanderPlas, J.T.},
  \bibinfo{author}{Bradley, L.D.}, \bibinfo{author}{{P{\'e}rez-Su{\'a}rez},
  D.}, \bibinfo{author}{{de Val-Borro}, M.}, \bibinfo{author}{Aldcroft, T.L.},
  \bibinfo{author}{Cruz, K.L.}, \bibinfo{author}{Robitaille, T.P.},
  \bibinfo{author}{Tollerud, E.J.}, \bibinfo{author}{Ardelean, C.},
  \bibinfo{author}{Babej, T.}, \bibinfo{author}{Bach, Y.P.},
  \bibinfo{author}{Bachetti, M.}, \bibinfo{author}{Bakanov, A.V.},
  \bibinfo{author}{Bamford, S.P.}, \bibinfo{author}{Barentsen, G.},
  \bibinfo{author}{Barmby, P.}, \bibinfo{author}{Baumbach, A.},
  \bibinfo{author}{Berry, K.L.}, \bibinfo{author}{Biscani, F.},
  \bibinfo{author}{Boquien, M.}, \bibinfo{author}{Bostroem, K.A.},
  \bibinfo{author}{Bouma, L.G.}, \bibinfo{author}{Brammer, G.B.},
  \bibinfo{author}{Bray, E.M.}, \bibinfo{author}{Breytenbach, H.},
  \bibinfo{author}{Buddelmeijer, H.}, \bibinfo{author}{Burke, D.J.},
  \bibinfo{author}{Calderone, G.}, \bibinfo{author}{Cano~Rodr{\'i}guez, J.L.},
  \bibinfo{author}{Cara, M.}, \bibinfo{author}{Cardoso, J.V.M.},
  \bibinfo{author}{Cheedella, S.}, \bibinfo{author}{Copin, Y.},
  \bibinfo{author}{Corrales, L.}, \bibinfo{author}{Crichton, D.},
  \bibinfo{author}{D'Avella, D.}, \bibinfo{author}{Deil, C.},
  \bibinfo{author}{Depagne, {\'E}.}, \bibinfo{author}{Dietrich, J.P.},
  \bibinfo{author}{Donath, A.}, \bibinfo{author}{Droettboom, M.},
  \bibinfo{author}{Earl, N.}, \bibinfo{author}{Erben, T.},
  \bibinfo{author}{Fabbro, S.}, \bibinfo{author}{Ferreira, L.A.},
  \bibinfo{author}{Finethy, T.}, \bibinfo{author}{Fox, R.T.},
  \bibinfo{author}{Garrison, L.H.}, \bibinfo{author}{Gibbons, S.L.J.},
  \bibinfo{author}{Goldstein, D.A.}, \bibinfo{author}{Gommers, R.},
  \bibinfo{author}{Greco, J.P.}, \bibinfo{author}{Greenfield, P.},
  \bibinfo{author}{Groener, A.M.}, \bibinfo{author}{Grollier, F.},
  \bibinfo{author}{Hagen, A.}, \bibinfo{author}{Hirst, P.},
  \bibinfo{author}{Homeier, D.}, \bibinfo{author}{Horton, A.J.},
  \bibinfo{author}{Hosseinzadeh, G.}, \bibinfo{author}{Hu, L.},
  \bibinfo{author}{Hunkeler, J.S.}, \bibinfo{author}{Ivezi{\'c}, {\v Z}.},
  \bibinfo{author}{Jain, A.}, \bibinfo{author}{Jenness, T.},
  \bibinfo{author}{Kanarek, G.}, \bibinfo{author}{Kendrew, S.},
  \bibinfo{author}{Kern, N.S.}, \bibinfo{author}{Kerzendorf, W.E.},
  \bibinfo{author}{Khvalko, A.}, \bibinfo{author}{King, J.},
  \bibinfo{author}{Kirkby, D.}, \bibinfo{author}{Kulkarni, A.M.},
  \bibinfo{author}{Kumar, A.}, \bibinfo{author}{Lee, A.},
  \bibinfo{author}{Lenz, D.}, \bibinfo{author}{Littlefair, S.P.},
  \bibinfo{author}{Ma, Z.}, \bibinfo{author}{Macleod, D.M.},
  \bibinfo{author}{Mastropietro, M.}, \bibinfo{author}{McCully, C.},
  \bibinfo{author}{Montagnac, S.}, \bibinfo{author}{Morris, B.M.},
  \bibinfo{author}{Mueller, M.}, \bibinfo{author}{Mumford, S.J.},
  \bibinfo{author}{Muna, D.}, \bibinfo{author}{Murphy, N.A.},
  \bibinfo{author}{Nelson, S.}, \bibinfo{author}{Nguyen, G.H.},
  \bibinfo{author}{Ninan, J.P.}, \bibinfo{author}{N{\"o}the, M.},
  \bibinfo{author}{Ogaz, S.}, \bibinfo{author}{Oh, S.},
  \bibinfo{author}{Parejko, J.K.}, \bibinfo{author}{Parley, N.},
  \bibinfo{author}{Pascual, S.}, \bibinfo{author}{Patil, R.},
  \bibinfo{author}{Patil, A.A.}, \bibinfo{author}{Plunkett, A.L.},
  \bibinfo{author}{Prochaska, J.X.}, \bibinfo{author}{Rastogi, T.},
  \bibinfo{author}{Reddy~Janga, V.}, \bibinfo{author}{Sabater, J.},
  \bibinfo{author}{Sakurikar, P.}, \bibinfo{author}{Seifert, M.},
  \bibinfo{author}{Sherbert, L.E.}, \bibinfo{author}{{Sherwood-Taylor}, H.},
  \bibinfo{author}{Shih, A.Y.}, \bibinfo{author}{Sick, J.},
  \bibinfo{author}{Silbiger, M.T.}, \bibinfo{author}{Singanamalla, S.},
  \bibinfo{author}{Singer, L.P.}, \bibinfo{author}{Sladen, P.H.},
  \bibinfo{author}{Sooley, K.A.}, \bibinfo{author}{Sornarajah, S.},
  \bibinfo{author}{Streicher, O.}, \bibinfo{author}{Teuben, P.},
  \bibinfo{author}{Thomas, S.W.}, \bibinfo{author}{Tremblay, G.R.},
  \bibinfo{author}{Turner, J.E.H.}, \bibinfo{author}{Terr{\'o}n, V.},
  \bibinfo{author}{{van Kerkwijk}, M.H.}, \bibinfo{author}{{de la Vega}, A.},
  \bibinfo{author}{Watkins, L.L.}, \bibinfo{author}{Weaver, B.A.},
  \bibinfo{author}{Whitmore, J.B.}, \bibinfo{author}{Woillez, J.},
  \bibinfo{author}{Zabalza, V.}, \bibinfo{author}{{Astropy Contributors}},
  \bibinfo{year}{2018}.
\newblock \bibinfo{title}{The {{Astropy Project}}: {{Building}} an
  {{Open}}-science {{Project}} and {{Status}} of the v2.0 {{Core Package}}}.
\newblock \bibinfo{journal}{The Astronomical Journal} \bibinfo{volume}{156},
  \bibinfo{pages}{123}.
\newblock \DOIprefix\doi{10.3847/1538-3881/aabc4f}.
\bibitem[{Bardeen et~al.(1986)Bardeen, Bond, Kaiser and Szalay}]{Bardeen1986}
\bibinfo{author}{Bardeen, J.M.}, \bibinfo{author}{Bond, J.R.},
  \bibinfo{author}{Kaiser, N.}, \bibinfo{author}{Szalay, A.S.},
  \bibinfo{year}{1986}.
\newblock \bibinfo{title}{The statistics of peaks of {{Gaussian}} random
  fields}.
\newblock \bibinfo{journal}{The Astrophysical Journal} \bibinfo{volume}{304},
  \bibinfo{pages}{15--15}.
\newblock \DOIprefix\doi{10.1086/164143}.
\bibitem[{Behroozi et~al.(2013)Behroozi, Wechsler and Conroy}]{Behroozi2013a}
\bibinfo{author}{Behroozi, P.S.}, \bibinfo{author}{Wechsler, R.H.},
  \bibinfo{author}{Conroy, C.}, \bibinfo{year}{2013}.
\newblock \bibinfo{title}{{{THE AVERAGE STAR FORMATION HISTORIES OF GALAXIES IN
  DARK MATTER HALOS FROM}} z = 0-8}.
\newblock \bibinfo{journal}{The Astrophysical Journal} \bibinfo{volume}{770},
  \bibinfo{pages}{57--57}.
\newblock \DOIprefix\doi{10.1088/0004-637X/770/1/57}.
\bibitem[{Benson et~al.(2012)Benson, Farahi, Cole, a.~Moustakas, Jenkins,
  Lovell, Kennedy, Helly and Frenk}]{Benson2012}
\bibinfo{author}{Benson, A.J.}, \bibinfo{author}{Farahi, a.},
  \bibinfo{author}{Cole, S.}, \bibinfo{author}{a.~Moustakas, L.},
  \bibinfo{author}{Jenkins, a.}, \bibinfo{author}{Lovell, M.},
  \bibinfo{author}{Kennedy, R.}, \bibinfo{author}{Helly, J.},
  \bibinfo{author}{Frenk, C.}, \bibinfo{year}{2012}.
\newblock \bibinfo{title}{Dark matter halo merger histories beyond cold dark
  matter - {{I}}. {{Methods}} and application to warm dark matter}.
\newblock \bibinfo{journal}{Monthly Notices of the Royal Astronomical Society}
  \bibinfo{volume}{428}, \bibinfo{pages}{1774--1789}.
\newblock \DOIprefix\doi{10.1093/mnras/sts159}.
\bibitem[{Berlind et~al.(2003)Berlind, Weinberg, Benson, Baugh, Cole, Frenk,
  Jenkins, Katz and Lacey}]{Berlind2003}
\bibinfo{author}{Berlind, A.A.}, \bibinfo{author}{Weinberg, D.H.},
  \bibinfo{author}{Benson, A.J.}, \bibinfo{author}{Baugh, C.M.},
  \bibinfo{author}{Cole, S.}, \bibinfo{author}{Frenk, C.S.},
  \bibinfo{author}{Jenkins, A.}, \bibinfo{author}{Katz, N.},
  \bibinfo{author}{Lacey, C.G.}, \bibinfo{year}{2003}.
\newblock \bibinfo{title}{The {{Halo Occupation Distribution And The Physics}}
  of {{Galaxy Formation}}}.
\newblock \bibinfo{journal}{The Astrophysical Journal} \bibinfo{volume}{593},
  \bibinfo{pages}{1--25}.
\bibitem[{Beutler et~al.(2013)Beutler, Blake, Colless, Jones, {Staveley-Smith},
  Campbell, Parker, Saunders and Watson}]{Beutler2013}
\bibinfo{author}{Beutler, F.}, \bibinfo{author}{Blake, C.},
  \bibinfo{author}{Colless, M.}, \bibinfo{author}{Jones, D.H.},
  \bibinfo{author}{{Staveley-Smith}, L.}, \bibinfo{author}{Campbell, L.A.},
  \bibinfo{author}{Parker, Q.}, \bibinfo{author}{Saunders, W.},
  \bibinfo{author}{Watson, F.}, \bibinfo{year}{2013}.
\newblock \bibinfo{title}{The {{6dF Galaxy Survey}}: Dependence of halo
  occupation on stellar mass}.
\newblock \bibinfo{journal}{Monthly Notices of the Royal Astronomical Society}
  \bibinfo{volume}{429}, \bibinfo{pages}{3604--3618}.
\newblock \DOIprefix\doi{10.1093/mnras/sts637}.
\bibitem[{Bhattacharya et~al.(2013)Bhattacharya, Habib, Heitmann and
  Vikhlinin}]{Bhattacharya2013}
\bibinfo{author}{Bhattacharya, S.}, \bibinfo{author}{Habib, S.},
  \bibinfo{author}{Heitmann, K.}, \bibinfo{author}{Vikhlinin, A.},
  \bibinfo{year}{2013}.
\newblock \bibinfo{title}{Dark {{Matter Halo Profiles}} of {{Massive
  Clusters}}: {{Theory}} versus {{Observations}}}.
\newblock \bibinfo{journal}{The Astrophysical Journal} \bibinfo{volume}{766},
  \bibinfo{pages}{32}.
\newblock \DOIprefix\doi{10.1088/0004-637X/766/1/32}.
\bibitem[{Bhattacharya et~al.(2011)Bhattacharya, Heitmann, White, Luki{\'c},
  Wagner and Habib}]{Bhattacharya2011}
\bibinfo{author}{Bhattacharya, S.}, \bibinfo{author}{Heitmann, K.},
  \bibinfo{author}{White, M.}, \bibinfo{author}{Luki{\'c}, Z.},
  \bibinfo{author}{Wagner, C.}, \bibinfo{author}{Habib, S.},
  \bibinfo{year}{2011}.
\newblock \bibinfo{title}{{{MASS FUNCTION PREDICTIONS BEYOND $\Lambda$CDM}}}.
\newblock \bibinfo{journal}{The Astrophysical Journal} \bibinfo{volume}{732},
  \bibinfo{pages}{122--122}.
\newblock \DOIprefix\doi{10.1088/0004-637X/732/2/122}.
\bibitem[{Blake et~al.(2008)Blake, Collister and Lahav}]{Blake2008}
\bibinfo{author}{Blake, C.}, \bibinfo{author}{Collister, A.},
  \bibinfo{author}{Lahav, O.}, \bibinfo{year}{2008}.
\newblock \bibinfo{title}{Halo-model signatures from 380 000 {{Sloan Digital
  Sky Survey}} luminous red galaxies with photometric redshifts}.
\newblock \bibinfo{journal}{Monthly Notices of the Royal Astronomical Society}
  \bibinfo{volume}{385}, \bibinfo{pages}{1257--1269}.
\newblock \DOIprefix\doi{10.1111/j.1365-2966.2007.11925.x}.
\bibitem[{Blas et~al.(2011)Blas, Lesgourgues and Tram}]{Blas2011}
\bibinfo{author}{Blas, D.}, \bibinfo{author}{Lesgourgues, J.},
  \bibinfo{author}{Tram, T.}, \bibinfo{year}{2011}.
\newblock \bibinfo{title}{The {{Cosmic Linear Anisotropy Solving System}}
  ({{CLASS}}). {{Part II}}: {{Approximation}} schemes}.
\newblock \bibinfo{journal}{Journal of Cosmology and Astroparticle Physics}
  \bibinfo{volume}{2011}, \bibinfo{pages}{034--034}.
\newblock \DOIprefix\doi{10.1088/1475-7516/2011/07/034}.
\bibitem[{Bocquet et~al.(2016)Bocquet, Saro, Dolag and Mohr}]{Bocquet2016}
\bibinfo{author}{Bocquet, S.}, \bibinfo{author}{Saro, A.},
  \bibinfo{author}{Dolag, K.}, \bibinfo{author}{Mohr, J.J.},
  \bibinfo{year}{2016}.
\newblock \bibinfo{title}{Halo mass function: Baryon impact, fitting formulae,
  and implications for cluster cosmology}.
\newblock \bibinfo{journal}{Monthly Notices of the Royal Astronomical Society}
  \bibinfo{volume}{456}, \bibinfo{pages}{2361--2373}.
\newblock \DOIprefix\doi{10.1093/mnras/stv2657}.
\bibitem[{Bond et~al.(1991)Bond, Cole, Efstathiou and Kaiser}]{Bond1991}
\bibinfo{author}{Bond, J.R.}, \bibinfo{author}{Cole, S.},
  \bibinfo{author}{Efstathiou, G.}, \bibinfo{author}{Kaiser, N.},
  \bibinfo{year}{1991}.
\newblock \bibinfo{title}{Excursion set mass functions for hierarchical
  {{Gaussian}} fluctuations}.
\newblock \bibinfo{journal}{The Astrophysical Journal} \bibinfo{volume}{379},
  \bibinfo{pages}{440--460}.
\newblock \DOIprefix\doi{10.1086/170520}.
\bibitem[{Bond and Efstathiou(1984)}]{Bond1984}
\bibinfo{author}{Bond, J.R.}, \bibinfo{author}{Efstathiou, G.},
  \bibinfo{year}{1984}.
\newblock \bibinfo{title}{Cosmic background radiation anisotropies in universes
  dominated by nonbaryonic dark matter}.
\newblock \bibinfo{journal}{The Astrophysical Journal} \bibinfo{volume}{285},
  \bibinfo{pages}{L45--L45}.
\newblock \DOIprefix\doi{10.1086/184362}.
\bibitem[{Bosch(2002)}]{Bosch2002}
\bibinfo{author}{Bosch, F.C.V.D.}, \bibinfo{year}{2002}.
\newblock \bibinfo{title}{The universal mass accretion history of cold dark
  matter haloes}.
\newblock \bibinfo{journal}{Monthly Notices of the Royal Astronomical Society}
  \bibinfo{volume}{331}, \bibinfo{pages}{98--110}.
\newblock \DOIprefix\doi{10.1046/j.1365-8711.2002.05171.x}.
\bibitem[{Bryan and Norman(1998)}]{Bryan1998}
\bibinfo{author}{Bryan, G.L.}, \bibinfo{author}{Norman, M.L.},
  \bibinfo{year}{1998}.
\newblock \bibinfo{title}{Statistical {{Properties}} of {{X}}-{{Ray Clusters}}:
  {{Analytic}} and {{Numerical Comparisons}}}.
\newblock \bibinfo{journal}{The Astrophysical Journal} \bibinfo{volume}{495},
  \bibinfo{pages}{80--99}.
\newblock \DOIprefix\doi{10.1086/305262}.
\bibitem[{Bullock(2001)}]{Bullock2001a}
\bibinfo{author}{Bullock, J.S.}, \bibinfo{year}{2001}.
\newblock \bibinfo{title}{Shapes of dark matter halos}.
\newblock \bibinfo{journal}{Arxiv e-prints} .
\bibitem[{Bullock et~al.(2001)Bullock, Kolatt, Sigad, Somerville, Kravtsov,
  a.~Klypin, Primack and Dekel}]{Bullock2001}
\bibinfo{author}{Bullock, J.S.}, \bibinfo{author}{Kolatt, T.S.},
  \bibinfo{author}{Sigad, Y.}, \bibinfo{author}{Somerville, R.S.},
  \bibinfo{author}{Kravtsov, a.V.}, \bibinfo{author}{a.~Klypin, a.},
  \bibinfo{author}{Primack, J.R.}, \bibinfo{author}{Dekel, A.},
  \bibinfo{year}{2001}.
\newblock \bibinfo{title}{Profiles of dark haloes: Evolution, scatter and
  environment}.
\newblock \bibinfo{journal}{Monthly Notices of the Royal Astronomical Society}
  \bibinfo{volume}{321}, \bibinfo{pages}{559--575}.
\newblock \DOIprefix\doi{10.1046/j.1365-8711.2001.04068.x}.
\bibitem[{Bullock et~al.(2002)Bullock, Wechsler and Somerville}]{Bullock2002}
\bibinfo{author}{Bullock, J.S.}, \bibinfo{author}{Wechsler, R.H.},
  \bibinfo{author}{Somerville, R.S.}, \bibinfo{year}{2002}.
\newblock \bibinfo{title}{Galaxy halo occupation at high redshift}.
\newblock \bibinfo{journal}{Monthly Notices of the Royal Astronomical Society}
  \bibinfo{volume}{329}, \bibinfo{pages}{246--256}.
\bibitem[{Cacciato et~al.(2012)Cacciato, Lahav, {van den Bosch}, Hoekstra and
  Dekel}]{Cacciato2012}
\bibinfo{author}{Cacciato, M.}, \bibinfo{author}{Lahav, O.},
  \bibinfo{author}{{van den Bosch}, F.C.}, \bibinfo{author}{Hoekstra, H.},
  \bibinfo{author}{Dekel, A.}, \bibinfo{year}{2012}.
\newblock \bibinfo{title}{On combining galaxy clustering and weak lensing to
  unveil galaxy biasing via the halo model}.
\newblock \bibinfo{journal}{Monthly Notices of the Royal Astronomical Society}
  \bibinfo{volume}{426}, \bibinfo{pages}{566--587}.
\newblock \DOIprefix\doi{10.1111/j.1365-2966.2012.21762.x}.
\bibitem[{Carretero et~al.(2015)Carretero, Castander, Gazta{\~n}aga, Crocce and
  Fosalba}]{Carretero2015}
\bibinfo{author}{Carretero, J.}, \bibinfo{author}{Castander, F.J.},
  \bibinfo{author}{Gazta{\~n}aga, E.}, \bibinfo{author}{Crocce, M.},
  \bibinfo{author}{Fosalba, P.}, \bibinfo{year}{2015}.
\newblock \bibinfo{title}{An algorithm to build mock galaxy catalogues using
  {{MICE}} simulations}.
\newblock \bibinfo{journal}{Monthly Notices of the Royal Astronomical Society}
  \bibinfo{volume}{447}, \bibinfo{pages}{646--670}.
\newblock \DOIprefix\doi{10.1093/mnras/stu2402}.
\bibitem[{Chan et~al.(2017)Chan, Sheth and Scoccimarro}]{Chan2017}
\bibinfo{author}{Chan, K.C.}, \bibinfo{author}{Sheth, R.K.},
  \bibinfo{author}{Scoccimarro, R.}, \bibinfo{year}{2017}.
\newblock \bibinfo{title}{Effective window function for {{Lagrangian}} halos}.
\newblock \bibinfo{journal}{Physical Review D} \bibinfo{volume}{96},
  \bibinfo{pages}{103543}.
\newblock \DOIprefix\doi{10.1103/PhysRevD.96.103543}.
\bibitem[{Chen et~al.(2021)Chen, Wolz, Spinelli and Murray}]{Chen2021}
\bibinfo{author}{Chen, Z.}, \bibinfo{author}{Wolz, L.},
  \bibinfo{author}{Spinelli, M.}, \bibinfo{author}{Murray, S.G.},
  \bibinfo{year}{2021}.
\newblock \bibinfo{title}{Extracting {{HI Astrophysics}} from {{Interferometric
  Intensity Mapping}}}.
\newblock \bibinfo{journal}{MNRAS} \bibinfo{volume}{502},
  \bibinfo{pages}{5259--5276}.
\newblock \DOIprefix\doi{10.1093/mnras/stab386}.
\bibitem[{Child et~al.(2018)Child, Habib, Heitmann, Frontiere, Finkel, Pope and
  Morozov}]{Child2018}
\bibinfo{author}{Child, H.L.}, \bibinfo{author}{Habib, S.},
  \bibinfo{author}{Heitmann, K.}, \bibinfo{author}{Frontiere, N.},
  \bibinfo{author}{Finkel, H.}, \bibinfo{author}{Pope, A.},
  \bibinfo{author}{Morozov, V.}, \bibinfo{year}{2018}.
\newblock \bibinfo{title}{Halo {{Profiles}} and the {{Concentration}}-{{Mass
  Relation}} for a {{$\Lambda$CDM Universe}}}.
\newblock \bibinfo{journal}{The Astrophysical Journal} \bibinfo{volume}{859},
  \bibinfo{pages}{55}.
\newblock \DOIprefix\doi{10.3847/1538-4357/aabf95}.
\bibitem[{Clampitt et~al.(2017)Clampitt, S{\'a}nchez, Kwan, Krause, MacCrann,
  Park, Troxel, Jain, Rozo, Rykoff, Wechsler, Blazek, Bonnett, Crocce, Fang,
  Gaztanaga, Gruen, Jarvis, Miquel, Prat, Ross, Sheldon, Zuntz, Abbott,
  Abdalla, Armstrong, Becker, {Benoit-L{\'e}vy}, Bernstein, Bertin, Brooks,
  Burke, Carnero~Rosell, Carrasco~Kind, Cunha, D'Andrea, {da Costa}, Desai,
  Diehl, Dietrich, Doel, Estrada, Evrard, Fausti~Neto, Flaugher, Fosalba,
  Frieman, Gruendl, Honscheid, James, Kuehn, Kuropatkin, Lahav, Lima, March,
  Marshall, Martini, Melchior, Mohr, Nichol, Nord, Plazas, Romer, Sanchez,
  Scarpine, Schubnell, {Sevilla-Noarbe}, Smith, {Soares-Santos}, Sobreira,
  Suchyta, Swanson, Tarle, Thomas, Vikram and Walker}]{Clampitt2017}
\bibinfo{author}{Clampitt, J.}, \bibinfo{author}{S{\'a}nchez, C.},
  \bibinfo{author}{Kwan, J.}, \bibinfo{author}{Krause, E.},
  \bibinfo{author}{MacCrann, N.}, \bibinfo{author}{Park, Y.},
  \bibinfo{author}{Troxel, M.A.}, \bibinfo{author}{Jain, B.},
  \bibinfo{author}{Rozo, E.}, \bibinfo{author}{Rykoff, E.S.},
  \bibinfo{author}{Wechsler, R.H.}, \bibinfo{author}{Blazek, J.},
  \bibinfo{author}{Bonnett, C.}, \bibinfo{author}{Crocce, M.},
  \bibinfo{author}{Fang, Y.}, \bibinfo{author}{Gaztanaga, E.},
  \bibinfo{author}{Gruen, D.}, \bibinfo{author}{Jarvis, M.},
  \bibinfo{author}{Miquel, R.}, \bibinfo{author}{Prat, J.},
  \bibinfo{author}{Ross, A.J.}, \bibinfo{author}{Sheldon, E.},
  \bibinfo{author}{Zuntz, J.}, \bibinfo{author}{Abbott, T.M.C.},
  \bibinfo{author}{Abdalla, F.B.}, \bibinfo{author}{Armstrong, R.},
  \bibinfo{author}{Becker, M.R.}, \bibinfo{author}{{Benoit-L{\'e}vy}, A.},
  \bibinfo{author}{Bernstein, G.M.}, \bibinfo{author}{Bertin, E.},
  \bibinfo{author}{Brooks, D.}, \bibinfo{author}{Burke, D.L.},
  \bibinfo{author}{Carnero~Rosell, A.}, \bibinfo{author}{Carrasco~Kind, M.},
  \bibinfo{author}{Cunha, C.E.}, \bibinfo{author}{D'Andrea, C.B.},
  \bibinfo{author}{{da Costa}, L.N.}, \bibinfo{author}{Desai, S.},
  \bibinfo{author}{Diehl, H.T.}, \bibinfo{author}{Dietrich, J.P.},
  \bibinfo{author}{Doel, P.}, \bibinfo{author}{Estrada, J.},
  \bibinfo{author}{Evrard, A.E.}, \bibinfo{author}{Fausti~Neto, A.},
  \bibinfo{author}{Flaugher, B.}, \bibinfo{author}{Fosalba, P.},
  \bibinfo{author}{Frieman, J.}, \bibinfo{author}{Gruendl, R.A.},
  \bibinfo{author}{Honscheid, K.}, \bibinfo{author}{James, D.J.},
  \bibinfo{author}{Kuehn, K.}, \bibinfo{author}{Kuropatkin, N.},
  \bibinfo{author}{Lahav, O.}, \bibinfo{author}{Lima, M.},
  \bibinfo{author}{March, M.}, \bibinfo{author}{Marshall, J.L.},
  \bibinfo{author}{Martini, P.}, \bibinfo{author}{Melchior, P.},
  \bibinfo{author}{Mohr, J.J.}, \bibinfo{author}{Nichol, R.C.},
  \bibinfo{author}{Nord, B.}, \bibinfo{author}{Plazas, A.A.},
  \bibinfo{author}{Romer, A.K.}, \bibinfo{author}{Sanchez, E.},
  \bibinfo{author}{Scarpine, V.}, \bibinfo{author}{Schubnell, M.},
  \bibinfo{author}{{Sevilla-Noarbe}, I.}, \bibinfo{author}{Smith, R.C.},
  \bibinfo{author}{{Soares-Santos}, M.}, \bibinfo{author}{Sobreira, F.},
  \bibinfo{author}{Suchyta, E.}, \bibinfo{author}{Swanson, M.E.C.},
  \bibinfo{author}{Tarle, G.}, \bibinfo{author}{Thomas, D.},
  \bibinfo{author}{Vikram, V.}, \bibinfo{author}{Walker, A.R.},
  \bibinfo{year}{2017}.
\newblock \bibinfo{title}{Galaxy\textendash galaxy lensing in the {{Dark Energy
  Survey Science Verification}} data}.
\newblock \bibinfo{journal}{Monthly Notices of the Royal Astronomical Society}
  \bibinfo{volume}{465}, \bibinfo{pages}{4204--4218}.
\newblock \DOIprefix\doi{10.1093/mnras/stw2988}.
\bibitem[{Cole and Kaiser(1989)}]{Cole1989}
\bibinfo{author}{Cole, S.}, \bibinfo{author}{Kaiser, N.}, \bibinfo{year}{1989}.
\newblock \bibinfo{title}{Biased clustering in the cold dark matter cosmogony}.
\newblock \bibinfo{journal}{Monthly Notices of the Royal Astronomical Society}
  \bibinfo{volume}{237}, \bibinfo{pages}{1127--1127}.
\bibitem[{Comparat et~al.(2017)Comparat, Prada, Yepes and
  Klypin}]{Comparat2017}
\bibinfo{author}{Comparat, J.}, \bibinfo{author}{Prada, F.},
  \bibinfo{author}{Yepes, G.}, \bibinfo{author}{Klypin, A.},
  \bibinfo{year}{2017}.
\newblock \bibinfo{title}{Accurate mass and velocity functions of dark matter
  haloes}.
\newblock \bibinfo{journal}{Monthly Notices of the Royal Astronomical Society}
  \bibinfo{volume}{469}, \bibinfo{pages}{4157--4174}.
\newblock \DOIprefix\doi{10.1093/mnras/stx1183}.
\bibitem[{Contreras et~al.(2013)Contreras, Baugh, Norberg and
  Padilla}]{Contreras2013}
\bibinfo{author}{Contreras, S.}, \bibinfo{author}{Baugh, C.M.},
  \bibinfo{author}{Norberg, P.}, \bibinfo{author}{Padilla, N.},
  \bibinfo{year}{2013}.
\newblock \bibinfo{title}{How robust are predictions of galaxy clustering?}
\newblock \bibinfo{journal}{Monthly Notices of the Royal Astronomical Society}
  \bibinfo{volume}{432}, \bibinfo{pages}{2717--2730}.
\newblock \DOIprefix\doi{10.1093/mnras/stt629}.
\bibitem[{Cooray and Sheth(2002)}]{Cooray2002}
\bibinfo{author}{Cooray, A.}, \bibinfo{author}{Sheth, R.},
  \bibinfo{year}{2002}.
\newblock \bibinfo{title}{Halo models of large scale structure}.
\newblock \bibinfo{journal}{Physics Reports} \bibinfo{volume}{327},
  \bibinfo{pages}{1--129}.
\bibitem[{Corasaniti and Achitouv(2011)}]{Corasaniti2011}
\bibinfo{author}{Corasaniti, P.S.}, \bibinfo{author}{Achitouv, I.},
  \bibinfo{year}{2011}.
\newblock \bibinfo{title}{Excursion set halo mass function and bias in a
  stochastic barrier model of ellipsoidal collapse}.
\newblock \bibinfo{journal}{Physical Review D} \bibinfo{volume}{84},
  \bibinfo{pages}{023009--023009}.
\newblock \DOIprefix\doi{10.1103/PhysRevD.84.023009}.
\bibitem[{Correa et~al.(2015)Correa, Wyithe, Schaye and Duffy}]{Correa2015b}
\bibinfo{author}{Correa, C.A.}, \bibinfo{author}{Wyithe, J.S.B.},
  \bibinfo{author}{Schaye, J.}, \bibinfo{author}{Duffy, A.R.},
  \bibinfo{year}{2015}.
\newblock \bibinfo{title}{The accretion history of dark matter haloes
  \textendash{} {{III}}. {{A}} physical model for the concentration\textendash
  mass relation}.
\newblock \bibinfo{journal}{Monthly Notices of the Royal Astronomical Society}
  \bibinfo{volume}{452}, \bibinfo{pages}{1217--1232}.
\newblock \DOIprefix\doi{10.1093/mnras/stv1363}.
\bibitem[{Coupon et~al.(2015)Coupon, Arnouts, {van Waerbeke}, Moutard, Ilbert,
  {van Uitert}, Erben, Garilli, Guzzo, Heymans, Hildebrandt, Hoekstra,
  Kilbinger, Kitching, Mellier, Miller, Scodeggio, Bonnett, Branchini,
  Davidzon, De~Lucia, Fritz, Fu, Hudelot, Hudson, Kuijken, Leauthaud,
  F{\`e}vre, McCracken, Moscardini, Rowe, Schrabback, Semboloni, Velander,
  Le~Fevre, McCracken, Moscardini, Rowe, Schrabback, Semboloni and
  Velander}]{Coupon2015}
\bibinfo{author}{Coupon, J.}, \bibinfo{author}{Arnouts, S.},
  \bibinfo{author}{{van Waerbeke}, L.}, \bibinfo{author}{Moutard, T.},
  \bibinfo{author}{Ilbert, O.}, \bibinfo{author}{{van Uitert}, E.},
  \bibinfo{author}{Erben, T.}, \bibinfo{author}{Garilli, B.},
  \bibinfo{author}{Guzzo, L.}, \bibinfo{author}{Heymans, C.},
  \bibinfo{author}{Hildebrandt, H.}, \bibinfo{author}{Hoekstra, H.},
  \bibinfo{author}{Kilbinger, M.}, \bibinfo{author}{Kitching, T.},
  \bibinfo{author}{Mellier, Y.}, \bibinfo{author}{Miller, L.},
  \bibinfo{author}{Scodeggio, M.}, \bibinfo{author}{Bonnett, C.},
  \bibinfo{author}{Branchini, E.}, \bibinfo{author}{Davidzon, I.},
  \bibinfo{author}{De~Lucia, G.}, \bibinfo{author}{Fritz, a.},
  \bibinfo{author}{Fu, L.}, \bibinfo{author}{Hudelot, P.},
  \bibinfo{author}{Hudson, M.J.}, \bibinfo{author}{Kuijken, K.},
  \bibinfo{author}{Leauthaud, a.}, \bibinfo{author}{F{\`e}vre, O.L.},
  \bibinfo{author}{McCracken, H.J.}, \bibinfo{author}{Moscardini, L.},
  \bibinfo{author}{Rowe, B.T.P.}, \bibinfo{author}{Schrabback, T.},
  \bibinfo{author}{Semboloni, E.}, \bibinfo{author}{Velander, M.},
  \bibinfo{author}{Le~Fevre, O.}, \bibinfo{author}{McCracken, H.J.},
  \bibinfo{author}{Moscardini, L.}, \bibinfo{author}{Rowe, B.T.P.},
  \bibinfo{author}{Schrabback, T.}, \bibinfo{author}{Semboloni, E.},
  \bibinfo{author}{Velander, M.}, \bibinfo{year}{2015}.
\newblock \bibinfo{title}{The galaxy-halo connection from a joint lensing,
  clustering and abundance analysis in the {{CFHTLenS}}/{{VIPERS}} field}.
\newblock \bibinfo{journal}{Monthly Notices of the Royal Astronomical Society}
  \bibinfo{volume}{449}, \bibinfo{pages}{1352--1379}.
\newblock \DOIprefix\doi{10.1093/mnras/stv276}.
\bibitem[{Dalal et~al.(2010)Dalal, Lithwick and Kuhlen}]{Dalal2010}
\bibinfo{author}{Dalal, N.}, \bibinfo{author}{Lithwick, Y.},
  \bibinfo{author}{Kuhlen, M.}, \bibinfo{year}{2010}.
\newblock \bibinfo{title}{The {{Origin}} of {{Dark Matter Halo Profiles}}}.
\newblock \bibinfo{journal}{arXiv e-prints} \bibinfo{volume}{1010},
  \bibinfo{pages}{arXiv:1010.2539}.
\bibitem[{Diemer(2018)}]{Diemer2018}
\bibinfo{author}{Diemer, B.}, \bibinfo{year}{2018}.
\newblock \bibinfo{title}{{{COLOSSUS}}: {{A Python Toolkit}} for {{Cosmology}},
  {{Large}}-scale {{Structure}}, and {{Dark Matter Halos}}}.
\newblock \bibinfo{journal}{The Astrophysical Journal Supplement Series}
  \bibinfo{volume}{239}, \bibinfo{pages}{35}.
\newblock \DOIprefix\doi{10.3847/1538-4365/aaee8c}.
\bibitem[{Diemer and Joyce(2019)}]{Diemer2019}
\bibinfo{author}{Diemer, B.}, \bibinfo{author}{Joyce, M.},
  \bibinfo{year}{2019}.
\newblock \bibinfo{title}{An {{Accurate Physical Model}} for {{Halo
  Concentrations}}}.
\newblock \bibinfo{journal}{The Astrophysical Journal} \bibinfo{volume}{871},
  \bibinfo{pages}{168}.
\newblock \DOIprefix\doi{10.3847/1538-4357/aafad6}.
\bibitem[{Diemer and Kravtsov(2015)}]{Diemer2015}
\bibinfo{author}{Diemer, B.}, \bibinfo{author}{Kravtsov, A.V.},
  \bibinfo{year}{2015}.
\newblock \bibinfo{title}{A {{Universal Model}} for {{Halo Concentrations}}}.
\newblock \bibinfo{journal}{The Astrophysical Journal} \bibinfo{volume}{799},
  \bibinfo{pages}{108}.
\newblock \DOIprefix\doi{10.1088/0004-637X/799/1/108}.
\bibitem[{Diemer et~al.(2013)Diemer, More and Kravtsov}]{Diemer2013}
\bibinfo{author}{Diemer, B.}, \bibinfo{author}{More, S.},
  \bibinfo{author}{Kravtsov, A.V.}, \bibinfo{year}{2013}.
\newblock \bibinfo{title}{{{THE PSEUDO}}-{{EVOLUTION OF HALO MASS}}}.
\newblock \bibinfo{journal}{The Astrophysical Journal} \bibinfo{volume}{766},
  \bibinfo{pages}{25--25}.
\newblock \DOIprefix\doi{10.1088/0004-637X/766/1/25}.
\bibitem[{Dolag et~al.(2004)Dolag, Bartelmann, Perrotta, Baccigalupi,
  Moscardini, Meneghetti and Tormen}]{Dolag2004}
\bibinfo{author}{Dolag, K.}, \bibinfo{author}{Bartelmann, M.},
  \bibinfo{author}{Perrotta, F.}, \bibinfo{author}{Baccigalupi, C.},
  \bibinfo{author}{Moscardini, L.}, \bibinfo{author}{Meneghetti, M.},
  \bibinfo{author}{Tormen, G.}, \bibinfo{year}{2004}.
\newblock \bibinfo{title}{Numerical study of halo concentrations in dark-energy
  cosmologies}.
\newblock \bibinfo{journal}{Astronomy \& Astrophysics} \bibinfo{volume}{416},
  \bibinfo{pages}{853--864}.
\newblock \DOIprefix\doi{10.1051/0004-6361:20031757}.
\bibitem[{Duffy et~al.(2008)Duffy, Schaye, Kay and Dalla~Vecchia}]{Duffy2008}
\bibinfo{author}{Duffy, A.R.}, \bibinfo{author}{Schaye, J.},
  \bibinfo{author}{Kay, S.T.}, \bibinfo{author}{Dalla~Vecchia, C.},
  \bibinfo{year}{2008}.
\newblock \bibinfo{title}{Dark matter halo concentrations in the {{Wilkinson
  Microwave Anisotropy Probe}} year 5 cosmology}.
\newblock \bibinfo{journal}{Monthly Notices of the Royal Astronomical Society:
  Letters} \bibinfo{volume}{390}, \bibinfo{pages}{L64--L68}.
\newblock \DOIprefix\doi{10.1111/j.1745-3933.2008.00537.x}.
\bibitem[{Dutton and Macci{\`o}(2014)}]{Dutton2014}
\bibinfo{author}{Dutton, A.A.}, \bibinfo{author}{Macci{\`o}, A.V.},
  \bibinfo{year}{2014}.
\newblock \bibinfo{title}{Cold dark matter haloes in the {{Planck}} era:
  Evolution of structural parameters for {{Einasto}} and {{NFW}} profiles}.
\newblock \bibinfo{journal}{Monthly Notices of the Royal Astronomical Society}
  \bibinfo{volume}{441}, \bibinfo{pages}{3359--3374}.
\newblock \DOIprefix\doi{10.1093/mnras/stu742}.
\bibitem[{Dvornik et~al.(2018)Dvornik, Hoekstra, Kuijken, Schneider, Amon,
  Nakajima, Viola, Choi, Erben, Farrow, Heymans, Hildebrandt, Sif{\'o}n and
  Wang}]{Dvornik2018}
\bibinfo{author}{Dvornik, A.}, \bibinfo{author}{Hoekstra, H.},
  \bibinfo{author}{Kuijken, K.}, \bibinfo{author}{Schneider, P.},
  \bibinfo{author}{Amon, A.}, \bibinfo{author}{Nakajima, R.},
  \bibinfo{author}{Viola, M.}, \bibinfo{author}{Choi, A.},
  \bibinfo{author}{Erben, T.}, \bibinfo{author}{Farrow, D.J.},
  \bibinfo{author}{Heymans, C.}, \bibinfo{author}{Hildebrandt, H.},
  \bibinfo{author}{Sif{\'o}n, C.}, \bibinfo{author}{Wang, L.},
  \bibinfo{year}{2018}.
\newblock \bibinfo{title}{Unveiling galaxy bias via the halo model, {{KiDS}},
  and {{GAMA}}}.
\newblock \bibinfo{journal}{Monthly Notices of the Royal Astronomical Society}
  \bibinfo{volume}{479}, \bibinfo{pages}{1240--1259}.
\newblock \DOIprefix\doi{10.1093/mnras/sty1502}.
\bibitem[{Einasto(1965)}]{Einasto1965}
\bibinfo{author}{Einasto, J.}, \bibinfo{year}{1965}.
\newblock \bibinfo{title}{On the {{Construction}} of a {{Composite Model}} for
  the {{Galaxy}} and on the {{Determination}} of the {{System}} of {{Galactic
  Parameters}}}.
\newblock \bibinfo{journal}{Trudy Astrofizicheskogo Instituta Alma-Ata}
  \bibinfo{volume}{5}, \bibinfo{pages}{87--100}.
\bibitem[{Eisenstein and Hu(1998)}]{Eisenstein1998}
\bibinfo{author}{Eisenstein, D.J.}, \bibinfo{author}{Hu, W.},
  \bibinfo{year}{1998}.
\newblock \bibinfo{title}{Baryonic {{Features}} in the {{Matter Transfer
  Function}}}.
\newblock \bibinfo{journal}{The Astrophysical Journal} \bibinfo{volume}{496},
  \bibinfo{pages}{605--614}.
\newblock \DOIprefix\doi{10.1086/305424}.
\bibitem[{Eke et~al.(2001)Eke, Navarro and Steinmetz}]{Eke2001}
\bibinfo{author}{Eke, V.R.}, \bibinfo{author}{Navarro, J.F.},
  \bibinfo{author}{Steinmetz, M.}, \bibinfo{year}{2001}.
\newblock \bibinfo{title}{The {{Power Spectrum Dependence}} of {{Dark Matter
  Halo Concentrations}}}.
\newblock \bibinfo{journal}{The Astrophysical Journal} \bibinfo{volume}{554},
  \bibinfo{pages}{114--125}.
\newblock \DOIprefix\doi{10.1086/321345}.
\bibitem[{Frenk and White(2012)}]{Frenk2012}
\bibinfo{author}{Frenk, C.}, \bibinfo{author}{White, S.}, \bibinfo{year}{2012}.
\newblock \bibinfo{title}{Dark matter and cosmic structure}.
\newblock \bibinfo{journal}{Annalen der Physik} \bibinfo{volume}{524},
  \bibinfo{pages}{507--534}.
\newblock \DOIprefix\doi{10.1002/andp.201200212}.
\bibitem[{Garcia et~al.(2020)Garcia, Rozo, Becker and More}]{Garcia2020}
\bibinfo{author}{Garcia, R.}, \bibinfo{author}{Rozo, E.},
  \bibinfo{author}{Becker, M.R.}, \bibinfo{author}{More, S.},
  \bibinfo{year}{2020}.
\newblock \bibinfo{title}{A {{Redefinition}} of the {{Halo Boundary Leads}} to
  a {{Simple}} yet {{Accurate Halo Model}} of {{Large Scale Structure}}}.
\newblock \bibinfo{journal}{arXiv e-prints} \bibinfo{volume}{2006},
  \bibinfo{pages}{arXiv:2006.12751}.
\bibitem[{Geach et~al.(2012)Geach, Sobral, Hickox, a.~Wake, Smail, Best, Baugh
  and Stott}]{Geach2012}
\bibinfo{author}{Geach, J.E.}, \bibinfo{author}{Sobral, D.},
  \bibinfo{author}{Hickox, R.C.}, \bibinfo{author}{a.~Wake, D.},
  \bibinfo{author}{Smail, I.}, \bibinfo{author}{Best, P.N.},
  \bibinfo{author}{Baugh, C.M.}, \bibinfo{author}{Stott, J.P.},
  \bibinfo{year}{2012}.
\newblock \bibinfo{title}{The clustering of {{H$\alpha$}} emitters at z =2.23
  from {{HiZELS}}}.
\newblock \bibinfo{journal}{Monthly Notices of the Royal Astronomical Society}
  \bibinfo{volume}{426}, \bibinfo{pages}{679--689}.
\newblock \DOIprefix\doi{10.1111/j.1365-2966.2012.21725.x}.
\bibitem[{Ginzburg et~al.(2017)Ginzburg, Desjacques and Chan}]{Ginzburg2017}
\bibinfo{author}{Ginzburg, D.}, \bibinfo{author}{Desjacques, V.},
  \bibinfo{author}{Chan, K.C.}, \bibinfo{year}{2017}.
\newblock \bibinfo{title}{Shot noise and biased tracers: {{A}} new look at the
  halo model}.
\newblock \bibinfo{journal}{Physical Review D} \bibinfo{volume}{96},
  \bibinfo{pages}{083528}.
\newblock \DOIprefix\doi{10.1103/PhysRevD.96.083528}.
\bibitem[{Giocoli et~al.(2010)Giocoli, Bartelmann, Sheth and
  Cacciato}]{Giocoli2010}
\bibinfo{author}{Giocoli, C.}, \bibinfo{author}{Bartelmann, M.},
  \bibinfo{author}{Sheth, R.K.}, \bibinfo{author}{Cacciato, M.},
  \bibinfo{year}{2010}.
\newblock \bibinfo{title}{Halo model description of the non-linear dark matter
  power spectrum at k{$\gg$} 1\hspace{1em}{{Mpc}}-1}.
\newblock \bibinfo{journal}{Monthly Notices of the Royal Astronomical Society}
  \bibinfo{volume}{408}, \bibinfo{pages}{300--313}.
\newblock \DOIprefix\doi{10.1111/j.1365-2966.2010.17108.x}.
\bibitem[{Giocoli et~al.(2012)Giocoli, Tormen and Sheth}]{Giocoli2012}
\bibinfo{author}{Giocoli, C.}, \bibinfo{author}{Tormen, G.},
  \bibinfo{author}{Sheth, R.K.}, \bibinfo{year}{2012}.
\newblock \bibinfo{title}{Formation times, mass growth histories and
  concentrations of dark matter haloes}.
\newblock \bibinfo{journal}{Monthly Notices of the Royal Astronomical Society}
  \bibinfo{volume}{422}, \bibinfo{pages}{185--198}.
\newblock \DOIprefix\doi{10.1111/j.1365-2966.2012.20594.x}.
\bibitem[{Hernquist(1990)}]{Hernquist1990}
\bibinfo{author}{Hernquist, L.}, \bibinfo{year}{1990}.
\newblock \bibinfo{title}{An analytical model for spherical galaxies and
  bulges}.
\newblock \bibinfo{journal}{The Astrophysical Journal} \bibinfo{volume}{356},
  \bibinfo{pages}{359--359}.
\newblock \DOIprefix\doi{10.1086/168845}.
\bibitem[{Hunter(2007)}]{Hunter2007}
\bibinfo{author}{Hunter, J.D.}, \bibinfo{year}{2007}.
\newblock \bibinfo{title}{Matplotlib: {{A 2D}} graphics environment}.
\newblock \bibinfo{journal}{Computing in Science \& Engineering}
  \bibinfo{volume}{9}, \bibinfo{pages}{90--95}.
\bibitem[{Ishikawa et~al.(2021)Ishikawa, Okumura, Oguri and Lin}]{Ishikawa2021}
\bibinfo{author}{Ishikawa, S.}, \bibinfo{author}{Okumura, T.},
  \bibinfo{author}{Oguri, M.}, \bibinfo{author}{Lin, S.C.},
  \bibinfo{year}{2021}.
\newblock \bibinfo{title}{Halo-model analysis of the clustering of photometric
  luminous red galaxies at \$0.10 \textbackslash leq z \textbackslash leq
  1.05\$ from the {{Subaru Hyper Suprime}}-{{Cam Survey}}}.
\newblock \bibinfo{journal}{arXiv e-prints} \bibinfo{volume}{2103},
  \bibinfo{pages}{arXiv:2103.08628}.
\bibitem[{Ishiyama et~al.(2015)Ishiyama, Enoki, Kobayashi, Makiya, Nagashima
  and Oogi}]{Ishiyama2015}
\bibinfo{author}{Ishiyama, T.}, \bibinfo{author}{Enoki, M.},
  \bibinfo{author}{Kobayashi, M.A.R.}, \bibinfo{author}{Makiya, R.},
  \bibinfo{author}{Nagashima, M.}, \bibinfo{author}{Oogi, T.},
  \bibinfo{year}{2015}.
\newblock \bibinfo{title}{The \${$\nu\$$} 2 {{GC Simulations}} :
  {{Quantifying}} the {{Dark Side}} of the {{Universe}} in the {{Planck
  Cosmology}}}.
\newblock \bibinfo{journal}{Publications of the Astronomical Society of Japan}
  \bibinfo{volume}{67}, \bibinfo{pages}{61--61}.
\bibitem[{Ishiyama et~al.(2020)Ishiyama, Prada, Klypin, Sinha, Metcalf, Jullo,
  Altieri, Cora, Croton, {de la Torre}, {Mill{\'a}n-Calero}, Oogi, Ruedas and
  {Vega-Mart{\'i}nez}}]{Ishiyama2020}
\bibinfo{author}{Ishiyama, T.}, \bibinfo{author}{Prada, F.},
  \bibinfo{author}{Klypin, A.A.}, \bibinfo{author}{Sinha, M.},
  \bibinfo{author}{Metcalf, R.B.}, \bibinfo{author}{Jullo, E.},
  \bibinfo{author}{Altieri, B.}, \bibinfo{author}{Cora, S.A.},
  \bibinfo{author}{Croton, D.}, \bibinfo{author}{{de la Torre}, S.},
  \bibinfo{author}{{Mill{\'a}n-Calero}, D.E.}, \bibinfo{author}{Oogi, T.},
  \bibinfo{author}{Ruedas, J.}, \bibinfo{author}{{Vega-Mart{\'i}nez}, C.A.},
  \bibinfo{year}{2020}.
\newblock \bibinfo{title}{The {{Uchuu Simulations}}: {{Data Release}} 1 and
  {{Dark Matter Halo Concentrations}}}.
\newblock \bibinfo{journal}{arXiv e-prints} \bibinfo{volume}{2007},
  \bibinfo{pages}{arXiv:2007.14720}.
\bibitem[{Jenkins et~al.(2001)Jenkins, Frenk, White, Colberg, Cole, Evrard,
  Couchman and Yoshida}]{Jenkins2001}
\bibinfo{author}{Jenkins, A.}, \bibinfo{author}{Frenk, C.S.},
  \bibinfo{author}{White, S.D.M.}, \bibinfo{author}{Colberg, J.M.},
  \bibinfo{author}{Cole, S.}, \bibinfo{author}{Evrard, A.E.},
  \bibinfo{author}{Couchman, H.M.P.}, \bibinfo{author}{Yoshida, N.},
  \bibinfo{year}{2001}.
\newblock \bibinfo{title}{The mass function of dark matter haloes}.
\newblock \bibinfo{journal}{Monthly Notices of the Royal Astronomical Society}
  \bibinfo{volume}{321}, \bibinfo{pages}{372--384}.
\newblock \DOIprefix\doi{10.1046/j.1365-8711.2001.04029.x}.
\bibitem[{Jing(1998)}]{Jing1998}
\bibinfo{author}{Jing, Y.P.}, \bibinfo{year}{1998}.
\newblock \bibinfo{title}{Accurate {{Fitting Formula}} for the {{Two}}-{{Point
  Correlation Function}} of {{Dark Matter Halos}}}.
\newblock \bibinfo{journal}{The Astrophysical Journal} \bibinfo{volume}{503},
  \bibinfo{pages}{L9--L13}.
\newblock \DOIprefix\doi{10.1086/311530}.
\bibitem[{Jing(1999)}]{Jing1999}
\bibinfo{author}{Jing, Y.P.}, \bibinfo{year}{1999}.
\newblock \bibinfo{title}{Accurate determination of the {{Lagrangian}} bias for
  the dark matter halos}.
\newblock \bibinfo{journal}{The Astrophysical Journal} \bibinfo{volume}{515},
  \bibinfo{pages}{L45--L48}.
\newblock \DOIprefix\doi{10.1086/311978}.
\bibitem[{Jing and Suto(2002)}]{Jing2002}
\bibinfo{author}{Jing, Y.P.}, \bibinfo{author}{Suto, Y.}, \bibinfo{year}{2002}.
\newblock \bibinfo{title}{Triaxial {{Modeling}} of {{Halo Density Profiles}}
  with {{High}}-{{Resolution N}} -{{Body Simulations}}}.
\newblock \bibinfo{journal}{The Astrophysical Journal} \bibinfo{volume}{574},
  \bibinfo{pages}{538--553}.
\newblock \DOIprefix\doi{10.1086/341065}.
\bibitem[{Kauffmann et~al.(1997)Kauffmann, Nusser and
  Steinmetz}]{Kauffmann1997}
\bibinfo{author}{Kauffmann, G.}, \bibinfo{author}{Nusser, A.},
  \bibinfo{author}{Steinmetz, M.}, \bibinfo{year}{1997}.
\newblock \bibinfo{title}{Galaxy formation and large-scale bias}.
\newblock \bibinfo{journal}{Monthly Notices of the Royal Astronomical Society}
  \bibinfo{volume}{286}, \bibinfo{pages}{795--811}.
\bibitem[{Kauffmann et~al.(2004)Kauffmann, White, Heckman, M{\'e}nard,
  Brinchmann, Charlot, Tremonti and Brinkmann}]{Kauffmann2004}
\bibinfo{author}{Kauffmann, G.}, \bibinfo{author}{White, S.D.M.},
  \bibinfo{author}{Heckman, T.M.}, \bibinfo{author}{M{\'e}nard, B.},
  \bibinfo{author}{Brinchmann, J.}, \bibinfo{author}{Charlot, S.},
  \bibinfo{author}{Tremonti, C.}, \bibinfo{author}{Brinkmann, J.},
  \bibinfo{year}{2004}.
\newblock \bibinfo{title}{The environmental dependence of the relations between
  stellar mass, structure, star formation and nuclear activity in galaxies}.
\newblock \bibinfo{journal}{Monthly Notices of the Royal Astronomical Society}
  \bibinfo{volume}{353}, \bibinfo{pages}{713--731}.
\newblock \DOIprefix\doi{10.1111/j.1365-2966.2004.08117.x}.
\bibitem[{Kim et~al.(2011)Kim, Baugh, Benson, Cole, Frenk, Lacey, Power and
  Schneider}]{Kim2011}
\bibinfo{author}{Kim, H.S.}, \bibinfo{author}{Baugh, C.M.},
  \bibinfo{author}{Benson, A.J.}, \bibinfo{author}{Cole, S.},
  \bibinfo{author}{Frenk, C.S.}, \bibinfo{author}{Lacey, C.G.},
  \bibinfo{author}{Power, C.}, \bibinfo{author}{Schneider, M.},
  \bibinfo{year}{2011}.
\newblock \bibinfo{title}{The spatial distribution of cold gas in hierarchical
  galaxy formation models}.
\newblock \bibinfo{journal}{Monthly Notices of the Royal Astronomical Society}
  \bibinfo{volume}{414}, \bibinfo{pages}{2367--2385}.
\newblock \DOIprefix\doi{10.1111/j.1365-2966.2011.18556.x}.
\bibitem[{Klypin et~al.(2016)Klypin, Yepes, Gottl{\"o}ber, Prada and
  He{\ss}}]{Klypin2016}
\bibinfo{author}{Klypin, A.}, \bibinfo{author}{Yepes, G.},
  \bibinfo{author}{Gottl{\"o}ber, S.}, \bibinfo{author}{Prada, F.},
  \bibinfo{author}{He{\ss}, S.}, \bibinfo{year}{2016}.
\newblock \bibinfo{title}{{{MultiDark}} simulations: The story of dark matter
  halo concentrations and density profiles}.
\newblock \bibinfo{journal}{Monthly Notices of the Royal Astronomical Society}
  \bibinfo{volume}{457}, \bibinfo{pages}{4340--4359}.
\newblock \DOIprefix\doi{10.1093/mnras/stw248}.
\bibitem[{Knebe et~al.(2011)Knebe, Knollmann, Muldrew, Pearce, {Aragon-Calvo},
  Ascasibar, Behroozi, Ceverino, Colombi, Diemand, Dolag, Falck, Fasel,
  Gardner, Gottl{\"o}ber, Hsu, Iannuzzi, Klypin, Luki{\'c}, Maciejewski,
  Mcbride, Neyrinck, Planelles, Potter, Quilis, Rasera, Read, Ricker, Roy,
  Springel, Stadel, Stinson, Sutter, Turchaninov, Tweed, Yepes and
  Zemp}]{Knebe2011}
\bibinfo{author}{Knebe, A.}, \bibinfo{author}{Knollmann, S.R.},
  \bibinfo{author}{Muldrew, S.I.}, \bibinfo{author}{Pearce, F.R.},
  \bibinfo{author}{{Aragon-Calvo}, M.A.}, \bibinfo{author}{Ascasibar, Y.},
  \bibinfo{author}{Behroozi, P.S.}, \bibinfo{author}{Ceverino, D.},
  \bibinfo{author}{Colombi, S.}, \bibinfo{author}{Diemand, J.},
  \bibinfo{author}{Dolag, K.}, \bibinfo{author}{Falck, B.L.},
  \bibinfo{author}{Fasel, P.}, \bibinfo{author}{Gardner, J.},
  \bibinfo{author}{Gottl{\"o}ber, S.}, \bibinfo{author}{Hsu, C.H.},
  \bibinfo{author}{Iannuzzi, F.}, \bibinfo{author}{Klypin, A.},
  \bibinfo{author}{Luki{\'c}, Z.}, \bibinfo{author}{Maciejewski, M.},
  \bibinfo{author}{Mcbride, C.}, \bibinfo{author}{Neyrinck, M.C.},
  \bibinfo{author}{Planelles, S.}, \bibinfo{author}{Potter, D.},
  \bibinfo{author}{Quilis, V.}, \bibinfo{author}{Rasera, Y.},
  \bibinfo{author}{Read, J.I.}, \bibinfo{author}{Ricker, P.M.},
  \bibinfo{author}{Roy, F.}, \bibinfo{author}{Springel, V.},
  \bibinfo{author}{Stadel, J.}, \bibinfo{author}{Stinson, G.},
  \bibinfo{author}{Sutter, P.M.}, \bibinfo{author}{Turchaninov, V.},
  \bibinfo{author}{Tweed, D.}, \bibinfo{author}{Yepes, G.},
  \bibinfo{author}{Zemp, M.}, \bibinfo{year}{2011}.
\newblock \bibinfo{title}{Haloes gone {{MAD}}: {{The Halo}}-{{Finder Comparison
  Project}}}.
\newblock \bibinfo{journal}{Monthly Notices of the Royal Astronomical Society}
  \bibinfo{volume}{415}, \bibinfo{pages}{2293--2318}.
\newblock \DOIprefix\doi{10.1111/j.1365-2966.2011.18858.x}.
\bibitem[{Krause et~al.(2012)Krause, Hirata, Martin, Neill and
  Wyder}]{Krause2012}
\bibinfo{author}{Krause, E.}, \bibinfo{author}{Hirata, C.M.},
  \bibinfo{author}{Martin, C.}, \bibinfo{author}{Neill, J.D.},
  \bibinfo{author}{Wyder, T.K.}, \bibinfo{year}{2012}.
\newblock \bibinfo{title}{Halo occupation distribution modelling of green
  valley galaxies}.
\newblock \bibinfo{journal}{Monthly Notices of the Royal Astronomical Society}
  \bibinfo{volume}{428}, \bibinfo{pages}{2548--2564}.
\newblock \DOIprefix\doi{10.1093/mnras/sts221}.
\bibitem[{Lapi and Danese(2014)}]{Lapi2014}
\bibinfo{author}{Lapi, A.}, \bibinfo{author}{Danese, L.}, \bibinfo{year}{2014}.
\newblock \bibinfo{title}{Statistics of {{Dark Matter Halos}} in the
  {{Excursion Set Peak Framework}}}.
\newblock \bibinfo{journal}{Journal of Cosmology and Astroparticle Physics}
  \bibinfo{volume}{7}, \bibinfo{pages}{44--44}.
\newblock \DOIprefix\doi{10.1088/1475-7516/2014/07/044}.
\bibitem[{Leauthaud et~al.(2011)Leauthaud, Tinker, Behroozi, Busha and
  Wechsler}]{Leauthaud2011}
\bibinfo{author}{Leauthaud, A.}, \bibinfo{author}{Tinker, J.},
  \bibinfo{author}{Behroozi, P.S.}, \bibinfo{author}{Busha, M.T.},
  \bibinfo{author}{Wechsler, R.H.}, \bibinfo{year}{2011}.
\newblock \bibinfo{title}{A {{THEORETICAL FRAMEWORK FOR COMBINING TECHNIQUES
  THAT PROBE THE LINK BETWEEN GALAXIES AND DARK MATTER}}}.
\newblock \bibinfo{journal}{The Astrophysical Journal} \bibinfo{volume}{738},
  \bibinfo{pages}{45--45}.
\newblock \DOIprefix\doi{10.1088/0004-637X/738/1/45}.
\bibitem[{Leauthaud et~al.(2012)Leauthaud, Tinker, Bundy, Behroozi, Massey,
  Rhodes, George, Kneib, Benson, Wechsler, Busha, Capak, Cort{\^e}s, Ilbert,
  Koekemoer, Le~F{\`e}vre, Lilly, McCracken, Salvato, Schrabback, Scoville,
  Smith and Taylor}]{Leauthaud2012}
\bibinfo{author}{Leauthaud, A.}, \bibinfo{author}{Tinker, J.},
  \bibinfo{author}{Bundy, K.}, \bibinfo{author}{Behroozi, P.S.},
  \bibinfo{author}{Massey, R.}, \bibinfo{author}{Rhodes, J.},
  \bibinfo{author}{George, M.R.}, \bibinfo{author}{Kneib, J.P.},
  \bibinfo{author}{Benson, A.J.}, \bibinfo{author}{Wechsler, R.H.},
  \bibinfo{author}{Busha, M.T.}, \bibinfo{author}{Capak, P.},
  \bibinfo{author}{Cort{\^e}s, M.}, \bibinfo{author}{Ilbert, O.},
  \bibinfo{author}{Koekemoer, A.M.}, \bibinfo{author}{Le~F{\`e}vre, O.},
  \bibinfo{author}{Lilly, S.}, \bibinfo{author}{McCracken, H.J.},
  \bibinfo{author}{Salvato, M.}, \bibinfo{author}{Schrabback, T.},
  \bibinfo{author}{Scoville, N.}, \bibinfo{author}{Smith, T.},
  \bibinfo{author}{Taylor, J.E.}, \bibinfo{year}{2012}.
\newblock \bibinfo{title}{{{NEW CONSTRAINTS ON THE EVOLUTION OF THE
  STELLAR}}-{{TO}}-{{DARK MATTER CONNECTION}}: {{A COMBINED ANALYSIS OF
  GALAXY}}-{{GALAXY LENSING}}, {{CLUSTERING}}, {{AND STELLAR MASS FUNCTIONS
  FROM}} z = 0.2 to z = 1}.
\newblock \bibinfo{journal}{The Astrophysical Journal} \bibinfo{volume}{744},
  \bibinfo{pages}{159--159}.
\newblock \DOIprefix\doi{10.1088/0004-637X/744/2/159}.
\bibitem[{Lesgourgues(2011)}]{Lesgourgues2011}
\bibinfo{author}{Lesgourgues, J.}, \bibinfo{year}{2011}.
\newblock \bibinfo{title}{The {{Cosmic Linear Anisotropy Solving System}}
  ({{CLASS}}) {{I}}: {{Overview}}}.
\newblock \bibinfo{journal}{arXiv e-prints} \bibinfo{volume}{1104},
  \bibinfo{pages}{arXiv:1104.2932}.
\bibitem[{Lewis et~al.(2000)Lewis, Challinor and Lasenby}]{Lewis2000}
\bibinfo{author}{Lewis, A.}, \bibinfo{author}{Challinor, A.},
  \bibinfo{author}{Lasenby, A.}, \bibinfo{year}{2000}.
\newblock \bibinfo{title}{Efficient {{Computation}} of {{Cosmic Microwave
  Background Anisotropies}} in {{Closed Friedmann}}-{{Robertson}}-{{Walker
  Models}}}.
\newblock \bibinfo{journal}{The Astrophysical Journal} \bibinfo{volume}{538},
  \bibinfo{pages}{473--476}.
\newblock \DOIprefix\doi{10.1086/309179}.
\bibitem[{Liddle and Lyth(2000)}]{Liddle2000}
\bibinfo{author}{Liddle, A.R.}, \bibinfo{author}{Lyth, D.H.},
  \bibinfo{year}{2000}.
\newblock \bibinfo{title}{Cosmological Inflation and Large-Scale Structure}.
\newblock \bibinfo{publisher}{{Cambridge University Press}},
  \bibinfo{address}{{Cambridge}}.
\bibitem[{Limber(1953)}]{Limber1953}
\bibinfo{author}{Limber, D.N.}, \bibinfo{year}{1953}.
\newblock \bibinfo{title}{The {{Analysis}} of {{Counts}} of the {{Extragalactic
  Nebulae}} in {{Terms}} of a {{Fluctuating Density Field}}.}
\newblock \bibinfo{journal}{The Astrophysical Journal} \bibinfo{volume}{117},
  \bibinfo{pages}{134}.
\newblock \DOIprefix\doi{10.1086/145672}.
\bibitem[{Ludlow et~al.(2016)Ludlow, Bose, Angulo, Wang, Hellwing, Navarro,
  Cole and Frenk}]{Ludlow2016}
\bibinfo{author}{Ludlow, A.D.}, \bibinfo{author}{Bose, S.},
  \bibinfo{author}{Angulo, R.E.}, \bibinfo{author}{Wang, L.},
  \bibinfo{author}{Hellwing, W.A.}, \bibinfo{author}{Navarro, J.F.},
  \bibinfo{author}{Cole, S.}, \bibinfo{author}{Frenk, C.S.},
  \bibinfo{year}{2016}.
\newblock \bibinfo{title}{The {{Mass}}-{{Concentration}}-{{Redshift Relation}}
  of {{Cold}} and {{Warm Dark Matter Halos}}}.
\newblock \bibinfo{journal}{Monthly Notices of the Royal Astronomical Society}
  \bibinfo{volume}{000}, \bibinfo{pages}{0--0}.
\newblock \DOIprefix\doi{10.1093/mnras/stw1046}.
\bibitem[{Ludlow et~al.(2014)Ludlow, Navarro, Angulo, {Boylan-kolchin},
  Springel, Frenk and White}]{Ludlow2014}
\bibinfo{author}{Ludlow, A.D.}, \bibinfo{author}{Navarro, J.F.},
  \bibinfo{author}{Angulo, R.E.}, \bibinfo{author}{{Boylan-kolchin}, M.},
  \bibinfo{author}{Springel, V.}, \bibinfo{author}{Frenk, C.},
  \bibinfo{author}{White, S.D.M.}, \bibinfo{year}{2014}.
\newblock \bibinfo{title}{The mass-concentration-redshift relation of cold dark
  matter haloes}.
\newblock \bibinfo{journal}{Monthly Notices of the Royal Astronomical Society}
  \bibinfo{volume}{441}, \bibinfo{pages}{378--388}.
\newblock \DOIprefix\doi{10.1093/mnras/stu483}.
\bibitem[{Ludlow et~al.(2013)Ludlow, Navarro, {Boylan-Kolchin}, Bett, Angulo,
  Li, White, Frenk and Springel}]{Ludlow2013}
\bibinfo{author}{Ludlow, A.D.}, \bibinfo{author}{Navarro, J.F.},
  \bibinfo{author}{{Boylan-Kolchin}, M.}, \bibinfo{author}{Bett, P.E.},
  \bibinfo{author}{Angulo, R.E.}, \bibinfo{author}{Li, M.},
  \bibinfo{author}{White, S.D.M.}, \bibinfo{author}{Frenk, C.},
  \bibinfo{author}{Springel, V.}, \bibinfo{year}{2013}.
\newblock \bibinfo{title}{The {{Mass Profile}} and {{Accretion History}} of
  {{Cold Dark Matter Halos}}}.
\newblock \bibinfo{journal}{Monthly Notices of the Royal Astronomical Society}
  \bibinfo{volume}{432}, \bibinfo{pages}{1103--1113}.
\newblock \DOIprefix\doi{10.1093/mnras/stt526}.
\bibitem[{Ma and Fry(2000)}]{Ma2000}
\bibinfo{author}{Ma, C.}, \bibinfo{author}{Fry, J.N.}, \bibinfo{year}{2000}.
\newblock \bibinfo{title}{Deriving the nonlinear cosmological power spectrum
  and bispectrum from analytic dark matter halo profiles and mass functions}.
\newblock \bibinfo{journal}{The Astrophysical Journal} \bibinfo{volume}{10},
  \bibinfo{pages}{503--513}.
\bibitem[{Ma et~al.(2011)Ma, Maggiore, Riotto and Zhang}]{Ma2011}
\bibinfo{author}{Ma, C.}, \bibinfo{author}{Maggiore, M.},
  \bibinfo{author}{Riotto, A.}, \bibinfo{author}{Zhang, J.},
  \bibinfo{year}{2011}.
\newblock \bibinfo{title}{The bias and mass function of dark matter haloes in
  non-{{Markovian}} extension of the excursion set theory}.
\newblock \bibinfo{journal}{Monthly Notices of the Royal Astronomical Society}
  \bibinfo{volume}{411}, \bibinfo{pages}{2644--2652}.
\newblock \DOIprefix\doi{10.1111/j.1365-2966.2010.17871.x}.
\bibitem[{Macci{\`o} et~al.(2008)Macci{\`o}, Dutton and {van den
  Bosch}}]{Maccio2008}
\bibinfo{author}{Macci{\`o}, A.V.}, \bibinfo{author}{Dutton, A.A.},
  \bibinfo{author}{{van den Bosch}, F.C.}, \bibinfo{year}{2008}.
\newblock \bibinfo{title}{Concentration, spin and shape of dark matter haloes
  as a function of the cosmological model: {{WMAP}} 1, {{WMAP}} 3 and {{WMAP}}
  5 results}.
\newblock \bibinfo{journal}{Monthly Notices of the Royal Astronomical Society}
  \bibinfo{volume}{391}, \bibinfo{pages}{1940--1954}.
\newblock \DOIprefix\doi{10.1111/j.1365-2966.2008.14029.x}.
\bibitem[{Mandelbaum et~al.(2006)Mandelbaum, Seljak, Kauffmann, Hirata and
  Brinkmann}]{Mandelbaum2006}
\bibinfo{author}{Mandelbaum, R.}, \bibinfo{author}{Seljak, U.},
  \bibinfo{author}{Kauffmann, G.}, \bibinfo{author}{Hirata, C.M.},
  \bibinfo{author}{Brinkmann, J.}, \bibinfo{year}{2006}.
\newblock \bibinfo{title}{Galaxy halo masses and satellite fractions from
  galaxy-galaxy lensing in the {{Sloan Digital Sky Survey}}: {{Stellar}} mass,
  luminosity, morphology and environment dependencies}.
\newblock \bibinfo{journal}{Monthly Notices of the Royal Astronomical Society}
  \bibinfo{volume}{368}, \bibinfo{pages}{715--731}.
\newblock \DOIprefix\doi{10.1111/j.1365-2966.2006.10156.x}.
\bibitem[{Mandelbaum et~al.(2005)Mandelbaum, Tasitsiomi, Seljak, Kravtsov and
  Wechsler}]{Mandelbaum2005}
\bibinfo{author}{Mandelbaum, R.}, \bibinfo{author}{Tasitsiomi, A.},
  \bibinfo{author}{Seljak, U.}, \bibinfo{author}{Kravtsov, A.V.},
  \bibinfo{author}{Wechsler, R.H.}, \bibinfo{year}{2005}.
\newblock \bibinfo{title}{Galaxy-galaxy lensing: {{Dissipationless}}
  simulations versus the halo model}.
\newblock \bibinfo{journal}{Monthly Notices of the Royal Astronomical Society}
  \bibinfo{volume}{362}, \bibinfo{pages}{1451--1462}.
\newblock \DOIprefix\doi{10.1111/j.1365-2966.2005.09417.x}.
\bibitem[{Manera et~al.(2010)Manera, Sheth and Scoccimarro}]{Manera2010}
\bibinfo{author}{Manera, M.}, \bibinfo{author}{Sheth, R.K.},
  \bibinfo{author}{Scoccimarro, R.}, \bibinfo{year}{2010}.
\newblock \bibinfo{title}{Large-scale bias and the inaccuracy of the
  peak-background split}.
\newblock \bibinfo{journal}{Monthly Notices of the Royal Astronomical Society}
  \bibinfo{volume}{402}, \bibinfo{pages}{589--602}.
\newblock \DOIprefix\doi{10.1111/j.1365-2966.2009.15921.x}.
\bibitem[{Mead and Verde(2021)}]{Mead2021}
\bibinfo{author}{Mead, A.J.}, \bibinfo{author}{Verde, L.},
  \bibinfo{year}{2021}.
\newblock \bibinfo{title}{Including beyond-linear halo bias in halo models}.
\newblock \bibinfo{journal}{Monthly Notices of the Royal Astronomical Society}
  \bibinfo{volume}{503}, \bibinfo{pages}{3095--3111}.
\newblock \DOIprefix\doi{10.1093/mnras/stab748}.
\bibitem[{Miyatake et~al.(2020)Miyatake, Kobayashi, Takada, Nishimichi,
  Shirasaki, Sugiyama, Takahashi, Osato, More and Park}]{Miyatake2020}
\bibinfo{author}{Miyatake, H.}, \bibinfo{author}{Kobayashi, Y.},
  \bibinfo{author}{Takada, M.}, \bibinfo{author}{Nishimichi, T.},
  \bibinfo{author}{Shirasaki, M.}, \bibinfo{author}{Sugiyama, S.},
  \bibinfo{author}{Takahashi, R.}, \bibinfo{author}{Osato, K.},
  \bibinfo{author}{More, S.}, \bibinfo{author}{Park, Y.}, \bibinfo{year}{2020}.
\newblock \bibinfo{title}{Cosmological inference from emulator based halo model
  {{I}}: {{Validation}} tests with {{HSC}} and {{SDSS}} mock catalogs}.
\newblock \bibinfo{journal}{arXiv e-prints} \bibinfo{volume}{2101},
  \bibinfo{pages}{arXiv:2101.00113}.
\bibitem[{Mo and White(1996)}]{Mo1996}
\bibinfo{author}{Mo, H.J.}, \bibinfo{author}{White, S.D.M.},
  \bibinfo{year}{1996}.
\newblock \bibinfo{title}{An analytic model for the spatial clustering of dark
  matter haloes}.
\newblock \bibinfo{journal}{Monthly Notices of the Royal Astronomical Society}
  \bibinfo{volume}{282}, \bibinfo{pages}{347--361}.
\newblock \DOIprefix\doi{10.1093/mnras/282.2.347}.
\bibitem[{Moore et~al.(1998)Moore, Governato, Quinn, Stadel and
  Lake}]{Moore1998}
\bibinfo{author}{Moore, B.}, \bibinfo{author}{Governato, F.},
  \bibinfo{author}{Quinn, T.}, \bibinfo{author}{Stadel, J.},
  \bibinfo{author}{Lake, G.}, \bibinfo{year}{1998}.
\newblock \bibinfo{title}{Resolving the {{Structure}} of {{Cold Dark Matter
  Halos}}}.
\newblock \bibinfo{journal}{The Astrophysical Journal} \bibinfo{volume}{499},
  \bibinfo{pages}{L5--L8}.
\newblock \DOIprefix\doi{10.1086/311333}.
\bibitem[{More(2013)}]{More2013}
\bibinfo{author}{More, S.}, \bibinfo{year}{2013}.
\newblock \bibinfo{title}{Cosmological {{Dependence}} of the {{Measurements}}
  of {{Luminosity Function}}, {{Projected Clustering}} and {{Galaxy}}-{{Galaxy
  Lensing Signal}}}.
\newblock \bibinfo{journal}{The Astrophysical Journal} \bibinfo{volume}{777},
  \bibinfo{pages}{L26--L26}.
\newblock \DOIprefix\doi{10.1088/2041-8205/777/2/L26}.
\bibitem[{More et~al.(2011)More, Kravtsov, Dalal and Gottl{\"o}ber}]{More2011}
\bibinfo{author}{More, S.}, \bibinfo{author}{Kravtsov, A.V.},
  \bibinfo{author}{Dalal, N.}, \bibinfo{author}{Gottl{\"o}ber, S.},
  \bibinfo{year}{2011}.
\newblock \bibinfo{title}{The {{Overdensity}} and {{Masses}} of the
  {{Friends}}-of-friends {{Halos}} and {{Universality}} of {{Halo Mass
  Function}}}.
\newblock \bibinfo{journal}{The Astrophysical Journal Supplement Series}
  \bibinfo{volume}{195}, \bibinfo{pages}{4}.
\newblock \DOIprefix\doi{10.1088/0067-0049/195/1/4}.
\bibitem[{More et~al.(2015)More, Miyatake, Mandelbaum, Takada, Spergel,
  Brownstein and Schneider}]{More2015}
\bibinfo{author}{More, S.}, \bibinfo{author}{Miyatake, H.},
  \bibinfo{author}{Mandelbaum, R.}, \bibinfo{author}{Takada, M.},
  \bibinfo{author}{Spergel, D.N.}, \bibinfo{author}{Brownstein, J.R.},
  \bibinfo{author}{Schneider, D.P.}, \bibinfo{year}{2015}.
\newblock \bibinfo{title}{The {{Weak Lensing Signal}} and the {{Clustering}} of
  {{Boss Galaxies}}. {{Ii}}. {{Astrophysical}} and {{Cosmological
  Constraints}}}.
\newblock \bibinfo{journal}{The Astrophysical Journal} \bibinfo{volume}{806},
  \bibinfo{pages}{2--2}.
\newblock \DOIprefix\doi{10.1088/0004-637X/806/1/2}.
\bibitem[{Moustakas and Somerville(2002)}]{Moustakas2002}
\bibinfo{author}{Moustakas, L.A.}, \bibinfo{author}{Somerville, R.S.},
  \bibinfo{year}{2002}.
\newblock \bibinfo{title}{The {{Masses}}, {{Ancestors}} and {{Descendents}} of
  {{Extremely Red Objects}}: {{Constraints}} from {{Spatial Clustering}}}.
\newblock \bibinfo{journal}{The Astrophysical Journal} \bibinfo{volume}{577},
  \bibinfo{pages}{1--10}.
\newblock \DOIprefix\doi{10.1086/342133}.
\bibitem[{Murray and Poulin(2019)}]{Murray2019}
\bibinfo{author}{Murray, S.}, \bibinfo{author}{Poulin, F.},
  \bibinfo{year}{2019}.
\newblock \bibinfo{title}{Hankel: {{A Python}} library for performing simple
  and accurate {{Hankel}} transformations}.
\newblock \bibinfo{journal}{The Journal of Open Source Software}
  \bibinfo{volume}{4}, \bibinfo{pages}{1397}.
\newblock \DOIprefix\doi{10.21105/joss.01397}.
\bibitem[{Murray et~al.(2013a)Murray, {Power C.} and Robotham}]{Murray2013}
\bibinfo{author}{Murray, S.}, \bibinfo{author}{{Power C.}},
  \bibinfo{author}{Robotham, A.}, \bibinfo{year}{2013}a.
\newblock \bibinfo{title}{How well do we know the halo mass function?}
\newblock \bibinfo{journal}{Monthly Notices of the Royal Astronomical Society:
  Letters} \bibinfo{volume}{434}, \bibinfo{pages}{L61--L65}.
\newblock \DOIprefix\doi{10.1093/mnrasl/slt079}.
\bibitem[{Murray et~al.(2013b)Murray, Power and Robotham}]{Murray2013a}
\bibinfo{author}{Murray, S.G.}, \bibinfo{author}{Power, C.},
  \bibinfo{author}{Robotham, A.S.G.}, \bibinfo{year}{2013}b.
\newblock \bibinfo{title}{{{HMFcalc}}: {{An}} online tool for calculating dark
  matter halo mass functions}.
\newblock \bibinfo{journal}{Astronomy and Computing} \bibinfo{volume}{3},
  \bibinfo{pages}{23--34}.
\newblock \DOIprefix\doi{10.1016/j.ascom.2013.11.001}.
\bibitem[{Navarro et~al.(1997)Navarro, Frenk and White}]{Navarro1997}
\bibinfo{author}{Navarro, J.F.}, \bibinfo{author}{Frenk, C.S.},
  \bibinfo{author}{White, S.D.M.}, \bibinfo{year}{1997}.
\newblock \bibinfo{title}{A {{Universal Density Profile}} from {{Hierarchical
  Clustering}}}.
\newblock \bibinfo{journal}{The Astrophysical Journal} \bibinfo{volume}{490},
  \bibinfo{pages}{493--508}.
\newblock \DOIprefix\doi{10.1086/304888}.
\bibitem[{Neyman et~al.(1953)Neyman, Scott and Shane}]{Neyman1953}
\bibinfo{author}{Neyman, J.}, \bibinfo{author}{Scott, E.L.},
  \bibinfo{author}{Shane, C.D.}, \bibinfo{year}{1953}.
\newblock \bibinfo{title}{On the {{Spatial Distribution}} of {{Galaxies}}: A
  {{Specific Model}}.}
\newblock \bibinfo{journal}{The Astrophysical Journal} \bibinfo{volume}{117},
  \bibinfo{pages}{92--92}.
\newblock \DOIprefix\doi{10.1086/145671}.
\bibitem[{Nicola et~al.(2020)Nicola, Alonso, S{\'a}nchez, Slosar, Awan,
  Broussard, Dunkley, Gawiser, Gomes, Mandelbaum, Miyatake, Newman,
  {Sevilla-Noarbe}, Skinner and Wagoner}]{Nicola2020}
\bibinfo{author}{Nicola, A.}, \bibinfo{author}{Alonso, D.},
  \bibinfo{author}{S{\'a}nchez, J.}, \bibinfo{author}{Slosar, A.},
  \bibinfo{author}{Awan, H.}, \bibinfo{author}{Broussard, A.},
  \bibinfo{author}{Dunkley, J.}, \bibinfo{author}{Gawiser, E.},
  \bibinfo{author}{Gomes, Z.}, \bibinfo{author}{Mandelbaum, R.},
  \bibinfo{author}{Miyatake, H.}, \bibinfo{author}{Newman, J.A.},
  \bibinfo{author}{{Sevilla-Noarbe}, I.}, \bibinfo{author}{Skinner, S.},
  \bibinfo{author}{Wagoner, E.L.}, \bibinfo{year}{2020}.
\newblock \bibinfo{title}{Tomographic galaxy clustering with the {{Subaru Hyper
  Suprime}}-{{Cam}} first year public data release}.
\newblock \bibinfo{journal}{Journal of Cosmology and Astroparticle Physics}
  \bibinfo{volume}{03}, \bibinfo{pages}{044}.
\newblock \DOIprefix\doi{10.1088/1475-7516/2020/03/044}.
\bibitem[{Nishimichi et~al.(2021)Nishimichi, Takada, Takahashi, Osato,
  Shirasaki, Oogi, Miyatake, Oguri, Murata, Kobayashi and
  Yoshida}]{Nishimichi2021}
\bibinfo{author}{Nishimichi, T.}, \bibinfo{author}{Takada, M.},
  \bibinfo{author}{Takahashi, R.}, \bibinfo{author}{Osato, K.},
  \bibinfo{author}{Shirasaki, M.}, \bibinfo{author}{Oogi, T.},
  \bibinfo{author}{Miyatake, H.}, \bibinfo{author}{Oguri, M.},
  \bibinfo{author}{Murata, R.}, \bibinfo{author}{Kobayashi, Y.},
  \bibinfo{author}{Yoshida, N.}, \bibinfo{year}{2021}.
\newblock \bibinfo{title}{{{DarkEmulator}}: {{Cosmological}} emulation code for
  halo clustering statistics}.
\newblock \bibinfo{journal}{Astrophysics Source Code Library, record
  ascl:2103.009} , \bibinfo{pages}{ascl:2103.009}.
\bibitem[{Nusser and Tiwari(2015)}]{Nusser2015}
\bibinfo{author}{Nusser, A.}, \bibinfo{author}{Tiwari, P.},
  \bibinfo{year}{2015}.
\newblock \bibinfo{title}{{{THE CLUSTERING OF RADIO GALAXIES}}: {{BIASING AND
  EVOLUTION VERSUS STELLAR MASS}}}.
\newblock \bibinfo{journal}{The Astrophysical Journal} \bibinfo{volume}{812},
  \bibinfo{pages}{85--85}.
\newblock \DOIprefix\doi{10.1088/0004-637X/812/1/85}.
\bibitem[{Ogata(2005)}]{Ogata2005}
\bibinfo{author}{Ogata, H.}, \bibinfo{year}{2005}.
\newblock \bibinfo{title}{A {{Numerical Integration Formula Based}} on the
  {{Bessel Functions}}}.
\newblock \bibinfo{journal}{Publ. RIMS, Kyoto Univ.} \bibinfo{volume}{41},
  \bibinfo{pages}{949--970}.
\bibitem[{Padmanabhan et~al.(2016)Padmanabhan, Refregier and
  Amara}]{Padmanabhan2016b}
\bibinfo{author}{Padmanabhan, H.}, \bibinfo{author}{Refregier, A.},
  \bibinfo{author}{Amara, A.}, \bibinfo{year}{2016}.
\newblock \bibinfo{title}{A halo model for cosmological neutral hydrogen :
  Abundances and clustering}.
\newblock \bibinfo{journal}{eprint arXiv:1611.06235}
  \href{http://arxiv.org/abs/1611.06235}{\tt arXiv:1611.06235}.
\bibitem[{Paranjape et~al.(2013a)Paranjape, Sefusatti, Chan, Desjacques, Monaco
  and Sheth}]{Paranjape2013}
\bibinfo{author}{Paranjape, A.}, \bibinfo{author}{Sefusatti, E.},
  \bibinfo{author}{Chan, K.C.}, \bibinfo{author}{Desjacques, V.},
  \bibinfo{author}{Monaco, P.}, \bibinfo{author}{Sheth, R.K.},
  \bibinfo{year}{2013}a.
\newblock \bibinfo{title}{Bias deconstructed: Unravelling the scale dependence
  of halo bias using real-space measurements}.
\newblock \bibinfo{journal}{Monthly Notices of the Royal Astronomical Society}
  \bibinfo{volume}{436}, \bibinfo{pages}{449--459}.
\newblock \DOIprefix\doi{10.1093/mnras/stt1578}.
\bibitem[{Paranjape et~al.(2013b)Paranjape, Sheth and
  Desjacques}]{Paranjape2013a}
\bibinfo{author}{Paranjape, A.}, \bibinfo{author}{Sheth, R.K.},
  \bibinfo{author}{Desjacques, V.}, \bibinfo{year}{2013}b.
\newblock \bibinfo{title}{Excursion set peaks: A self-consistent model of dark
  halo abundances and clustering}.
\newblock \bibinfo{journal}{Monthly Notices of the Royal Astronomical Society}
  \bibinfo{volume}{431}, \bibinfo{pages}{1503--1512}.
\newblock \DOIprefix\doi{10.1093/mnras/stt267}.
\bibitem[{Peacock(2007)}]{Peacock2007}
\bibinfo{author}{Peacock, J.A.}, \bibinfo{year}{2007}.
\newblock \bibinfo{title}{Testing anthropic predictions for ~ and the cosmic
  microwave background temperature}.
\newblock \bibinfo{journal}{Monthly Notices of the Royal Astronomical Society}
  \bibinfo{volume}{379}, \bibinfo{pages}{1067--1074}.
\newblock \DOIprefix\doi{10.1111/j.1365-2966.2007.11978.x}.
\bibitem[{Peacock and Smith(2000)}]{Peacock2000}
\bibinfo{author}{Peacock, J.A.}, \bibinfo{author}{Smith, R.E.},
  \bibinfo{year}{2000}.
\newblock \bibinfo{title}{Halo occupation numbers and galaxy bias}.
\newblock \bibinfo{journal}{Monthly Notices of the Royal Astronomical Society}
  \bibinfo{volume}{318}, \bibinfo{pages}{1144--1156}.
\newblock \DOIprefix\doi{10.1046/j.1365-8711.2000.03779.x}.
\bibitem[{Pillepich et~al.(2010)Pillepich, Porciani and Hahn}]{Pillepich2010}
\bibinfo{author}{Pillepich, A.}, \bibinfo{author}{Porciani, C.},
  \bibinfo{author}{Hahn, O.}, \bibinfo{year}{2010}.
\newblock \bibinfo{title}{Halo mass function and scale-dependent bias from
  {{N}}-body simulations with non-{{Gaussian}} initial conditions}.
\newblock \bibinfo{journal}{Monthly Notices of the Royal Astronomical Society}
  \bibinfo{volume}{402}, \bibinfo{pages}{191--206}.
\newblock \DOIprefix\doi{10.1111/j.1365-2966.2009.15914.x}.
\bibitem[{{Planck Collaboration}(2015)}]{PlanckCollaboration2015}
\bibinfo{author}{{Planck Collaboration}}, \bibinfo{year}{2015}.
\newblock \bibinfo{title}{Planck 2015 reslts. {{XIII}}. {{Cosmological}}
  parameters}.
\newblock \bibinfo{journal}{Astronomy \& Astrophysics} ,
  \bibinfo{pages}{20--20}.
\bibitem[{Poole et~al.(2015)Poole, Blake, Marin, Power, Mutch, Croton, Colless,
  Couch, Drinkwater and Glazebrook}]{Poole2015}
\bibinfo{author}{Poole, G.B.}, \bibinfo{author}{Blake, C.},
  \bibinfo{author}{Marin, F.A.}, \bibinfo{author}{Power, C.},
  \bibinfo{author}{Mutch, S.J.}, \bibinfo{author}{Croton, D.J.},
  \bibinfo{author}{Colless, M.}, \bibinfo{author}{Couch, W.},
  \bibinfo{author}{Drinkwater, M.J.}, \bibinfo{author}{Glazebrook, K.},
  \bibinfo{year}{2015}.
\newblock \bibinfo{title}{The {{Gigaparsec WiggleZ}} simulations:
  Characterizing scale-dependant bias and associated systematics in growth of
  structure measurements}.
\newblock \bibinfo{journal}{Monthly Notices of the Royal Astronomical Society}
  \bibinfo{volume}{449}, \bibinfo{pages}{1454--1469}.
\newblock \DOIprefix\doi{10.1093/mnras/stv314}.
\bibitem[{Prada et~al.(2012)Prada, Klypin, Cuesta, {Betancort-Rijo} and
  Primack}]{Prada2012}
\bibinfo{author}{Prada, F.}, \bibinfo{author}{Klypin, A.A.},
  \bibinfo{author}{Cuesta, A.J.}, \bibinfo{author}{{Betancort-Rijo}, J.E.},
  \bibinfo{author}{Primack, J.}, \bibinfo{year}{2012}.
\newblock \bibinfo{title}{Halo concentrations in the standard {{$\Lambda$}}
  cold dark matter cosmology}.
\newblock \bibinfo{journal}{Monthly Notices of the Royal Astronomical Society}
  \bibinfo{volume}{423}, \bibinfo{pages}{3018--3030}.
\newblock \DOIprefix\doi{10.1111/j.1365-2966.2012.21007.x}.
\bibitem[{Press and Schechter(1974)}]{Press1974}
\bibinfo{author}{Press, W.H.}, \bibinfo{author}{Schechter, P.},
  \bibinfo{year}{1974}.
\newblock \bibinfo{title}{Formation of galaxies and clusters of galaxies by
  self-similar gravitational condensation}.
\newblock \bibinfo{journal}{The Astrophysical Journal} \bibinfo{volume}{187},
  \bibinfo{pages}{425--438}.
\bibitem[{{Rafiei-Ravandi} et~al.(2020){Rafiei-Ravandi}, Smith and
  Masui}]{Rafiei-Ravandi2020}
\bibinfo{author}{{Rafiei-Ravandi}, M.}, \bibinfo{author}{Smith, K.M.},
  \bibinfo{author}{Masui, K.W.}, \bibinfo{year}{2020}.
\newblock \bibinfo{title}{Characterizing fast radio bursts through statistical
  cross-correlations}.
\newblock \bibinfo{journal}{Physical Review D} \bibinfo{volume}{102},
  \bibinfo{pages}{023528}.
\newblock \DOIprefix\doi{10.1103/PhysRevD.102.023528}.
\bibitem[{Reddick et~al.(2014)Reddick, Tinker, Wechsler and Lu}]{Reddick2014}
\bibinfo{author}{Reddick, R.M.}, \bibinfo{author}{Tinker, J.L.},
  \bibinfo{author}{Wechsler, R.H.}, \bibinfo{author}{Lu, Y.},
  \bibinfo{year}{2014}.
\newblock \bibinfo{title}{{{COSMOLOGICAL CONSTRAINTS FROM GALAXY CLUSTERING AND
  THE MASS}}-{{TO}}-{{NUMBER RATIO OF GALAXY CLUSTERS}}: {{MARGINALIZING OVER
  THE PHYSICS OF GALAXY FORMATION}}}.
\newblock \bibinfo{journal}{The Astrophysical Journal} \bibinfo{volume}{783},
  \bibinfo{pages}{118--118}.
\newblock \DOIprefix\doi{10.1088/0004-637X/783/2/118}.
\bibitem[{Robitaille et~al.(2013)Robitaille, Tollerud, Greenfield, Droettboom,
  Bray, Aldcroft, Davis, Ginsburg, {Price-Whelan}, Kerzendorf, Conley,
  Crighton, Barbary, Muna, Ferguson, Grollier, Parikh, Nair, G{\"u}nther, Deil,
  Woillez, Conseil, Kramer, Turner, Singer, Fox, Weaver, Zabalza, Edwards,
  Azalee~Bostroem, Burke, Casey, Crawford, Dencheva, Ely, Jenness, Labrie, Lim,
  Pierfederici, Pontzen, Ptak, Refsdal, Servillat and
  Streicher}]{Robitaille2013}
\bibinfo{author}{Robitaille, T.P.}, \bibinfo{author}{Tollerud, E.J.},
  \bibinfo{author}{Greenfield, P.}, \bibinfo{author}{Droettboom, M.},
  \bibinfo{author}{Bray, E.}, \bibinfo{author}{Aldcroft, T.},
  \bibinfo{author}{Davis, M.}, \bibinfo{author}{Ginsburg, A.},
  \bibinfo{author}{{Price-Whelan}, A.M.}, \bibinfo{author}{Kerzendorf, W.E.},
  \bibinfo{author}{Conley, A.}, \bibinfo{author}{Crighton, N.},
  \bibinfo{author}{Barbary, K.}, \bibinfo{author}{Muna, D.},
  \bibinfo{author}{Ferguson, H.}, \bibinfo{author}{Grollier, F.},
  \bibinfo{author}{Parikh, M.M.}, \bibinfo{author}{Nair, P.H.},
  \bibinfo{author}{G{\"u}nther, H.M.}, \bibinfo{author}{Deil, C.},
  \bibinfo{author}{Woillez, J.}, \bibinfo{author}{Conseil, S.},
  \bibinfo{author}{Kramer, R.}, \bibinfo{author}{Turner, J.E.H.},
  \bibinfo{author}{Singer, L.}, \bibinfo{author}{Fox, R.},
  \bibinfo{author}{Weaver, B.A.}, \bibinfo{author}{Zabalza, V.},
  \bibinfo{author}{Edwards, Z.I.}, \bibinfo{author}{Azalee~Bostroem, K.},
  \bibinfo{author}{Burke, D.J.}, \bibinfo{author}{Casey, A.R.},
  \bibinfo{author}{Crawford, S.M.}, \bibinfo{author}{Dencheva, N.},
  \bibinfo{author}{Ely, J.}, \bibinfo{author}{Jenness, T.},
  \bibinfo{author}{Labrie, K.}, \bibinfo{author}{Lim, P.L.},
  \bibinfo{author}{Pierfederici, F.}, \bibinfo{author}{Pontzen, A.},
  \bibinfo{author}{Ptak, A.}, \bibinfo{author}{Refsdal, B.},
  \bibinfo{author}{Servillat, M.}, \bibinfo{author}{Streicher, O.},
  \bibinfo{year}{2013}.
\newblock \bibinfo{title}{Astropy: {{A}} community {{Python}} package for
  astronomy}.
\newblock \bibinfo{journal}{Astronomy \& Astrophysics} \bibinfo{volume}{558},
  \bibinfo{pages}{A33--A33}.
\newblock \DOIprefix\doi{10.1051/0004-6361/201322068}.
\bibitem[{{Rodr{\'i}guez-Torres} et~al.(2015){Rodr{\'i}guez-Torres}, Prada,
  Chuang, Guo, Klypin, Behroozi, Hahn, Comparat, Yepes, {Montero-Dorta},
  Brownstein, Maraston, McBride, Tinker, Gottl{\"o}ber, Favole, Shu, Kitaura,
  Bolton, Scoccimarro, Samushia, Schlegel, Schneider and
  Thomas}]{Rodriguez-Torres2015}
\bibinfo{author}{{Rodr{\'i}guez-Torres}, S.A.}, \bibinfo{author}{Prada, F.},
  \bibinfo{author}{Chuang, C.H.}, \bibinfo{author}{Guo, H.},
  \bibinfo{author}{Klypin, A.}, \bibinfo{author}{Behroozi, P.},
  \bibinfo{author}{Hahn, C.H.}, \bibinfo{author}{Comparat, J.},
  \bibinfo{author}{Yepes, G.}, \bibinfo{author}{{Montero-Dorta}, A.D.},
  \bibinfo{author}{Brownstein, J.R.}, \bibinfo{author}{Maraston, C.},
  \bibinfo{author}{McBride, C.K.}, \bibinfo{author}{Tinker, J.},
  \bibinfo{author}{Gottl{\"o}ber, S.}, \bibinfo{author}{Favole, G.},
  \bibinfo{author}{Shu, Y.}, \bibinfo{author}{Kitaura, F.S.},
  \bibinfo{author}{Bolton, A.}, \bibinfo{author}{Scoccimarro, R.},
  \bibinfo{author}{Samushia, L.}, \bibinfo{author}{Schlegel, D.},
  \bibinfo{author}{Schneider, D.P.}, \bibinfo{author}{Thomas, D.},
  \bibinfo{year}{2015}.
\newblock \bibinfo{title}{The clustering of galaxies in the {{SDSS}}-{{III
  Baryon Oscillation Spectroscopic Survey}}: {{Modeling}} the clustering and
  halo occupation distribution of {{BOSS}}-{{CMASS}} galaxies in the {{Final
  Data Release}}}.
\newblock \bibinfo{journal}{eprint arXiv:1509.06404}
  \href{http://arxiv.org/abs/1509.06404}{\tt arXiv:1509.06404}.
\bibitem[{Schaan and White(2021)}]{Schaan2021}
\bibinfo{author}{Schaan, E.}, \bibinfo{author}{White, M.},
  \bibinfo{year}{2021}.
\newblock \bibinfo{title}{Multi-tracer intensity mapping:
  {{Cross}}-correlations, {{Line}} noise \& {{Decorrelation}}}.
\newblock \bibinfo{journal}{arXiv e-prints} \bibinfo{volume}{2103},
  \bibinfo{pages}{arXiv:2103.01964}.
\bibitem[{Scherrer and Bertschinger(1991)}]{Scherrer1991}
\bibinfo{author}{Scherrer, R.J.}, \bibinfo{author}{Bertschinger, E.},
  \bibinfo{year}{1991}.
\newblock \bibinfo{title}{Statistics of primordial density perturbations from
  discrete seed masses}.
\newblock \bibinfo{journal}{The Astrophysical Journal} \bibinfo{volume}{381},
  \bibinfo{pages}{349--349}.
\newblock \DOIprefix\doi{10.1086/170658}.
\bibitem[{Schneider(2015)}]{Schneider2015}
\bibinfo{author}{Schneider, A.}, \bibinfo{year}{2015}.
\newblock \bibinfo{title}{Structure formation with suppressed small-scale
  perturbations}.
\newblock \bibinfo{journal}{Monthly Notices of the Royal Astronomical Society}
  \bibinfo{volume}{451}, \bibinfo{pages}{3117--3130}.
\newblock \DOIprefix\doi{10.1093/mnras/stv1169}.
\bibitem[{Schneider et~al.(2020)Schneider, Giri and Mirocha}]{Schneider2020}
\bibinfo{author}{Schneider, A.}, \bibinfo{author}{Giri, S.K.},
  \bibinfo{author}{Mirocha, J.}, \bibinfo{year}{2020}.
\newblock \bibinfo{title}{A halo model approach for the 21-cm power spectrum at
  cosmic dawn}.
\newblock \bibinfo{journal}{arXiv e-prints} \bibinfo{volume}{2011},
  \bibinfo{pages}{arXiv:2011.12308}.
\bibitem[{Schneider et~al.(2012)Schneider, Smith, Macci{\`o} and
  Moore}]{Schneider2012}
\bibinfo{author}{Schneider, A.}, \bibinfo{author}{Smith, R.E.},
  \bibinfo{author}{Macci{\`o}, A.V.}, \bibinfo{author}{Moore, B.},
  \bibinfo{year}{2012}.
\newblock \bibinfo{title}{Non-linear evolution of cosmological structures in
  warm dark matter models}.
\newblock \bibinfo{journal}{Monthly Notices of the Royal Astronomical Society}
  \bibinfo{volume}{424}, \bibinfo{pages}{684--698}.
\newblock \DOIprefix\doi{10.1111/j.1365-2966.2012.21252.x}.
\bibitem[{Schneider et~al.(2013)Schneider, Smith and Reed}]{Schneider2013}
\bibinfo{author}{Schneider, A.}, \bibinfo{author}{Smith, R.E.},
  \bibinfo{author}{Reed, D.S.}, \bibinfo{year}{2013}.
\newblock \bibinfo{title}{Halo {{Mass Function}} and the {{Free Streaming
  Scale}}}.
\newblock \bibinfo{journal}{Monthly Notices of the Royal Astronomical Society}
  \bibinfo{volume}{433}, \bibinfo{pages}{16--16}.
\newblock \DOIprefix\doi{10.1093/mnras/stt829}.
\bibitem[{Scoccimarro et~al.(2001)Scoccimarro, Sheth, Hui and
  Jain}]{Scoccimarro2001}
\bibinfo{author}{Scoccimarro, R.}, \bibinfo{author}{Sheth, R.K.},
  \bibinfo{author}{Hui, L.}, \bibinfo{author}{Jain, B.}, \bibinfo{year}{2001}.
\newblock \bibinfo{title}{How many galaxies fit in a halo? {{Constraints}} on
  galaxy formation efficiency from spatial clustering}.
\newblock \bibinfo{journal}{The Astrophysical Journal} \bibinfo{volume}{546},
  \bibinfo{pages}{20--34}.
\bibitem[{Seljak(2000)}]{Seljak2000}
\bibinfo{author}{Seljak, U.}, \bibinfo{year}{2000}.
\newblock \bibinfo{title}{Analytic model for galaxy and dark matter
  clustering}.
\newblock \bibinfo{journal}{Monthly Notices of the Royal Astronomical Society}
  \bibinfo{volume}{213}.
\bibitem[{Seljak and Vlah(2015)}]{Seljak2015}
\bibinfo{author}{Seljak, U.}, \bibinfo{author}{Vlah, Z.}, \bibinfo{year}{2015}.
\newblock \bibinfo{title}{Halo {{Zel}}'dovich model and perturbation theory:
  {{Dark}} matter power spectrum and correlation function}.
\newblock \bibinfo{journal}{Physical Review D} \bibinfo{volume}{91},
  \bibinfo{pages}{123516}.
\newblock \DOIprefix\doi{10.1103/PhysRevD.91.123516}.
\bibitem[{Seljak and Warren(2004)}]{Seljak2004}
\bibinfo{author}{Seljak, U.}, \bibinfo{author}{Warren, M.S.},
  \bibinfo{year}{2004}.
\newblock \bibinfo{title}{Large-scale bias and stochasticity of haloes and dark
  matter}.
\newblock \bibinfo{journal}{Monthly Notices of the Royal Astronomical Society}
  \bibinfo{volume}{355}, \bibinfo{pages}{129--136}.
\newblock \DOIprefix\doi{10.1111/j.1365-2966.2004.08297.x}.
\bibitem[{Shen et~al.(2013)Shen, McBride, White, Zheng, Myers, Guo,
  Kirkpatrick, Padmanabhan, Parejko, Ross, Schlegel, Schneider, Streblyanska,
  Swanson, Zehavi, Pan, Bizyaev, Brewington, Ebelke, Malanushenko,
  Malanushenko, Oravetz, Simmons and Snedden}]{Shen2013}
\bibinfo{author}{Shen, Y.}, \bibinfo{author}{McBride, C.K.},
  \bibinfo{author}{White, M.}, \bibinfo{author}{Zheng, Z.},
  \bibinfo{author}{Myers, A.D.}, \bibinfo{author}{Guo, H.},
  \bibinfo{author}{Kirkpatrick, J.A.}, \bibinfo{author}{Padmanabhan, N.},
  \bibinfo{author}{Parejko, J.K.}, \bibinfo{author}{Ross, N.P.},
  \bibinfo{author}{Schlegel, D.J.}, \bibinfo{author}{Schneider, D.P.},
  \bibinfo{author}{Streblyanska, A.}, \bibinfo{author}{Swanson, M.E.C.},
  \bibinfo{author}{Zehavi, I.}, \bibinfo{author}{Pan, K.},
  \bibinfo{author}{Bizyaev, D.}, \bibinfo{author}{Brewington, H.},
  \bibinfo{author}{Ebelke, G.}, \bibinfo{author}{Malanushenko, V.},
  \bibinfo{author}{Malanushenko, E.}, \bibinfo{author}{Oravetz, D.},
  \bibinfo{author}{Simmons, A.}, \bibinfo{author}{Snedden, S.},
  \bibinfo{year}{2013}.
\newblock \bibinfo{title}{{{CROSS}}-{{CORRELATION OF SDSS DR7 QUASARS AND DR10
  BOSS GALAXIES}}: {{THE WEAK LUMINOSITY DEPENDENCE OF QUASAR CLUSTERING AT}} z
  {$\sim$} 0.5}.
\newblock \bibinfo{journal}{The Astrophysical Journal} \bibinfo{volume}{778},
  \bibinfo{pages}{98--98}.
\newblock \DOIprefix\doi{10.1088/0004-637X/778/2/98}.
\bibitem[{Sheth et~al.(2001a)Sheth, Diaferio, Hui and Scoccimarro}]{Sheth2001}
\bibinfo{author}{Sheth, R.K.}, \bibinfo{author}{Diaferio, A.},
  \bibinfo{author}{Hui, L.}, \bibinfo{author}{Scoccimarro, R.},
  \bibinfo{year}{2001}a.
\newblock \bibinfo{title}{On the streaming motions of haloes and galaxies}.
\newblock \bibinfo{journal}{Monthly Notices of the Royal Astronomical Society}
  \bibinfo{volume}{326}, \bibinfo{pages}{463--472}.
\newblock \DOIprefix\doi{10.1046/j.1365-8711.2001.04457.x}.
\bibitem[{Sheth et~al.(2001b)Sheth, Hui, Diaferio and Scoccimarro}]{Sheth2001a}
\bibinfo{author}{Sheth, R.K.}, \bibinfo{author}{Hui, L.},
  \bibinfo{author}{Diaferio, A.}, \bibinfo{author}{Scoccimarro, R.},
  \bibinfo{year}{2001}b.
\newblock \bibinfo{title}{Linear and non-linear contributions to pairwise
  peculiar velocities}.
\newblock \bibinfo{journal}{Monthly Notices of the Royal Astronomical Society}
  \bibinfo{volume}{325}, \bibinfo{pages}{1288--1302}.
\newblock \DOIprefix\doi{10.1046/j.1365-8711.2001.04222.x}.
\bibitem[{Sheth and Tormen(1999)}]{Sheth1999}
\bibinfo{author}{Sheth, R.K.}, \bibinfo{author}{Tormen, G.},
  \bibinfo{year}{1999}.
\newblock \bibinfo{title}{Large scale bias and the peak background split}.
\newblock \bibinfo{journal}{Monthly Notices of the Royal Astronomical Society}
  \bibinfo{volume}{308}, \bibinfo{pages}{119--126}.
\newblock \DOIprefix\doi{10.1046/j.1365-8711.1999.02692.x}.
\bibitem[{Simon(2007)}]{Simon2007}
\bibinfo{author}{Simon, P.}, \bibinfo{year}{2007}.
\newblock \bibinfo{title}{How accurate is {{Limber}}'s equation?}
\newblock \bibinfo{journal}{Astronomy \& Astrophysics} \bibinfo{volume}{473},
  \bibinfo{pages}{711--714}.
\newblock \DOIprefix\doi{10.1051/0004-6361:20066352}.
\bibitem[{a.~Skibba et~al.(2015)a.~Skibba, Coil, Mendez, Blanton, Bray, Cool,
  Eisenstein, Guo, Miyaji, Moustakas and Zhu}]{Skibba2015}
\bibinfo{author}{a.~Skibba, R.}, \bibinfo{author}{Coil, A.L.},
  \bibinfo{author}{Mendez, A.J.}, \bibinfo{author}{Blanton, M.R.},
  \bibinfo{author}{Bray, A.D.}, \bibinfo{author}{Cool, R.J.},
  \bibinfo{author}{Eisenstein, D.J.}, \bibinfo{author}{Guo, H.},
  \bibinfo{author}{Miyaji, T.}, \bibinfo{author}{Moustakas, J.},
  \bibinfo{author}{Zhu, G.}, \bibinfo{year}{2015}.
\newblock \bibinfo{title}{{{DARK MATTER HALO MODELS OF STELLAR
  MASS}}-{{DEPENDENT GALAXY CLUSTERING IN PRIMUS}}+{{DEEP2 AT}} 0.2 {$<$} z
  {$<$} 1.2}.
\newblock \bibinfo{journal}{The Astrophysical Journal} \bibinfo{volume}{807},
  \bibinfo{pages}{152--152}.
\newblock \DOIprefix\doi{10.1088/0004-637X/807/2/152}.
\bibitem[{a.~Skibba and Sheth(2009)}]{Skibba2009}
\bibinfo{author}{a.~Skibba, R.}, \bibinfo{author}{Sheth, R.K.},
  \bibinfo{year}{2009}.
\newblock \bibinfo{title}{A halo model of galaxy colours and clustering in the
  {{Sloan Digital Sky Survey}}}.
\newblock \bibinfo{journal}{Monthly Notices of the Royal Astronomical Society}
  \bibinfo{volume}{392}, \bibinfo{pages}{1080--1091}.
\newblock \DOIprefix\doi{10.1111/j.1365-2966.2008.14007.x}.
\bibitem[{Smith et~al.(2011)Smith, Desjacques and Marian}]{Smith2011}
\bibinfo{author}{Smith, R.E.}, \bibinfo{author}{Desjacques, V.},
  \bibinfo{author}{Marian, L.}, \bibinfo{year}{2011}.
\newblock \bibinfo{title}{Nonlinear clustering in models with primordial
  non-{{Gaussianity}}: {{The}} halo model approach}.
\newblock \bibinfo{journal}{Physical Review D} \bibinfo{volume}{83},
  \bibinfo{pages}{043526--043526}.
\newblock \DOIprefix\doi{10.1103/PhysRevD.83.043526}.
\bibitem[{Smith and Markovi{\v c}(2011)}]{Smith2011a}
\bibinfo{author}{Smith, R.E.}, \bibinfo{author}{Markovi{\v c}, K.},
  \bibinfo{year}{2011}.
\newblock \bibinfo{title}{Testing the warm dark matter paradigm with
  large-scale structures}.
\newblock \bibinfo{journal}{Physical Review D} \bibinfo{volume}{4},
  \bibinfo{pages}{1--18}.
\bibitem[{Smith et~al.(2003)Smith, Peacock, Jenkins, White, Frenk, Pearce,
  Thomas, Efstathiou and Couchman}]{Smith2003}
\bibinfo{author}{Smith, R.E.}, \bibinfo{author}{Peacock, J.A.},
  \bibinfo{author}{Jenkins, A.}, \bibinfo{author}{White, S.D.M.},
  \bibinfo{author}{Frenk, C.S.}, \bibinfo{author}{Pearce, F.R.},
  \bibinfo{author}{Thomas, P.A.}, \bibinfo{author}{Efstathiou, G.},
  \bibinfo{author}{Couchman, H.M.P.}, \bibinfo{year}{2003}.
\newblock \bibinfo{title}{Stable clustering, the halo model and non-linear
  cosmological power spectra}.
\newblock \bibinfo{journal}{Monthly Notices of the Royal Astronomical Society}
  \bibinfo{volume}{341}, \bibinfo{pages}{1311--1332}.
\newblock \DOIprefix\doi{10.1046/j.1365-8711.2003.06503.x}.
\bibitem[{Sugiyama(1995)}]{Sugiyama1995}
\bibinfo{author}{Sugiyama, N.}, \bibinfo{year}{1995}.
\newblock \bibinfo{title}{Cosmic {{Background Anisotropies}} in {{Cold Dark
  Matter Cosmology}}}.
\newblock \bibinfo{journal}{The Astrophysical Journal Supplement Series}
  \bibinfo{volume}{100}, \bibinfo{pages}{281}.
\newblock \DOIprefix\doi{10.1086/192220}.
\bibitem[{Sunayama et~al.(2015)Sunayama, Hearin, Padmanabhan and
  Leauthaud}]{Sunayama2015}
\bibinfo{author}{Sunayama, T.}, \bibinfo{author}{Hearin, A.P.},
  \bibinfo{author}{Padmanabhan, N.}, \bibinfo{author}{Leauthaud, A.},
  \bibinfo{year}{2015}.
\newblock \bibinfo{title}{The {{Scale}}-{{Dependence}} of {{Halo Assembly
  Bias}}}.
\newblock \bibinfo{journal}{Arxiv e-prints} .
\bibitem[{Szapudi et~al.(2005)Szapudi, Pan, Prunet and
  Budav{\'a}ri}]{Szapudi2005}
\bibinfo{author}{Szapudi, I.}, \bibinfo{author}{Pan, J.},
  \bibinfo{author}{Prunet, S.}, \bibinfo{author}{Budav{\'a}ri, T.},
  \bibinfo{year}{2005}.
\newblock \bibinfo{title}{Fast {{Edge Corrected Measurement}} of the
  {{Two}}-{{Point Correlation Function}} and the {{Power Spectrum}}}.
\newblock \bibinfo{journal}{Arxiv e-prints} ,
  \bibinfo{pages}{4--4}\DOIprefix\doi{10.1086/496971}.
\bibitem[{Takahashi et~al.(2012)Takahashi, Sato, Nishimichi, Taruya and
  Oguri}]{Takahashi2012}
\bibinfo{author}{Takahashi, R.}, \bibinfo{author}{Sato, M.},
  \bibinfo{author}{Nishimichi, T.}, \bibinfo{author}{Taruya, A.},
  \bibinfo{author}{Oguri, M.}, \bibinfo{year}{2012}.
\newblock \bibinfo{title}{{{REVISING THE HALOFIT MODEL FOR THE NONLINEAR MATTER
  POWER SPECTRUM}}}.
\newblock \bibinfo{journal}{The Astrophysical Journal} \bibinfo{volume}{761},
  \bibinfo{pages}{152--152}.
\newblock \DOIprefix\doi{10.1088/0004-637X/761/2/152}.
\bibitem[{Taylor(2011)}]{Taylor2011}
\bibinfo{author}{Taylor, J.E.}, \bibinfo{year}{2011}.
\newblock \bibinfo{title}{Dark {{Matter Halos}} from the {{Inside Out}}}.
\newblock \bibinfo{journal}{Advances in Astronomy} \bibinfo{volume}{2011},
  \bibinfo{pages}{1--17}.
\newblock \DOIprefix\doi{10.1155/2011/604898}.
\bibitem[{Tinker et~al.(2008)Tinker, Kravtsov, Klypin, Abazajian, Warren,
  Yepes, Gottl{\"o}ber and Holz}]{Tinker2008}
\bibinfo{author}{Tinker, J.L.}, \bibinfo{author}{Kravtsov, A.V.},
  \bibinfo{author}{Klypin, A.}, \bibinfo{author}{Abazajian, K.N.},
  \bibinfo{author}{Warren, M.S.}, \bibinfo{author}{Yepes, G.},
  \bibinfo{author}{Gottl{\"o}ber, S.}, \bibinfo{author}{Holz, D.E.},
  \bibinfo{year}{2008}.
\newblock \bibinfo{title}{Toward a halo mass function for precision cosmology:
  The limits of universality}.
\newblock \bibinfo{journal}{The Astrophysical Journal} \bibinfo{volume}{688},
  \bibinfo{pages}{709--728}.
\bibitem[{Tinker et~al.(2010)Tinker, Robertson, Kravtsov, Klypin, Warren, Yepes
  and Gottl{\"o}ber}]{Tinker2010}
\bibinfo{author}{Tinker, J.L.}, \bibinfo{author}{Robertson, B.E.},
  \bibinfo{author}{Kravtsov, A.V.}, \bibinfo{author}{Klypin, A.},
  \bibinfo{author}{Warren, M.S.}, \bibinfo{author}{Yepes, G.},
  \bibinfo{author}{Gottl{\"o}ber, S.}, \bibinfo{year}{2010}.
\newblock \bibinfo{title}{The large-scale bias of dark matter halos: Numerical
  calibration and model tests}.
\newblock \bibinfo{journal}{The Astrophysical Journal} \bibinfo{volume}{724},
  \bibinfo{pages}{878--886}.
\newblock \DOIprefix\doi{10.1088/0004-637X/724/2/878}.
\bibitem[{Tinker and Weinberg(2005)}]{Tinker2005}
\bibinfo{author}{Tinker, J.L.}, \bibinfo{author}{Weinberg, D.},
  \bibinfo{year}{2005}.
\newblock \bibinfo{title}{On the mass-to-light ratio of large-scale structure}.
\newblock \bibinfo{journal}{The Astrophysical Journal} \bibinfo{volume}{631},
  \bibinfo{pages}{41--58}.
\bibitem[{To et~al.(2020)To, Krause, Rozo, Wu, Gruen, Wechsler, Eifler, Rykoff,
  Costanzi, Becker, Bernstein, Blazek, Bocquet, Bridle, Cawthon, Choi, Crocce,
  Davis, DeRose, {Drlica-Wagner}, {Elvin-Poole}, Fang, Farahi, Friedrich,
  Gatti, Gaztanaga, Giannantonio, Hartley, Hoyle, Jarvis, MacCrann, McClintock,
  Miranda, Pereira, Park, Porredon, Prat, Rau, Ross, Samuroff, S{\'a}nchez,
  {Sevilla-Noarbe}, Sheldon, Troxel, Varga, Vielzeuf, Zhang, Zuntz, Abbott,
  Aguena, Annis, Avila, Bertin, Bhargava, Brooks, Burke, Carnero~Rosell,
  Carrasco~Kind, Carretero, Chang, Conselice, {da Costa}, Davis, Desai, Diehl,
  Dietrich, Everett, Evrard, Ferrero, Flaugher, Fosalba, Frieman,
  {Garc{\'i}a-Bellido}, Gruendl, Gutierrez, Hinton, Hollowood, Huterer, James,
  Jeltema, Kron, Kuehn, Kuropatkin, Lima, Maia, Marshall, Menanteau, Miquel,
  Morgan, Muir, Myles, Palmese, {Paz-Chinch{\'o}n}, Plazas, Romer, Roodman,
  Sanchez, Santiago, Scarpine, Serrano, Smith, Suchyta, Swanson, Tarle, Thomas,
  Tucker, Weller and Wester}]{To2020}
\bibinfo{author}{To, C.}, \bibinfo{author}{Krause, E.}, \bibinfo{author}{Rozo,
  E.}, \bibinfo{author}{Wu, H.}, \bibinfo{author}{Gruen, D.},
  \bibinfo{author}{Wechsler, R.H.}, \bibinfo{author}{Eifler, T.F.},
  \bibinfo{author}{Rykoff, E.S.}, \bibinfo{author}{Costanzi, M.},
  \bibinfo{author}{Becker, M.R.}, \bibinfo{author}{Bernstein, G.M.},
  \bibinfo{author}{Blazek, J.}, \bibinfo{author}{Bocquet, S.},
  \bibinfo{author}{Bridle, S.L.}, \bibinfo{author}{Cawthon, R.},
  \bibinfo{author}{Choi, A.}, \bibinfo{author}{Crocce, M.},
  \bibinfo{author}{Davis, C.}, \bibinfo{author}{DeRose, J.},
  \bibinfo{author}{{Drlica-Wagner}, A.}, \bibinfo{author}{{Elvin-Poole}, J.},
  \bibinfo{author}{Fang, X.}, \bibinfo{author}{Farahi, A.},
  \bibinfo{author}{Friedrich, O.}, \bibinfo{author}{Gatti, M.},
  \bibinfo{author}{Gaztanaga, E.}, \bibinfo{author}{Giannantonio, T.},
  \bibinfo{author}{Hartley, W.G.}, \bibinfo{author}{Hoyle, B.},
  \bibinfo{author}{Jarvis, M.}, \bibinfo{author}{MacCrann, N.},
  \bibinfo{author}{McClintock, T.}, \bibinfo{author}{Miranda, V.},
  \bibinfo{author}{Pereira, M.E.S.}, \bibinfo{author}{Park, Y.},
  \bibinfo{author}{Porredon, A.}, \bibinfo{author}{Prat, J.},
  \bibinfo{author}{Rau, M.M.}, \bibinfo{author}{Ross, A.J.},
  \bibinfo{author}{Samuroff, S.}, \bibinfo{author}{S{\'a}nchez, C.},
  \bibinfo{author}{{Sevilla-Noarbe}, I.}, \bibinfo{author}{Sheldon, E.},
  \bibinfo{author}{Troxel, M.A.}, \bibinfo{author}{Varga, T.N.},
  \bibinfo{author}{Vielzeuf, P.}, \bibinfo{author}{Zhang, Y.},
  \bibinfo{author}{Zuntz, J.}, \bibinfo{author}{Abbott, T.M.C.},
  \bibinfo{author}{Aguena, M.}, \bibinfo{author}{Annis, J.},
  \bibinfo{author}{Avila, S.}, \bibinfo{author}{Bertin, E.},
  \bibinfo{author}{Bhargava, S.}, \bibinfo{author}{Brooks, D.},
  \bibinfo{author}{Burke, D.L.}, \bibinfo{author}{Carnero~Rosell, A.},
  \bibinfo{author}{Carrasco~Kind, M.}, \bibinfo{author}{Carretero, J.},
  \bibinfo{author}{Chang, C.}, \bibinfo{author}{Conselice, C.},
  \bibinfo{author}{{da Costa}, L.N.}, \bibinfo{author}{Davis, T.M.},
  \bibinfo{author}{Desai, S.}, \bibinfo{author}{Diehl, H.T.},
  \bibinfo{author}{Dietrich, J.P.}, \bibinfo{author}{Everett, S.},
  \bibinfo{author}{Evrard, A.E.}, \bibinfo{author}{Ferrero, I.},
  \bibinfo{author}{Flaugher, B.}, \bibinfo{author}{Fosalba, P.},
  \bibinfo{author}{Frieman, J.}, \bibinfo{author}{{Garc{\'i}a-Bellido}, J.},
  \bibinfo{author}{Gruendl, R.A.}, \bibinfo{author}{Gutierrez, G.},
  \bibinfo{author}{Hinton, S.R.}, \bibinfo{author}{Hollowood, D.L.},
  \bibinfo{author}{Huterer, D.}, \bibinfo{author}{James, D.J.},
  \bibinfo{author}{Jeltema, T.}, \bibinfo{author}{Kron, R.},
  \bibinfo{author}{Kuehn, K.}, \bibinfo{author}{Kuropatkin, N.},
  \bibinfo{author}{Lima, M.}, \bibinfo{author}{Maia, M.A.G.},
  \bibinfo{author}{Marshall, J.L.}, \bibinfo{author}{Menanteau, F.},
  \bibinfo{author}{Miquel, R.}, \bibinfo{author}{Morgan, R.},
  \bibinfo{author}{Muir, J.}, \bibinfo{author}{Myles, J.},
  \bibinfo{author}{Palmese, A.}, \bibinfo{author}{{Paz-Chinch{\'o}n}, F.},
  \bibinfo{author}{Plazas, A.A.}, \bibinfo{author}{Romer, A.K.},
  \bibinfo{author}{Roodman, A.}, \bibinfo{author}{Sanchez, E.},
  \bibinfo{author}{Santiago, B.}, \bibinfo{author}{Scarpine, V.},
  \bibinfo{author}{Serrano, S.}, \bibinfo{author}{Smith, M.},
  \bibinfo{author}{Suchyta, E.}, \bibinfo{author}{Swanson, M.E.C.},
  \bibinfo{author}{Tarle, G.}, \bibinfo{author}{Thomas, D.},
  \bibinfo{author}{Tucker, D.L.}, \bibinfo{author}{Weller, J.},
  \bibinfo{author}{Wester, W.}, \bibinfo{year}{2020}.
\newblock \bibinfo{title}{Dark {{Energy Survey Year}} 1 {{Results}}:
  {{Cosmological Constraints}} from {{Cluster Abundances}}, {{Weak Lensing}},
  and {{Galaxy Correlations}}}.
\newblock \bibinfo{journal}{arXiv e-prints} \bibinfo{volume}{2010},
  \bibinfo{pages}{arXiv:2010.01138}.
\bibitem[{To et~al.(2021)To, Krause, Rozo, Wu, Gruen, DeRose, Rykoff, Wechsler,
  Becker, Costanzi, Eifler, {da Silva Pereira}, Kokron and {DES
  Collaboration}}]{To2021}
\bibinfo{author}{To, C.H.}, \bibinfo{author}{Krause, E.},
  \bibinfo{author}{Rozo, E.}, \bibinfo{author}{Wu, H.Y.},
  \bibinfo{author}{Gruen, D.}, \bibinfo{author}{DeRose, J.},
  \bibinfo{author}{Rykoff, E.}, \bibinfo{author}{Wechsler, R.H.},
  \bibinfo{author}{Becker, M.}, \bibinfo{author}{Costanzi, M.},
  \bibinfo{author}{Eifler, T.}, \bibinfo{author}{{da Silva Pereira}, M.E.},
  \bibinfo{author}{Kokron, N.}, \bibinfo{author}{{DES Collaboration}},
  \bibinfo{year}{2021}.
\newblock \bibinfo{title}{Combination of cluster number counts and two-point
  correlations: Validation on mock {{Dark Energy Survey}}}.
\newblock \bibinfo{journal}{Monthly Notices of the Royal Astronomical Society}
  \bibinfo{volume}{502}, \bibinfo{pages}{4093--4111}.
\newblock \DOIprefix\doi{10.1093/mnras/stab239}.
\bibitem[{Umeh et~al.(2021)Umeh, Maartens, Padmanabhan and Camera}]{Umeh2021}
\bibinfo{author}{Umeh, O.}, \bibinfo{author}{Maartens, R.},
  \bibinfo{author}{Padmanabhan, H.}, \bibinfo{author}{Camera, S.},
  \bibinfo{year}{2021}.
\newblock \bibinfo{title}{The effect of finite halo size on the clustering of
  neutral hydrogen}.
\newblock \bibinfo{journal}{arXiv e-prints} \bibinfo{volume}{2102},
  \bibinfo{pages}{arXiv:2102.06116}.
\bibitem[{{van der Walt} et~al.(2011){van der Walt}, Colbert and
  Varoquaux}]{vanderWalt2011}
\bibinfo{author}{{van der Walt}, S.}, \bibinfo{author}{Colbert, S.C.},
  \bibinfo{author}{Varoquaux, G.}, \bibinfo{year}{2011}.
\newblock \bibinfo{title}{The {{NumPy Array}}: {{A Structure}} for {{Efficient
  Numerical Computation}}}.
\newblock \bibinfo{journal}{Computing in Science \& Engineering}
  \bibinfo{volume}{13}, \bibinfo{pages}{22--30}.
\newblock \DOIprefix\doi{10.1109/MCSE.2011.37}.
\bibitem[{Viel et~al.(2005)Viel, Lesgourgues, Haehnelt, Matarrese and
  Riotto}]{Viel2005}
\bibinfo{author}{Viel, M.}, \bibinfo{author}{Lesgourgues, J.},
  \bibinfo{author}{Haehnelt, M.G.}, \bibinfo{author}{Matarrese, S.},
  \bibinfo{author}{Riotto, A.}, \bibinfo{year}{2005}.
\newblock \bibinfo{title}{Constraining warm dark matter candidates including
  sterile neutrinos and light gravitinos with {{WMAP}} and the
  {{Lyman}}-{$\alpha$} forest}.
\newblock \bibinfo{journal}{Physical Review D} \bibinfo{volume}{71},
  \bibinfo{pages}{1--10}.
\newblock \DOIprefix\doi{10.1103/PhysRevD.71.063534}.
\bibitem[{Virtanen et~al.(2020)Virtanen, Gommers, Oliphant, Haberland, Reddy,
  Cournapeau, Burovski, Peterson, Weckesser, Bright, {van der Walt}, Brett,
  Wilson, Millman, Mayorov, Nelson, Jones, Kern, Larson, Carey, Polat, Feng,
  Moore, VanderPlas, Laxalde, Perktold, Cimrman, Henriksen, Quintero, Harris,
  Archibald, Ribeiro, Pedregosa and {van Mulbregt}}]{Virtanen2020}
\bibinfo{author}{Virtanen, P.}, \bibinfo{author}{Gommers, R.},
  \bibinfo{author}{Oliphant, T.E.}, \bibinfo{author}{Haberland, M.},
  \bibinfo{author}{Reddy, T.}, \bibinfo{author}{Cournapeau, D.},
  \bibinfo{author}{Burovski, E.}, \bibinfo{author}{Peterson, P.},
  \bibinfo{author}{Weckesser, W.}, \bibinfo{author}{Bright, J.},
  \bibinfo{author}{{van der Walt}, S.J.}, \bibinfo{author}{Brett, M.},
  \bibinfo{author}{Wilson, J.}, \bibinfo{author}{Millman, K.J.},
  \bibinfo{author}{Mayorov, N.}, \bibinfo{author}{Nelson, A.R.J.},
  \bibinfo{author}{Jones, E.}, \bibinfo{author}{Kern, R.},
  \bibinfo{author}{Larson, E.}, \bibinfo{author}{Carey, C.J.},
  \bibinfo{author}{Polat, {\.I}.}, \bibinfo{author}{Feng, Y.},
  \bibinfo{author}{Moore, E.W.}, \bibinfo{author}{VanderPlas, J.},
  \bibinfo{author}{Laxalde, D.}, \bibinfo{author}{Perktold, J.},
  \bibinfo{author}{Cimrman, R.}, \bibinfo{author}{Henriksen, I.},
  \bibinfo{author}{Quintero, E.A.}, \bibinfo{author}{Harris, C.R.},
  \bibinfo{author}{Archibald, A.M.}, \bibinfo{author}{Ribeiro, A.H.},
  \bibinfo{author}{Pedregosa, F.}, \bibinfo{author}{{van Mulbregt}, P.},
  \bibinfo{year}{2020}.
\newblock \bibinfo{title}{{{SciPy}} 1.0: Fundamental algorithms for scientific
  computing in {{Python}}}.
\newblock \bibinfo{journal}{Nature Methods} \bibinfo{volume}{17},
  \bibinfo{pages}{261--272}.
\newblock \DOIprefix\doi{10.1038/s41592-019-0686-2}.
\bibitem[{Wake et~al.(2008)Wake, Croom, Sadler and Johnston}]{Wake2008a}
\bibinfo{author}{Wake, D.A.}, \bibinfo{author}{Croom, S.M.},
  \bibinfo{author}{Sadler, E.M.}, \bibinfo{author}{Johnston, H.M.},
  \bibinfo{year}{2008}.
\newblock \bibinfo{title}{The clustering of radio galaxies at z = 0.55 from the
  {{2SLAQ LRG}} survey}.
\newblock \bibinfo{journal}{Monthly Notices of the Royal Astronomical Society}
  \bibinfo{volume}{391}, \bibinfo{pages}{1674--1684}.
\newblock \DOIprefix\doi{10.1111/j.1365-2966.2008.14039.x}.
\bibitem[{Wang et~al.(2020)Wang, Mao, Zentner, Lange, {van den Bosch} and
  Wechsler}]{Wang2020}
\bibinfo{author}{Wang, K.}, \bibinfo{author}{Mao, Y.Y.},
  \bibinfo{author}{Zentner, A.R.}, \bibinfo{author}{Lange, J.U.},
  \bibinfo{author}{{van den Bosch}, F.C.}, \bibinfo{author}{Wechsler, R.H.},
  \bibinfo{year}{2020}.
\newblock \bibinfo{title}{Concentrations of {{Dark Haloes Emerge}} from {{Their
  Merger Histories}}}.
\newblock \bibinfo{journal}{arXiv e-prints} \bibinfo{volume}{2004},
  \bibinfo{pages}{arXiv:2004.13732}.
\bibitem[{Watson et~al.(2013)Watson, Iliev, D'Aloisio, Knebe, Shapiro and
  Yepes}]{Watson2013}
\bibinfo{author}{Watson, W.A.}, \bibinfo{author}{Iliev, I.T.},
  \bibinfo{author}{D'Aloisio, A.}, \bibinfo{author}{Knebe, A.},
  \bibinfo{author}{Shapiro, P.R.}, \bibinfo{author}{Yepes, G.},
  \bibinfo{year}{2013}.
\newblock \bibinfo{title}{The halo mass function through the cosmic ages}.
\newblock \bibinfo{journal}{Monthly Notices of the Royal Astronomical Society}
  \bibinfo{volume}{433}, \bibinfo{pages}{1230--1245}.
\newblock \DOIprefix\doi{10.1093/mnras/stt791}.
\bibitem[{Wechsler et~al.(2002)Wechsler, Bullock, Primack, Kravtsov and
  Dekel}]{Wechsler2002}
\bibinfo{author}{Wechsler, R.H.}, \bibinfo{author}{Bullock, J.S.},
  \bibinfo{author}{Primack, J.R.}, \bibinfo{author}{Kravtsov, A.V.},
  \bibinfo{author}{Dekel, A.}, \bibinfo{year}{2002}.
\newblock \bibinfo{title}{Concentrations of {{Dark Halos}} from {{Their
  Assembly Histories}}}.
\newblock \bibinfo{journal}{The Astrophysical Journal} \bibinfo{volume}{568},
  \bibinfo{pages}{52--70}.
\newblock \DOIprefix\doi{10.1086/338765}.
\bibitem[{White et~al.(2001)White, Hernquist and Springel}]{White2001}
\bibinfo{author}{White, M.}, \bibinfo{author}{Hernquist, L.},
  \bibinfo{author}{Springel, V.}, \bibinfo{year}{2001}.
\newblock \bibinfo{title}{The {{Halo Model}} and {{Numerical Simulations}}}.
\newblock \bibinfo{journal}{The Astrophysical Journal} \bibinfo{volume}{550},
  \bibinfo{pages}{L129--L132}.
\newblock \DOIprefix\doi{10.1086/319644}.
\bibitem[{Wolz et~al.(2019)Wolz, Murray, Blake and Wyithe}]{Wolz2019}
\bibinfo{author}{Wolz, L.}, \bibinfo{author}{Murray, S.G.},
  \bibinfo{author}{Blake, C.}, \bibinfo{author}{Wyithe, J.S.},
  \bibinfo{year}{2019}.
\newblock \bibinfo{title}{Intensity mapping cross-correlations {{II}}: {{HI}}
  halo models including shot noise}.
\newblock \bibinfo{journal}{Monthly Notices of the Royal Astronomical Society}
  \bibinfo{volume}{484}, \bibinfo{pages}{1007}.
\newblock \DOIprefix\doi{10.1093/mnras/sty3142}.
\bibitem[{Zaldarriaga and Seljak(2000)}]{Zaldarriaga2000}
\bibinfo{author}{Zaldarriaga, M.}, \bibinfo{author}{Seljak, U.},
  \bibinfo{year}{2000}.
\newblock \bibinfo{title}{{{CMBFAST}} for {{Spatially Closed Universes}}}.
\newblock \bibinfo{journal}{The Astrophysical Journal} \bibinfo{volume}{129},
  \bibinfo{pages}{431--434}.
\bibitem[{Zehavi et~al.(2011)Zehavi, Zheng, Weinberg, Blanton, a.~Bahcall,
  a.~Berlind, Brinkmann, a.~Frieman, Gunn, Lupton, Nichol, Percival, Schneider,
  a.~Skibba, a.~Strauss, Tegmark and York}]{Zehavi2011}
\bibinfo{author}{Zehavi, I.}, \bibinfo{author}{Zheng, Z.},
  \bibinfo{author}{Weinberg, D.H.}, \bibinfo{author}{Blanton, M.R.},
  \bibinfo{author}{a.~Bahcall, N.}, \bibinfo{author}{a.~Berlind, A.},
  \bibinfo{author}{Brinkmann, J.}, \bibinfo{author}{a.~Frieman, J.},
  \bibinfo{author}{Gunn, J.E.}, \bibinfo{author}{Lupton, R.H.},
  \bibinfo{author}{Nichol, R.C.}, \bibinfo{author}{Percival, W.J.},
  \bibinfo{author}{Schneider, D.P.}, \bibinfo{author}{a.~Skibba, R.},
  \bibinfo{author}{a.~Strauss, M.}, \bibinfo{author}{Tegmark, M.},
  \bibinfo{author}{York, D.G.}, \bibinfo{year}{2011}.
\newblock \bibinfo{title}{Galaxy {{Clustering}} in the {{Completed Sdss
  Redshift Survey}}: The {{Dependence}} on {{Color}} and {{Luminosity}}}.
\newblock \bibinfo{journal}{The Astrophysical Journal} \bibinfo{volume}{736},
  \bibinfo{pages}{59--59}.
\newblock \DOIprefix\doi{10.1088/0004-637X/736/1/59}.
\bibitem[{Zehavi et~al.(2005)Zehavi, Zheng, Weinberg, Frieman, Berlind,
  Blanton, Scoccimarro, Sheth, Strauss, Kayo, Suto, Fukugita, Nakamura,
  Bahcall, Brinkmann, Gunn, Hennessy, Ivezi{\'c}, Knapp, Loveday, Meiksin,
  Schlegel, Schneider, Szapudi, Tegmark, Vogeley and York}]{Zehavi2005}
\bibinfo{author}{Zehavi, I.}, \bibinfo{author}{Zheng, Z.},
  \bibinfo{author}{Weinberg, D.H.}, \bibinfo{author}{Frieman, J.A.},
  \bibinfo{author}{Berlind, A.A.}, \bibinfo{author}{Blanton, M.R.},
  \bibinfo{author}{Scoccimarro, R.}, \bibinfo{author}{Sheth, R.K.},
  \bibinfo{author}{Strauss, M.A.}, \bibinfo{author}{Kayo, I.},
  \bibinfo{author}{Suto, Y.}, \bibinfo{author}{Fukugita, M.},
  \bibinfo{author}{Nakamura, O.}, \bibinfo{author}{Bahcall, N.A.},
  \bibinfo{author}{Brinkmann, J.}, \bibinfo{author}{Gunn, J.E.},
  \bibinfo{author}{Hennessy, G.S.}, \bibinfo{author}{Ivezi{\'c}, {\v Z}.},
  \bibinfo{author}{Knapp, G.R.}, \bibinfo{author}{Loveday, J.},
  \bibinfo{author}{Meiksin, A.}, \bibinfo{author}{Schlegel, D.J.},
  \bibinfo{author}{Schneider, D.P.}, \bibinfo{author}{Szapudi, I.},
  \bibinfo{author}{Tegmark, M.}, \bibinfo{author}{Vogeley, M.S.},
  \bibinfo{author}{York, D.G.}, \bibinfo{year}{2005}.
\newblock \bibinfo{title}{The {{Luminosity}} and {{Color Dependence}} of the
  {{Galaxy Correlation Function}}}.
\newblock \bibinfo{journal}{The Astrophysical Journal} \bibinfo{volume}{630},
  \bibinfo{pages}{1--27}.
\newblock \DOIprefix\doi{10.1086/431891}.
\bibitem[{Zemp et~al.(2011)Zemp, Gnedin, Gnedin and Kravtsov}]{Zemp2011}
\bibinfo{author}{Zemp, M.}, \bibinfo{author}{Gnedin, O.Y.},
  \bibinfo{author}{Gnedin, N.Y.}, \bibinfo{author}{Kravtsov, A.V.},
  \bibinfo{year}{2011}.
\newblock \bibinfo{title}{On {{Determining The Shape Of Matter
  Distributions}}}.
\newblock \bibinfo{journal}{The Astrophysical Journal Supplement Series}
  \bibinfo{volume}{197}, \bibinfo{pages}{30--30}.
\newblock \DOIprefix\doi{10.1088/0067-0049/197/2/30}.
\bibitem[{Zentner(2007)}]{Zentner2007}
\bibinfo{author}{Zentner, A.R.}, \bibinfo{year}{2007}.
\newblock \bibinfo{title}{The {{Excursion Set Theory}} of {{Halo Mass
  Functions}}, {{Halo Clustering}}, and {{Halo Growth}}}.
\newblock \bibinfo{journal}{International Journal of Modern Physics D}
  \bibinfo{volume}{16}, \bibinfo{pages}{763--815}.
\newblock \DOIprefix\doi{10.1142/S0218271807010511}.
\bibitem[{Zhao et~al.(2009)Zhao, Jing, Mo and B{\"o}rner}]{Zhao2009}
\bibinfo{author}{Zhao, D.H.}, \bibinfo{author}{Jing, Y.P.},
  \bibinfo{author}{Mo, H.J.}, \bibinfo{author}{B{\"o}rner, G.},
  \bibinfo{year}{2009}.
\newblock \bibinfo{title}{{{ACCURATE UNIVERSAL MODELS FOR THE MASS ACCRETION
  HISTORIES AND CONCENTRATIONS OF DARK MATTER HALOS}}}.
\newblock \bibinfo{journal}{The Astrophysical Journal} \bibinfo{volume}{707},
  \bibinfo{pages}{354--369}.
\newblock \DOIprefix\doi{10.1088/0004-637X/707/1/354}.
\bibitem[{Zhao et~al.(2003)Zhao, Mo, Jing and Borner}]{Zhao2003}
\bibinfo{author}{Zhao, D.H.}, \bibinfo{author}{Mo, H.J.},
  \bibinfo{author}{Jing, Y.P.}, \bibinfo{author}{Borner, G.},
  \bibinfo{year}{2003}.
\newblock \bibinfo{title}{The growth and structure of dark matter haloes}.
\newblock \bibinfo{journal}{Monthly Notices of the Royal Astronomical Society}
  \bibinfo{volume}{339}, \bibinfo{pages}{12--24}.
\newblock \DOIprefix\doi{10.1046/j.1365-8711.2003.06135.x}.
\bibitem[{Zheng(2004)}]{Zheng2004}
\bibinfo{author}{Zheng, Z.}, \bibinfo{year}{2004}.
\newblock \bibinfo{title}{Interpreting the {{Observed Clustering}} of {{Red
  Galaxies}} at z\textasciitilde 3}.
\newblock \bibinfo{journal}{The Astrophysical Journal} \bibinfo{volume}{610},
  \bibinfo{pages}{61--68}.
\newblock \DOIprefix\doi{10.1086/421542}.
\bibitem[{Zheng et~al.(2005)Zheng, Berlind, Weinberg, Benson, Baugh, Frenk,
  Katz, Lacey, Cole and Dave}]{Zheng2005}
\bibinfo{author}{Zheng, Z.}, \bibinfo{author}{Berlind, A.A.},
  \bibinfo{author}{Weinberg, D.H.}, \bibinfo{author}{Benson, A.J.},
  \bibinfo{author}{Baugh, C.M.}, \bibinfo{author}{Frenk, C.S.},
  \bibinfo{author}{Katz, N.}, \bibinfo{author}{Lacey, C.G.},
  \bibinfo{author}{Cole, S.}, \bibinfo{author}{Dave, R.}, \bibinfo{year}{2005}.
\newblock \bibinfo{title}{Theoretical {{Models}} of the {{Halo Occupation
  Distribution}}: {{Separating Central}} and {{Satellite Galaxies}}}.
\bibitem[{Zheng and Guo(2015)}]{Zheng2015}
\bibinfo{author}{Zheng, Z.}, \bibinfo{author}{Guo, H.}, \bibinfo{year}{2015}.
\newblock \bibinfo{title}{Accurate and {{Efficient Halo}}-based {{Galaxy
  Clustering Modelling}} with {{Simulations}}}.
\newblock \bibinfo{journal}{Arxiv e-prints} .
\bibitem[{Zhou et~al.(2021)Zhou, Newman, Mao, Meisner, Moustakas, Myers,
  Prakash, Zentner, Brooks, Duan, Landriau, Levi, Prada and Tarle}]{Zhou2021}
\bibinfo{author}{Zhou, R.}, \bibinfo{author}{Newman, J.A.},
  \bibinfo{author}{Mao, Y.Y.}, \bibinfo{author}{Meisner, A.},
  \bibinfo{author}{Moustakas, J.}, \bibinfo{author}{Myers, A.D.},
  \bibinfo{author}{Prakash, A.}, \bibinfo{author}{Zentner, A.R.},
  \bibinfo{author}{Brooks, D.}, \bibinfo{author}{Duan, Y.},
  \bibinfo{author}{Landriau, M.}, \bibinfo{author}{Levi, M.E.},
  \bibinfo{author}{Prada, F.}, \bibinfo{author}{Tarle, G.},
  \bibinfo{year}{2021}.
\newblock \bibinfo{title}{The clustering of {{DESI}}-like luminous red galaxies
  using photometric redshifts}.
\newblock \bibinfo{journal}{Monthly Notices of the Royal Astronomical Society}
  \bibinfo{volume}{501}, \bibinfo{pages}{3309--3331}.
\newblock \DOIprefix\doi{10.1093/mnras/staa3764}.

\end{thebibliography}

\appendix

\section{\textsc{halomod}'s Caching System}
\label{app:caching}

The \textsc{hmf} package has always included some form of caching, whereby derived quantities are stored for later retrieval, and only updated when absolutely necessary. At the time of its initial publication, \textsc{hmf} relied on each parameter being hard-coded into an \verb|update| method in a specific way so as to trigger re-calculation of all child properties. This required great care when modifying the code, and especially when adding new parameters, as breaking the chain of dependencies could cause subtle errors. 

We have since updated the caching system of \textsc{hmf} to be much more robust. We use \python's function decorators, a kind of functional programming technology, to abstract the caching logic from the main \framework. There is now a specific module containing two methods and a single base class which contain all the machinery for the cache system. This increases the readability of the main code, and maximises the usage of the DRY (Don't Repeat Yourself) principle. 

The caching system makes heavy use of the general distinction between a \parameter\ and \cached. All parameters receive a \verb|@parameter| decorator, which uses function closure to modify its behaviour. Likewise, all quantities receive a \verb|@cached_quantity| decorator.
Every time a \parameter\ or \cached\ is accessed, though it may seem that a simple attribute is being returned, an entire wrapping function is called (though it is built to add very little overhead).

The \verb|@cached_property| decorator performs several operations each time its function is called. Firstly, it is useful to understand that four private tables are stored on the \framework. There are two dictionaries which map parameter names to quantities they affect and vice versa (call these \verb|par_prop| and \verb|prop_par| respectively), a dynamic list of quantities that are \textit{currently being evaluated} called \verb|activeq| (there can be more than one because quantities can be nested), and a simple dictionary mapping quantities to a boolean value which specifies whether it needs to be recalculated, called \verb|recalc|.

The first thing required when accessing a quantity is to update any other active quantities' \verb|prop_par| dictionaries with its own parameter dependencies. This will only do anything if this particular quantity has already been calculated and cached (and thus we know its parameter dependencies). 
Following this, if the quantity is in \verb|recalc| and it doesn't need to be recalculated, just return the cached value.
Otherwise, if it's in \verb|recalc| and \textit{does} need to be recalculated, call the the evaluation function, cache it, switch \verb|recalc| to false for this parameter, and return.
Finally, if it's not in \verb|recalc| at all, this must be the first time we've ever tried to access this quantity. In this case, we first add it to the \verb|activeq|, so that any parameters used while trying to evaluate this quantity will be added to its dependents (in \verb|prop_par|). 
We then evaluate the function and cache it, then invert the (now full) \verb|prop_par| dictionary to obtain \verb|par_prop|. We then remove it from the \verb|activeq| and set its value in \verb|recalc| to false so it isn't recalculated next time, and then return the value.

The missing ingredient to this is how accessing the \texttt{parameter}s adds itself to the \verb|prop_par| of all quantities in \verb|activeq|. This is of course achieved
by all \texttt{parameter}s also being \python descriptors (i.e. decorated), and it is simple for the \parameter\ to iterate through the \verb|activeq| list and add itself to each quantity's \verb|prop_par| entry.

This method is as efficient as possible for such a generalised process. No re-calculations are performed unless necessary. However, higher efficiency could be achieved by simply using low-level functions chained together in an appropriate way, storing only the variables that needed storing for a particular application. This is due to the extra memory usage required to store all variables in the caching system, and the minimal overhead of re-checking the cache. 

Importantly, this process is completely transparent to the user, who merely accesses quantities as if they were pre-stored variables. Furthermore, it is simple for the developer, who merely needs to attach a \verb|@parameter| or \verb|@cached_quantity| to any new parameters or quantities they define. 
\section{Hankel Transform}
 \label{app:hankel}

The transformation from 3D power spectrum to real-space correlation function is given by \cref{eq:hankel}, and is a zeroth-order spherical Hankel transformation.
It is necessary to perform this transformation to calculate the two-halo correlations, and also for the one-halo term if an analytic form for the self-convolution of the density profile does not exist. 

The transformation poses a challenge \citep[eg.][]{Diemer2018}, as it is highly oscillatory in log-space, in which it is necessary to perform the integration.

\verb|halomod| largely avoids this issue by using a technique outlined in \cite{Szapudi2005}, developed in \cite{Ogata2005}, and implemented in \cite{Murray2019}. 
The technique uses a quadrature rule based on the roots of the Bessel function and a double-exponential transformation. 

The general rule is defined as
\begin{equation}
	\label{eq:hankeltrans}
	\int_0^\infty f(x)J_\nu(x) {\rm d}x \approx \pi \sum_{k=1}^\infty w_{\nu k} f(y_{\nu k})J_\nu(y_{\nu k})\psi'(hr_{\nu k}),
\end{equation}
where $h$ is an integration step-size (similar to the trapezoidal rule) and
\begin{eqnarray}
	y_{\nu k} &=& \pi \psi(hr_{\nu k})/h \\
	\psi(t) &=& t\tanh(\pi \sinh(t)/2) \\
	\psi'(t) &=& \frac{\pi t \cosh(t) + \sinh(\pi \sinh(t))}{ 1 + \cosh(\pi \sinh(t))} \\
	w_{\nu k} &=& \frac{Y_\nu(\pi r_{\nu k})}{J_{\nu+1}(\pi r_{\nu k})}.
\end{eqnarray}
Here $J_\nu(x)$ represents a Bessel function of the first kind of order $\nu$, and $Y_\nu(x)$ a Bessel function of the second kind. Lastly, $r_{\nu k}$ are the roots of the Bessel function $J_\nu$, divided by $\pi$ (alternatively, the roots of $J_\nu(\pi x)$).

In the case of the zeroth-order spherical Hankel transform, we may make some simplifications. We firstly take \cref{eq:hankel} and make the substitution $z=kr$:
\begin{equation}
	\label{eq:hankelsub}
	\xi(r) = \frac{1}{2\pi^2r^3}\int_0^\infty P(z/r)z^2j_0(z){\rm d}z,
\end{equation} 
where $j_0$ is the spherical Bessel function of order zero. We also note that the spherical Bessel functions are related to the standard functions by
\begin{equation}
	\label{eq:sphbessel}
	j_\nu(x) = \sqrt{\frac{\pi}{2x}}J_{\nu+1/2}(x),
\end{equation}
so we may update \cref{eq:hankelsub} to
\begin{equation}
	\label{eq:finalhankel}
	\xi(r) = \frac{1}{2\pi^2r^3}\int_0^\infty P(z/r) z^2\sqrt{\frac{\pi}{2z}}J_{1/2}(z){\rm d}z.
\end{equation}
\cref{eq:finalhankel} is in the form required to implement the method of \cite{Ogata2005}, with $\nu=1/2$. In this case, the zeros of the function, $r_{\nu k} = 1,2,3...$. 

We may make one further optimization by noticing that 
\begin{equation}
	\label{eq:weq1}
	w_{1/2 k} = \frac{y_0(x_k)}{j_1(x_k)} = \frac{-\cos(x_k)/x_k}{\sin(x_k)/x_k^2 - \cos(x_k)/x_k} = 1,
\end{equation}
where the last equality holds because $x_k = \pi r_{1/2,k}$ are defined as the roots of $\sin(x)$.
Thus we finally arrive at the equation
\begin{equation}
	\label{eq:hankel_final_identity}
	\xi(r) \approx \pi \sum_{k=1}^\infty f(y_{1/2, k})J_{1/2}(y_{1/2, k})\psi'(hr_{1/2, k}),
\end{equation}
with
\begin{equation}
	\label{eq:f}
	f(x) = P(x/r)x^2\sqrt{\frac{\pi}{2x}}.
\end{equation}

In practice, the infinite sum is truncated at some finite value $N$. 
The benefit of this method, due to the double-exponential transformation, is that this $N$ can be quite small for a very accurate result. 
Since the values, $y_{\nu k}$, at which the sum is evaluated approach the zeros of the bessel function as $k \rightarrow 0$, \citet{Murray2019} found that a maximum of $N=3.2/h$ steps are required (for reasonably small $h$).

 The required value of $h$ is mostly dictated by the existence of information in the integrand at low $x$. As expected, a smaller value of $h$ will in general give more accurate results (modulo accumulated numerical addition error). The requisite value of $h$ will be different when transforming 
 different functions. 
 
 To avoid potential memory issues, we sum \cref{eq:hankel_final_identity} in chunks of 100 terms, checking for convergence after each chunk.
 We find in general that using $h=0.005$ for the $P(k) \rightarrow \xi(r)$ transformation is sufficient. 
 
 A consistent way to test the accuracy of the transform is to perform a full round-trip transform from $P(k) \rightarrow \xi(r) \rightarrow P(k)$ and check whether the result is close to the input.
 Since the transform uses values of $f(x)$ for $x$ as small as $\sim 10h$ and as large as $\sim 10/h$, corresponding to $k$ values between $(10 h / r_{\rm max}, 10/(h r_{\rm min})$ (and similarly for $r$),
 the vectorized $P(k_i)$ and $\xi(r_i)$ need to be extrapolated -- a procedure that must be done carefully, but will still yield some error for points outside the original domain.
 Thus the accuracy of the round-trip transform fundamentally depends on the choice of $r$-vector: $r_{\rm min}$, $r_{\rm max}$ and $d\log r$, as well as $h_k$ and $h_r$ (the resolution of the integration in the forward and backwards transforms). 
 We find that for the linear power spectrum, we achieve an error of less than 1\% over scales $k \in (10^{-2}, 10^4)$ using the parameters $r_{\rm min} = 10^{-3}$, $r_{\rm max} = 10^{3.5}$, $d\log r = 0.01$, $h_k = 10^{-4}$ and $h_r = 10^{-3}$.
 
 For scales $k < 10^{-2}$, the 2-halo term may be very well approximated by the linear power multiplied by the effective bias (\cref{eq:beff}), and we switch to that approximation on these scales (otherwise $r_{\rm max}$ would need to grow significantly, reducing performance).

\section{Projected Correlation Function Limits}
\label{app:proj}
The finite upper limit can be handled empirically, keeping in mind that the correlation function may become negative at large scales (to a small degree). 
To perform this convergence test, we cumulatively integrate the real-space correlation function at several scales $r_p$, determining where the contribution to the integral becomes less than a percent. 
While there is no continuous relation which describes this location as a function of $r_p$, we deduce the following stepwise prescription:
\begin{equation}
	r_{\rm max} = {\rm max}(80.5,5r_p).
\end{equation}
This prescription captures the idea that for small $r_p$, we do not want to integrate out to negative correlation, while for larger $r_p$, we must integrate well past the point at which the correlation becomes negative. In practice, measurements of the projected correlation function usually set a finite upper limit of $\sim 60 h^{-1}$Mpc, and we allow the user to set this arbitrarily if desired.

The finite lower limit has more difficult convergence properties. We make the substitution $y=r-r_p$ in \cref{eq:wprp} and multiply by $y$ to estimate the contribution of each (logarithmic) scale to the integral. We then proceed to perform an approximate analytic analysis of the convergence by assuming that as $y \rightarrow 0$, the correlation can be expressed as a power law $\xi(r) \approx r^{-a}$. This yields the following expression for the contribution of each scale $r$:
 \begin{equation}
 	\label{eq:projcorrcontr}
 	C(y,a,r_p) = \frac{y(y+r_p)^{(1-a)}}{\sqrt{y(y+2r_p)}}.
 \end{equation} 
This equation shows a characteristic shape, with a peak around the scale $r_p$. 
This indicates that the integral limit must be less than the scale of this peak.
In fact, we deduce the scale at which the contribution is 1\% of the value at the peak, to ensure sub-percent accuracy.

To do this, we solve ${\rm d} C(y,a,r_p)/{\rm d}y = 0$ for $y=r_{\rm peak}$, and calculate $C(r_{\rm peak},a,r_p) = C_{\rm max}$. The next step is to solve $C(y,a,r_p) = f_{\rm peak}C_{\rm max}$, where $f_{\rm peak}$ is the fraction of the peak value we wish to locate (we set this to 0.01 in our implementation). However this is not analytically solvable. We make the valid assumption that at the solution, $y \ll r_p$ (since the peak is around $r_p$). 
This enables us to solve the simple equation 
\begin{equation}
	\label{eq:projcorrsolve}
	r_p^{1-a}\sqrt{y_{\rm cut}/(2r_p)} = f_{\rm peak}C_{\rm max},
\end{equation}
which has the solution
\begin{equation}
	\label{eq:projcorrsolution}
	y_{\rm cut} = \theta(a)r_pf_{\rm peak}^2,
\end{equation}
where
\begin{align}
	\label{eq:theta}
	\theta(a) &= \frac{-2 a^3+\left(p+9\right) a^2-\left(3p+13\right) a+3p+7}{2(a-1)^2 \left[p+22(a-1)\right]\left(\frac{p+1}{a-1}\right)^{2 a}} \\
	p &= \sqrt{4 a^2-8 a+5}
\end{align}
$\theta(a)$ is a weakly but monotonically decreasing function, of the order $\approx 0.1$ in the range of $a$ we are interested in. In \verb|halomod|, $y_{\rm cut}$ is calculated for each $r_p$, ensuring accuracy at any scale.

\end{document}